\documentclass[journal]{new-aiaa}

\usepackage[utf8]{inputenc}
\usepackage{xcolor}
\usepackage{graphicx}
\usepackage{hhline,booktabs,amsmath,mathtools}
\usepackage[version=4]{mhchem}
\usepackage{siunitx}
\usepackage{longtable,tabularx}
 \usepackage{subfigure}
 \usepackage{subfigmat}
\usepackage{multirow}
\usepackage{multicol}

\usepackage{algorithm}
\usepackage{algpseudocode}
\usepackage{tikz}
\usetikzlibrary{positioning,shapes,arrows,decorations.pathreplacing}
\usetikzlibrary{shapes.geometric,fit,backgrounds,matrix,calc,patterns}
\usetikzlibrary{patterns,plotmarks}
\usepackage{pgfplots}
\pgfplotsset{compat=1.18}
\usepackage{pgfplotstable}
\usepackage{float}

\definecolor{myblue}{RGB}{31,119,180}
\definecolor{myorange}{RGB}{255,127,14}
\definecolor{mygreen}{RGB}{44,160,44}
\definecolor{myred}{RGB}{214,39,40}
\definecolor{mypurple}{RGB}{148,103,189}

\usepackage{rotating}
\usepackage{listings}
\usepackage{tcolorbox}
\tcbuselibrary{listings,skins,breakable}

\definecolor{codegreen}{rgb}{0,0.6,0}
\definecolor{codegray}{rgb}{0.5,0.5,0.5}
\definecolor{backcolour}{rgb}{0.97,0.97,0.97}
\definecolor{keywordblue}{rgb}{0.13,0.13,1}

\lstdefinestyle{cstyle}{
    backgroundcolor=\color{backcolour},   
    commentstyle=\color{codegreen},
    keywordstyle=\color{keywordblue}\bfseries,
    numberstyle=\tiny\color{codegray},
    basicstyle=\ttfamily\footnotesize,
    breaklines=true, numbers=left, numbersep=5pt, tabsize=2, language=C,
    morekeywords={PetscCall,PetscInt,PetscScalar,Mat,Vec,EPS,ST,PetscViewer}
}
\usepackage{noto}
\title{Kinetic Linear Stability Theory for High-Speed Compressible Flows: \\
A High Performance Computing Framework}

\author{Irmak T. ~Karpuzcu and Deborah A. Levin}
\affil{Aerospace Engineering, University of Illinois, Urbana-Champaign, IL 61801,USA}

\author{Vassilis Theofilis}
\affil{Faculty of Aerospace Engineering, Technion - Israel Institute of Technology, Haifa 32000, Israel}

\begin{document}

\maketitle
\section*{Abstract}

Shock waves in high-speed compressible flows contain finite-thickness, high-gradient regions where the continuum assumption becomes increasingly questionable and strong non-equilibrium effects occur such as the non-Maxwellian distribution of the micro-velocities.  Classical linear stability analyses of shocks are almost exclusively based on Navier--Stokes or moment-equation closures, and therefore cannot directly retain the translational non-equilibrium and bi-modal velocity distributions that develop inside the shock layer.  In this work, we develop and apply for the first time a kinetic linear stability theory (kLST) framework for one-dimensional normal shocks by linearizing the Boltzmann equation with the Bhatnagar--Gross--Krook (BGK) collision operator directly about kinetic BE-BGK base flows.  The perturbation problem is formulated in terms of the reduced distribution functions, while the macroscopic perturbations are recovered through velocity-space moments, so that the stability operator acts on the velocity distribution function rather than on a closed set of continuum variables.  

The formulation is first verified against compressible Couette-flow eigenvalue benchmarks under near-continuum conditions, showing excellent agreement and validating the kinetic matrix assembly.  The verified framework is then applied to argon normal shocks at $M_\infty=1.2$, $3.0$, and $4.0$, where the base flows are obtained from a BE-BGK solver and used directly in the stability operator.  For the low-Mach case, where the BE-BGK and Gilbarg--Paolucci profiles are nearly identical and the flow remains close to equilibrium, the computed spectra recover the expected stable continuous branches.  As the Mach number increases, the non-equilibrium velocity distribution inside the shock layer becomes much stronger, and comparing Maxwellian-equilibrium and non-equilibrium VDF based eigenspectra constructions shows that the kinetic effects systematically shift the eigenspectra toward less stable regions.  These results indicate that continuum-based shock stability predictions can miss important changes in the spectrum even when the macroscopic profiles appear well resolved for high Ma number shock dominated flows.

To make these computations practical at high Mach number, where the micro-velocity resolution required for convergence leads to matrices with $O(10^5)$ unknowns and up to billions of nonzero entries, we also develop a massively parallel SLEPc/PETSc solver infrastructure.  Two solver strategies are investigated: shift-and-invert Arnoldi with direct MUMPS LU factorization for moderately sized cases, and Jacobi--Davidson (JD) with block-Jacobi ILU preconditioning for the largest systems. The performance of both of these methods were tested and shown to be scaling well compared to a baseline case performed with MATLAB. The sparsity patterns of kLST matrices are shown to be challenging because of the coupled spatial and micro-velocity discretizations, producing severe LU fill-in and making direct factorization memory-limited for direct solvers, thus necessitating a solver like JD to be utilized.  With the parallel framework, the present study extends kLST calculations to a $M_\infty=4.0$ normal shock with $281{,}088$ unknowns, which to our knowledge is the highest-Mach-number kinetic linear stability calculation reported to date for isolated finite-thickness shock layers.

\section*{Nomenclature}\label{sec:nom}

{\renewcommand\arraystretch{1.05}

\subsection*{Latin Alphabet}
\begin{multicols}{2}
\noindent
$\mathsf{A}$ = system matrix \\
$A_1$ = relaxation coefficient \\
$A_{g1}, A_{h1}$ = relaxation terms \\
$a_n$ = Chebyshev expansion coefficients \\
$B(\Sigma, \xi^*)$ = collision kernel (differential cross-section) \\
$\mathsf{B}$ = mass matrix \\
$c$ = phase velocity ($\omega/\alpha$) \\
$c_i, c_j$ = Chebyshev differentiation coefficients \\
$C$ = viscosity constant; also shifted system matrix $(\mathsf{A}-\sigma\mathsf{B})$ in ILU context \\
$C_i$ = local row-block of $C$ owned by MPI rank $i$ \\
$C_{g\rho}, C_{gu}, C_{gv}, C_{gE}$ = VDF linearization coefficients \\
$C_{h\rho}, C_{hu}, C_{hv}, C_{hE}$ = energy dist. linearization coefficients \\
$D$ = Chebyshev derivative matrix \\
$D_{jm}$ = Chebyshev derivative matrix elements \\
$d_{Ar}$ = Argon collision diameter \\
$D_x, D_{xx}$ = mapped first/second derivative matrices \\
$D_y, D_{yy}$ = unmapped Chebyshev derivative matrices \\
$\Delta$ = shock thickness \\
$\Delta f$ = VDF deviation ($f - f^e$) \\
$\Delta x_i$ = spatial cell width (FVM) \\
$e_m$ = $m$-th standard basis vector \\
$E$ = total energy \\
$\mathcal{E}_{\text{neq}}(x)$ = non-equilibrium $L_2$ norm \\
$e$ = specific internal energy \\
$f$ = velocity distribution function \\
$f^e$ = equilibrium VDF \\
$f_m$ = residual vector (IRAM) \\
$F(\xi, \xi^*, \Sigma)$ = loss term in collision integral \\
$g$ = reduced VDF \\
$g^e$ = equilibrium reduced VDF \\
$G(\xi, \xi^*, \Sigma)$ = gain term in collision integral \\
$G_\rho, G_u, G_v, G_E$ = linearized VDF coefficients \\
$h$ = reduced energy dist. function \\
$h^e$ = equilibrium reduced energy dist. \\
$H_m$ = Hessenberg matrix (IRAM) \\
$H_n$ = Hermite polynomial \\
$H_\rho, H_u, H_v, H_E$ = linearized energy coefficients \\
$\mathrm{i}$ = imaginary unit $(\sqrt{-1})$ \\
$i$ = generic index \\
$I_m$ = identity matrix of size $m$ \\
$j$ = spatial point index (0-based in equations) \\
$k$ = velocity node index / requested number of eigenvalues \\
$k_1, k_2, k_3, k_4$ = Runge-Kutta stages \\
$k_{\min}$ = minimum directions retained on SLEPc restart \\
$k_x, k_y$ = velocity grid indices in $(\xi_x, \xi_y)$ \\
$K$ = dimensionality constant \\
$Kn$ = Knudsen number \\
$l_\infty$ = upstream mean free path \\
$L$ = half domain length \\
$\mathbf{L}$ = lower triangular factor \\
$\tilde{\mathbf{L}}$ = incomplete lower triangular factor (ILU) \\
$L_{ref}$ = reference length \\
$L_s$ = shock thickness \\
$m$ = spatial point index (summation) / mass flux \\
$m_{Ar}$ = Argon atomic mass \\
$M$ = Mach number \\
$Ma$ = Mach number (alternate notation) \\
$\mathbf{M}$ = mass/quadrature weight matrix \\
$\mathbf{M}_m$ = projected matrix $V_m^H A V_m$ (Jacobi-Davidson) \\
$\mathbf{M}^{-1}_{\mathrm{BJ}}$ = block-Jacobi ILU preconditioner \\
$\bar{M}_\infty$ = reference Mach number \\
$M_{\infty}$ = free-stream Mach number \\
$n$ = number density / generic index \\
$n_p$ = number of MPI ranks/processes \\
$n_p^\star$ = estimated optimal MPI rank count \\
$n_{\mathrm{loc}}^\star$ = optimal local rows per MPI rank \\
$\mathbf{n}$ = wall-normal unit vector \\
$\text{nnz}$ = number of non-zeros in a sparse matrix \\
$N$ = number of Chebyshev intervals \\
$\mathbf{N}_m$ = projected matrix $V_m^H B V_m$ (Jacobi-Davidson) \\
$N_y$ = total spatial points ($N+1$) \\
$\texttt{ncv}$ = maximum SLEPc search-space dimension \\
$p$ = velocity node index (summation) \\
$P$ = total pressure \\
$\mathbf{P}, \mathbf{Q}$ = fill-reducing permutation matrices \\
$P_\infty$ = free stream pressure \\
$\texttt{PetscInt}$ = PETSc integer type \\
$\texttt{PetscScalar}$ = PETSc scalar type (complex in this work) \\
$Pr$ = Prandtl number \\
$\mathbf{q}$ = eigenvector $[\hat{g},\hat{h}]^\top$ \\
$Q$ = GH quadrature points per direction \\
$r$ = slope-ratio argument in the MUSCL limiter \\
$r^{\pm}$ = ratio of adjacent slopes (MUSCL) \\
$\text{RHS}$ = right-hand-side operator \\
$\mathbf{r}$ = residual vector in Jacobi--Davidson \\
$\boldsymbol{r}$ = position vector \\
$R$ = gas constant; also remainder matrix in ILU $(\tilde L\tilde U = C-R)$ \\
$Re$ = Reynolds number \\
$s$ = viscosity exponent \\
$s_\infty$ = far-field truncation location (domain half-width) \\
$\mathbf{S}$ = shifted matrix ($\mathsf{A} - \sigma\mathsf{B}$) \\
$\mathcal{S}$ = sparsity pattern of a matrix \\
$t$ = time \\
$\mathbf{t}$ = correction direction in Jacobi-Davidson \\
$T$ = temperature \\
$\mathbf{T}$ = shift-and-invert operator \\
$T_n$ = Jacobi tridiagonal matrix (GH) \\
$T_\infty$ = free-stream temperature \\
$\texttt{tol}$ = eigenpair convergence tolerance \\
$u$ = $x$-velocity \\
$\mathbf{U}$ = upper triangular factor \\
$\tilde{\mathbf{U}}$ = incomplete upper triangular factor (ILU) \\
$u_m$ = Chebyshev polynomial of degree $m$ \\
$U$ = bulk velocity \\
$u_{\text{ref}}$ = reference thermal speed \\
$U_w$ = wall velocity \\
$v$ = $y$-velocity \\
$\mathbf{V}$ = eigenvector matrix \\
$v_1$ = initial vector (IRAM) \\
$\mathbf{v}_j$ = individual eigenvector \\
$V_m$ = orthonormal matrix (IRAM) \\
$w$ = $z$-velocity \\
$\mathbf{w}$ = intermediate vector in triangular solve \\
$W_k$ = GH quadrature weights \\
$x,y,z$ = spatial coordinates \\
$\tilde{\mathbf{x}}$ = Ritz (approximate eigenvector) in Jacobi-Davidson \\
$X$ = $\mathsf{A}_{gg}$ block \\
$x_i$ = Chebyshev collocation point \\
$X2$ = $\mathsf{A}_{gh}$ block \\
$y$ = eigenvector (Arnoldi) \\
$Y$ = $\mathsf{A}_{hh}$ block \\
$Y1$ = $\mathsf{A}_{hg}$ block \\
$z$ = normalized coordinate \\
$Z$ = diagonal BC matrix \\
$\mathbf{z}$ = intermediate vector in triangular solve \\
$z_j$ = approximate eigenvector (IRAM) \\
\end{multicols}

\subsection*{Greek Symbols}
\begin{multicols}{2}
\noindent
$\alpha$ = streamwise wavenumber / shock strength \\
$\beta$ = spanwise wavenumber / viscosity constant \\
$\beta_k$ = Jacobi matrix elements \\
$\gamma$ = specific heats ratio \\
$\delta$ = mean free path coefficient \\
$\delta_{mn}$ = Kronecker delta \\
$\zeta$ = Chebyshev coordinate \\
$\zeta_j$ = Chebyshev collocation point at index $j$ \\
$\theta$ = normalized temperature / Ritz value \\
$\vartheta$ = Chebyshev angle \\
$\mathbf{\Lambda}$ = diagonal matrix of eigenvalues \\
$\lambda$ = wavelength/mean free path / eigenvalue \\
$\tilde\lambda$ = harmonic Ritz eigenvalue (Jacobi-Davidson) \\
$\mu$ = dynamic viscosity / shifted target eigenvalue \\
$\mu_\infty$ = reference viscosity (at 300 K) \\
$\xi$ = microvelocity \\
$\xi^*$ = collision partner microvelocity \\
$\xi_x, \xi_y, \xi_z$ = microvelocity components \\
$\boldsymbol{\xi}$ = microvelocity vector \\
$\nu$ = collision frequency ($p/\mu$) \\
$\Phi_n$ = Chebyshev basis function \\
$\phi$ = generic variable in FVM ($g$ or $h$) \\
$\phi(\lambda)$ = polynomial filter (IRAM) \\
$\psi(r)$ = Minmod slope limiter function \\
$\mathcal{K}_m$ = Krylov subspace of dimension $m$ \\
$\rho$ = density \\
$\rho_\infty$ = free-stream density \\
$\rho_{\mathrm{fill}}$ = global LU fill-in ratio \\
$\rho^{\mathrm{loc}}_{\mathrm{fill}}$ = local (per-rank) ILU fill-in ratio \\
$\sigma$ = Prandtl number / shift parameter \\
$\sigma(F)$ = collision integral (RHS of Boltzmann equation) \\
$\Sigma$ = solid angle for collision orientation \\
$\tau$ = shear stress/relaxation time \\
$\Omega(f)$ = collision operator \\
$\varpi$ = simplified temporal frequency \\
$\Omega$ = normalized velocity (shock formulation) \\
$\omega$ = temporal eigenvalue \\
$\omega_i$ = imaginary part of $\omega$ \\
$\omega_r$ = real part of $\omega$ \\
\end{multicols}

\begin{multicols}{2}
\subsection*{Subscripts \& Superscripts}
\noindent
$(\cdot)_c$ = base flow \\
$(\cdot)_\infty$ = upstream/free-stream/reference \\
$(\cdot)_1$ = downstream \\
$(\cdot)_w$ = wall \\
$(\cdot)_j$ = spatial point index \\
$(\cdot)_k$ = velocity node index \\
$(\cdot)_m$ = spatial summation index \\
$(\cdot)_p$ = velocity summation index \\
$(\cdot)_{\text{eq}}$ = equation (0-based) index \\

$\tilde{(\cdot)}$ = total perturbation \\
$\hat{(\cdot)}$ = Fourier-mode amplitude \\
$(\cdot)^e$ = equilibrium \\
$(\cdot)^*$ = complex conjugate \\
$(\cdot)^H$ = Hermitian conjugate (conjugate transpose) \\
$(\cdot)^\dagger$ = conjugate transpose (alternate notation) \\
$(\cdot)^\top$ = transpose \\
$(\cdot)_{\text{io}}$ = inlet or outlet \\
$(\cdot)_{\text{neq}}$ = non-equilibrium \\

\subsection*{Abbreviations}
\noindent
AMD = Approximate Minimum Degree \\
AIJ = PETSc compressed sparse row matrix format \\
ARPACK = Arnoldi package \\
AXPY = scalar $a$ times vector $x$ plus vector $y$ \\
BC = boundary condition \\
BE = Boltzmann equation \\
BE-BGK = Boltzmann equation with BGK collision operator \\
BGK = Bhatnagar--Gross--Krook (collision operator) \\
BJ = block-Jacobi (preconditioner) \\
BLAS = Basic Linear Algebra Subprograms \\
BLR = Block Low-Rank \\
cLST = continuum linear stability theory \\
CSR = Compressed Sparse Row \\
D\&B = Duck \& Balakumar \\
DSMC = Direct Simulation Monte Carlo \\
EVP = eigenvalue problem \\
FEAST = contour-integration eigensolver package \\
FVM = Finite Volume Method \\
G\&P = Gilbarg \& Paolucci \\
GH = Gauss--Hermite \\
GMRES = Generalized Minimal Residual method \\
HPC = High-Performance Computing \\
ILU = Incomplete LU factorization \\
ILU-JD = ILU-preconditioned Jacobi--Davidson \\
IRAM = Implicitly Restarted Arnoldi Method \\
JD = Jacobi--Davidson (eigensolver) \\
kLST = kinetic linear stability theory \\
KSP = Krylov Solver Package (PETSc) \\
LNS = linearized Navier--Stokes \\
LNSE = linearized Navier--Stokes equations \\
LU = Lower-Upper factorization \\
LU-AR = LU-preconditioned Arnoldi \\
METIS = Matrix fill-reducing Ordering library \\
MKL = Math Kernel Library (Intel) \\
MPI = Message Passing Interface \\
MUMPS = Multifrontal Massively Parallel Solver \\
MUSCL = Monotonic Upstream-Centered Scheme for Conservation Laws \\
NEP = nonlinear eigenvalue problem \\
NS = Navier--Stokes \\
ODE = Ordinary Differential Equation \\
OMP = Open Multi-Processing (OpenMP) \\
PC = Preconditioner (PETSc) \\
PEP = polynomial eigenvalue problem \\
PETSc = Portable, Extensible Toolkit for Scientific Computation \\
QZ = QZ algorithm for generalized eigenvalue problems \\
RK4 = Fourth-order Runge-Kutta \\
RMS = Root Mean Square \\
RSVD = randomized singular value decomposition \\
SLEPc = Scalable Library for Eigenvalue Problem Computations \\
SpMV = Sparse Matrix-Vector product \\
TACC = Texas Advanced Computing Center \\
VDF = velocity distribution function \\
\end{multicols}
}

\section{The Need for a High Fidelity Linear Stability Theory for Compressible High Speed Flows}

Shock waves are ubiquitous features of high-speed flows, appearing in various forms such as 1D normal shocks, 2D oblique shocks, 3D bow shocks, and result in complex shock-shock or shock-boundary layer interactions. Rather than static discontinuities, these structures are fundamental to the aerodynamic performance and stability of hypersonic vehicles due to the finite thickness high gradient layers they form. They generate steep flow gradients that trigger continuum breakdown and significantly influence boundary layer stability and transition. Furthermore, shocks can act as sources of intrinsic unsteadiness, particularly through Edney-like interactions \cite{Edney}. Recent studies by Sawant et al. \cite{sawant_etal_2022} and Karpuzcu et al. \cite{karpuzcu2024linear} have demonstrated that even weak oblique shocks play a critical role in driving stability mechanisms.

Identifying the specific instability mechanisms in shock-dominated flows is a formidable task, particularly when shocks interact with complex structures like boundary layers, shear layers, or expansion fans. To resolve these underlying physics, we isolate the shock in its most fundamental form: a one-dimensional (1D) normal shock with asymptotic boundaries. By simplifying the system, we can decouple the shockâ€™s intrinsic stability characteristics from external flow interference. The earlier investigations of 1D shock layers included both experimental~\cite{Schmidt1969,Alsmeyer1976} and numerical analyses~\cite{Giddens1971}. The comparison of experimental measurements of the mono-atomic (argon) and molecular (nitrogen) gases at several Mach numbers with the numerical results of different methods such as continuum Navier-Stokes (NS), Boltzmann equation with Bhatnagar-Gross-Krook (BE-BGK)~\cite{BGK,HirokiBGK,chigullapalli2009modeling} or direct simulation Monte Carlo (DSMC)~\cite{Bird} revealed that the DSMC gives the best comparison followed by BE-BGK even though its Prandtl (Pr) number dependence is not accurate, and then, followed by the NS solutions. These studies highlighted the significant departure of the base flows from Navier-Stokes predictions as the Mach number increases, demonstrating that continuum-based equations fail to accurately resolve the internal shock structure, particularly the translational non-equilibrium and the bi-modal velocity distribution function (VDF) observed at higher Mach numbers. 

To study global instabilities in shock dominated flows, modal linear stability analysis is performed traditionally using the NS-based equations~\cite{sidharth2018onset,hao2021occurrence,karpuzcu2024linear}. However,  the linearized Navier-Stokes equations (LNSE) have the same limitations as the NS base flow equations because of the continuum assumption; e.g., not being able to capture translational non-equilibrium flow effects such as the bi-modal velocity distribution in the high gradient shock regions. The literature on the investigation of the linear stability of 1D normal shocks is quite limited; Duck \& Balakumar~\cite{DuckBalakumarNormalShock} were the first to investigate the stability of the finite thickness shocks with the LNSE where they utilized the formulations provided by Gilbarg and Paolucci (G\&P)~\cite{GILBARGPAOLUCCI1953} for the base flows. Their analysis resulted in only stable and continuous mode branches. Recent work by Sawant et al.~\cite{sawantPhDThesis} improved upon Navier-Stokes formulations by employing the Anisotropic Conservation Equations (ACE) for linear stability analysis. Their findings demonstrated that accounting for anisotropic stresses predicts a less stable system. In a follow-up study, Sawant et al.~\cite{sawant_POF} utilized the kinetic DSMC method to show that shocks in stagnation flows can exhibit intrinsic instabilities directly within the shock layer. There has been a considerable amount of effort on the stability of high-pressure shock fronts, such as in detonations but the shock thickness is ignored. However those systems are beyond the scope of this work~\cite{NMKuznetsov_1989,SwanFowlesShock,SwanFowlesShock2,Erpenbeck1,Erpenbeck2,Gardner}.

A more sophisticated approach for investigating shock stability involves solving the Boltzmann equation from first principles to provide a higher-fidelity representation of the underlying physics. The stability of the Boltzmann equation itself was of interest extensively~\cite{DiPernaLions1989,Ukai1974,Guo2003VMB}, however the use of linearized Boltzmann equation to investigate the stability of canonical flows is relatively new~\cite{StabilityBGK,zou_bi_zhong_yuan_tang_2023,MaBELSE}.  Zou et al.~\cite{StabilityBGK,zou_bi_zhong_yuan_tang_2023} initially proposed a methodology for incompressible flows showing great agreement with the literature and then extended it to compressible flows. Specifically they studied the compressible Couette flow~\cite{ZouetalCompCouetteBGK2024}, where they show the effect of increasing Knudsen number on the eigenvalue spectra. However their main focus was for rarefied channel flows and they did not investigate the effect of non-equilibrium due to the bi-modal distribution of the VDFs in high gradient regions.

The present work closes exactly this gap. For the first time, we perform a fully kinetic linear stability analysis of finite-thickness 1D normal shocks by linearizing the Boltzmann--BGK equation directly about a kinetic (BE-BGK) base flow, so that \emph{both} the mean state and the perturbations are modeled at the level of the velocity distribution function. Unlike all prior NS-, LNSE-, and ACE-based shock stability studies, and unlike the existing linearized-BE efforts, which have been confined to rarefied channel flows, our formulation retains the bi-modal, translationally non-equilibrium structure of the shock interior throughout the stability problem rather than averaging it with a moment closure. This enables the first direct, quantitative comparison of equilibrium (Maxwellian) versus genuinely non-equilibrium eigenspectra for a shock layer, thereby isolating the imprint of microscale kinetic physics on macroscopic shock stability. 

To make such analyses tractable for even Mach numbers up to $M_\infty=4$, where dense micro-velocity grids drive the generalized eigenproblem to hundreds of thousands of unknowns, we further develop a massively parallel SLEPc/PETSc~\cite{slepc,petsc} solver infrastructure. To the best of our knowledge, this pushes the kinetic LST to the highest Mach number ever reported in the literature. Parallel sparse eigenvalue solvers have been applied to a wide range of large-scale stability and modal-analysis problems, but not previously to kinetic shock LST at the problem sizes considered here ($n\sim10^5$--$10^6$ complex unknowns). Hernandez et al. performed a broad algorithmic surveys of sparse eigensolvers, including Arnoldi, Lanczos, Davidson, and Jacobi--Davidson methods~\cite{hernandez2009survey}. Ramalingam et al. and Jezequel et al. made performance tests for PETSc-based libraries on large sparse linear systems and kernel specialization for PDE applications~\cite{ramalingam2012improving,jezequel2012solving}. SLEPc then extended these capabilities to parallel Generalized Davidson and Jacobi--Davidson solvers for standard and generalized eigenvalue problems~\cite{romero2014parallel}, and later to polynomial and nonlinear eigenvalue problems relevant to high-order PDE, delay-differential, and wave-propagation models~\cite{campos2016parallel,campos2021nep}. More recent applications have used PETSc/SLEPc or their Python interfaces for stability and resolvent analyses in electrical-grid, porous-media, and very-large-scale linear-system settings~\cite{tzounas2020comparison,shivaraj2025stability,padovan2025resolvent4py}. Together, these studies establish PETSc/SLEPc as a standard platform for large sparse eigenvalue problems in fluids, networks, and PDE-based stability analysis; the present work applies the same infrastructure to BE-BGK kinetic operators whose dimension and velocity-space coupling exceed typical continuum LST formulations.

The remainder of this paper is organized as follows. Section~\ref{sec:kLSTformulation} presents the kLST formulation for one-dimensional shock layers by linearizing the Boltzmann--BGK equation about the BE-BGK base flow and constructing the generalized eigenvalue problem in terms of the reduced distribution functions. Section~\ref{sec:NumMethods} describes the numerical discretization, including Gauss--Hermite quadrature in micro-velocity space, Chebyshev collocation with algebraic spatial mapping, and the BE-BGK base-flow solver used to generate the non-equilibrium shock profiles. Section~\ref{sec:ParallelSolver} provides details of the massively parallel SLEPc/PETSc implementation, including the direct LU--Arnoldi and ILU-preconditioned Jacobi--Davidson strategies used for the largest matrices as well as quantifies the matrix sizes, sparsity patterns, and solver performance on single- and multi-node Frontera runs. Section~\ref{sec:1DShockAnalysis} then applies the framework to $M_\infty=1.2$, $3.0$, and $4.0$ normal shocks and compares equilibrium and non-equilibrium eigenspectra. The main conclusions are summarized in Section~\ref{sec:Conclusions}.

The appendices provide supporting verification material for this paper. Appendix~A verifies the kLST matrix assembly against compressible Couette-flow eigenvalue benchmarks from the literature. Appendix~B documents the base-flow and eigenspectrum mesh-convergence studies used for the one-dimensional shock calculations. Finally Appendix~C lists the complete SLEPc/PETSc solver settings used in the HPC benchmark studies.

\section{Kinetic Linear Stability Theory Formulation and Implementation for the 1D Shocks}\label{sec:kLSTformulation}

In this section we discuss the linearization of the Boltzmann Equation with the use of the BGK operator (BE-BGK) and construct linear stability formulations for the temporal stability of planar 1D shocks as well as the base flow solver implementation. Temporal stability analysis investigates the long-time behavior as $t \to \infty$, where spatial wavenumbers ($\beta$) are chosen real and we solve for complex temporal wavenumbers ($\omega=i\omega_i+\omega_r$). For the considerations in this study, an eigenvalue with positive imaginary part indicates a growing mode and thus the system will undergo transition. For 1D shocks, this would manifest as spanwise variations that can eventually lead to bifurcations. 

\begin{figure}[h!]
\centering
\includegraphics{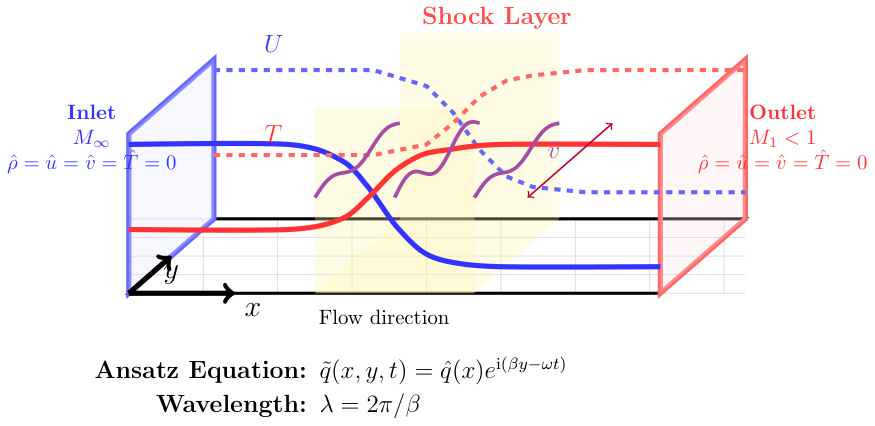}
\caption{Three-dimensional schematic of 1D normal shock flow with linear stability perturbations. The base flow profiles $u_c(x)$ and $T_c(x)$ vary only in the flow direction ($x$) as shown by the parallel lines across the $y$-direction, representing the one-dimensional shock structure. Perturbations $\tilde{v}(x,y,t)$ vary in the spanwise $y$-direction with wavelength $\lambda=2\pi/\beta$. Inlet and outlet boundaries enforce zero perturbation conditions.}
\label{fig:1DShockSetup}
\end{figure}

The 1D normal shock flow setup is shown in Fig.~\ref{fig:1DShockSetup}. The flow domain extends in the streamwise ($x$) direction with upstream (inlet) and downstream (outlet) boundaries. The shock structure is centered in the domain, with upstream conditions at Mach number $M_\infty$ transitioning to downstream subsonic conditions. The base flow is one-dimensional in $x$ but perturbations are assumed to vary in the spanwise ($y$) direction for the 1D shock case. The boundary conditions at inlet and outlet is defined as Dirichlet boundary conditions for the base flows, i.e. assuming equilibrium at the inlet and outlet with the pre-calculated Rankine-Hugoniot conditions and specify that perturbations decay to zero far from the shock, i.e., $\hat{u}, \hat{v}, \hat{\rho}, \hat{T} \to 0$ as $x \to \pm\infty$ for the linear stability analysis.

\subsection{Base Flow Solutions for 1D Shocks}\label{sec:BaseFlowSolvers}

In this study we will employ two methods to calculate the base flows for the finite thickness 1D normal shocks. The first one is the well-known Gilbarg and Paolucci (G\&P)~\cite{GILBARGPAOLUCCI1953} method where the one-dimensional steady-state shock structure is governed by the normalized equations derived from continuum mechanics and thermodynamic relations. This first method represents the continuum approach to the 1D shock profiles. The following definitions and relations were taken from G\&P~\cite{GILBARGPAOLUCCI1953} that constitutes the basis for the calculations:

\begin{align}
\rho_\infty u_\infty &= \rho_1 u_1 = m \\
p_\infty + \rho_\infty u_\infty^2 &= p_1 + \rho_1 u_1^2 = P 
\end{align}

\begin{align}
e_\infty + \frac{1}{2}u_\infty^2 + \frac{p_\infty}{\rho_\infty} = e_1 + \frac{1}{2}u_1^2 + \frac{p_1}{\rho_1} = E
\end{align}

Normalized velocity $\Omega$ and temperature $\theta$ is defined as:
\begin{align}
\Omega &= \frac{m}{P} u, \\
\theta &= \frac{m}{P} T
\end{align}

The upstream mean free path is defined as;
\begin{equation}
l_\infty=\frac{16}{\delta} \frac{\mu_\infty}{\rho_\infty \sqrt{2 \pi R T_\infty}}
\end{equation}
and the spatial direction is normalized as;
\begin{equation}
z=x/l_\infty
\end{equation}

Using this, the shock strength is defined as,
\begin{equation}
\alpha = \frac{2Em^2}{P^2} - 1
\end{equation}
and the temperature and streamwise velocity is related through~\cite{GILBARGPAOLUCCI1953};
\begin{equation}
\frac{d \theta}{d \Omega} = \frac{8}{15} \cdot \frac{\Omega\left\{\theta - \frac{1}{3} \left[ (1 - \Omega)^2 + \alpha \right] \right\}}{\Omega^2 - \Omega + \theta}
\end{equation}

From Equation (25) of G\&P~\cite{GILBARGPAOLUCCI1953}, the relation between $\Omega$, $\theta$ and the spatial variable $z$ is:

\begin{equation}
\begin{aligned}
&\begin{gathered}
\frac{\beta}{\bar{M}_\infty}\left(\frac{\theta}{\theta_\infty}\right)^s \frac{d \Omega}{d z}=\Omega+\frac{\theta}{\Omega}-1 \\
\frac{5}{2} \frac{\beta}{M_\infty}\left(\frac{\theta}{\theta_\infty}\right)^s \frac{d \theta}{d z}=\theta-\frac{1}{3}\left[(1-\Omega)^2+\alpha\right]
\end{gathered}
\end{aligned}
\label{eqn:GB1DShockeqn}
\end{equation}
where $\beta$ is a constant related to viscosity, ${M}_\infty$ is the reference Mach number, $s$ is the viscosity temperature exponent.

The shock thickness will be used as the reference length scale throughout the one-dimensional normal shock results. This is defined as:
\begin{equation}
\Delta=\frac{u_\infty-u_1}{\left|\frac{d u}{d x}\right|_{\max }}
\label{eqn:ShockThickness}
\end{equation}
where $u_\infty$ and $u_1$ are the upstream and downstream velocities, respectively. We will use this shock thickness definition $\Delta$ as the length scale in the subsequent analyses.

The system of equations given in Eqn.~\ref{eqn:GB1DShockeqn} was solved numerically using an ODE solver. Since the downstream point is a more stable point to start the numerical solution process, the integration is carried from downstream to upstream. Upstream conditions are input, and downstream conditions are calculated using the Rankine-Hugoniot relations. 

The second method that we will use to represent the base flows is the Boltzmann equation, which is a kinetic method. The Boltzmann equation is the most comprehensive equation to describe fluid motion and is given as;

\begin{equation}\label{eqn:Boltzmann}
\frac{\partial f}{\partial t}+\xi \cdot \frac{\partial f}{\partial \boldsymbol{r}}=\Omega (f)
\end{equation}
where
\[\Omega (f)(\xi ) = \int_{{\xi ^*}} d {\xi ^*}\int_\Sigma  {B(\Sigma ,{\xi ^*})[G(\xi ,{\xi ^*},\Sigma ) - F(\xi ,{\xi ^*},\Sigma )]} d\Sigma \]
and $f(\xi,r,t)$ is the velocity distribution function for particles in the velocity space $\xi=(\xi_1,\xi_2,....)$ and $r$ is the position at time $t$. The right-hand side collision term $\Omega(f)$ (also denoted as $\sigma(F)$ in some literature) represents the rate of change of the distribution function due to binary molecular collisions. In this collision integral, $\xi$ represents the velocity of the particle of interest, while $\xi^*$ denotes the velocity of the collision partner particle. The term $\Sigma$ represents the solid angle that describes the orientation of the collision, and the double integral integrates over all possible collision partner velocities and all possible collision orientations. The collision kernel $B(\Sigma, \xi^*)$ (also called the differential cross-section) characterizes the probability of collisions occurring with a specific geometry and depends on the intermolecular potential. The term $G(\xi, \xi^*, \Sigma)$ represents the gain term, which accounts for particles entering the velocity state $\xi$ as a result of collisions, while $F(\xi, \xi^*, \Sigma)$ represents the loss term, which accounts for particles leaving the velocity state $\xi$ due to collisions. The difference $[G(\xi ,{\xi ^*},\Sigma ) - F(\xi ,{\xi ^*},\Sigma )]$ thus gives the net rate of change of particles at velocity $\xi$ due to all possible binary collisions. The most challenging task in obtaining a solution to the Boltzmann equation is computing this collision term. There are several methods and approximations made to simplify this term, one of which is known as the Bhatnagar-Gross-Krook (BGK) approximation where the Boltzmann equation~\cite{BGK,HirokiBGK,chigullapalli2009modeling} collision operator is approximated as follows;
\[\Omega (f) =  - \frac{1}{\tau }\left[ {f - {f^e}} \right]\] 
with $\tau=p/\mu$ so Eqn.~\ref{eqn:Boltzmann} then becomes;

\begin{equation}\label{eqn:BoltzmannBGK}
\frac{\partial f}{\partial t}+\xi \cdot \frac{\partial f}{\partial \boldsymbol{r}}=\frac{f^e-f}{\tau}
\end{equation}

\subsection{Linearization and Discretization of the BE-BGK}\label{sec:LinearizeBGK}

The process for the linearization of the Boltzmann transport equation will be detailed in this section. In this work, we are following the methodology of Zou et al.~\cite{ZouetalCompCouetteBGK2024}, so the macroparameters are normalized accordingly, temperatures with $T_\infty$, density with $\rho_\infty$, velocity by the thermal speed $u_{ref}=\sqrt{2RT_\infty}$, pressure by $\rho_\infty u_{ref}^2$, viscosity by $\rho_\infty u_{ref} L_{ref}$, and velocity distribution functions by $\rho_\infty/u_{ref}^2$ and time by $L_{ref}/u_{ref}$. For the 1D shock flow $L_{ref}$ is the shock thickness from the G\&P shock profiles. In this section all the given macroparameters are thus in their non-dimensional form. The Reynolds, Mach and Kn numbers are defined using the upper wall conditions for the flow parameters, indicated by the subscript $_\infty$ and defined as;
$Re_\Delta=\rho_\infty u_\infty L_{ref} / \mu_\infty$, $Ma=u_\infty/\sqrt{\gamma R T_\infty}$ and $Kn_\Delta = \frac{{2(5 - 2s )(7 - 2s )Ma}}{{15Re_\Delta}}\sqrt {\frac{\gamma }{{2\pi }}} .$ with $s=0.5$ for hard sphere molecules. 

We start with the Boltzmann equation with BGK operator in 2 spatial dimensions and 3 micro velocity components in Eqn.~\ref{eqn:BoltzmannBGK}. Considering the fact that there would be many more microvelocity nodes needed for the compressible flow cases, keeping the problem three dimensional in microvelocity nodes is not feasible. For that reason reduced velocity distribution functions are introduced again as before in the base flow calculations;
\begin{equation}
\begin{aligned}
g\left(x, y, \xi_x, \xi_y\right) & =\int_{-\infty}^{\infty} f\left(x, y, \xi_x, \xi_y, \xi_z\right) d \xi_z \\
h\left(x, y, \xi_x, \xi_y\right) & =\int_{-\infty}^{\infty} \xi^2 f\left(x, y, \xi_x, \xi_y, \xi_z\right) d \xi_z
\end{aligned}
\end{equation}
$h\left(x, y, \xi_x, \xi_y\right)$ is introduced so that the energy conservation equation can be constructed. With that the reduced Boltzmann equations are given as;
\begin{equation}
\begin{aligned}
 \frac{\partial g}{\partial t}+\xi \cdot \frac{\partial g}{\partial \boldsymbol{r}}=\frac{g^e-g}{\tau} \\
\frac{\partial h}{\partial t}+\boldsymbol{\xi} \cdot \frac{\partial h}{\partial \boldsymbol{r}}=\frac{h^e-h}{\tau}
\end{aligned}
\end{equation}
the second equation is for the conservation of energy since we are using the reduced velocity distribution functions, i.e. VDFs integrated over the third direction (z direction) and now only have two microvelocity components. The equilibrium distribution functions are defined as;
\begin{equation}
\begin{gathered}
g^e=\frac{\rho}{2 \pi R T} \exp \left(-\frac{(\xi-u)^2}{2 R T}\right), \\
h^e=(K+1) R T g^e,
\end{gathered}
\end{equation}
Now assuming that the both g and h has a mean and a perturbation component $g = g_c  + \tilde g$, $h =h_c  + \tilde h$ where subscript c is for the mean component, known from the base flow solution;
\begin{equation}
\large
\begin{aligned}
\label{eqn:CompBGKtilde}
&\frac{\partial \tilde{g}}{\partial t}+\xi \cdot \frac{\partial \tilde{g}}{\partial \boldsymbol{x}}=\frac{\mu_c P_c\left(\tilde{g}^e-\tilde{g}\right)+\mu_c\left(g_c^e-g_c\right) \tilde{P}-P_c\left(g_c^e-g_c\right) \tilde{\mu}}{\mu_c^2},\\
&\frac{\partial \tilde{h}}{\partial t}+\xi \cdot \frac{\partial \tilde{h}}{\partial \boldsymbol{x}}=\frac{\mu_c P_c\left(\tilde{h}^e-\tilde{h}\right)+\mu_c\left(h_c^e-h_c\right) \tilde{P}-P_c\left(h_c^e-h_c\right) \tilde{\mu}}{\mu_c^2},
\end{aligned}
\end{equation}
Next, assuming an ansatz for $\tilde g$ and $\tilde h$ and the macroparameters. For the 1D normal shock case where derivatives are in the same direction as the main flow, the ansatz becomes:
\begin{equation}
\large
\begin{gathered}
\left.[\tilde{\rho}, \tilde{u}, \tilde{v}, \tilde{T}]^{\mathrm{tr}}=[\hat{\rho}(x), \hat{u}(x), \hat{v}(x), \hat{T}(x))\right]^{\mathrm{tr}} \times e^{\mathrm{i}(\beta y-\omega t)} \\
{[\tilde{g}, \tilde{h}]^{\mathrm{tr}}=\left[\hat{g}\left(\xi_x, \xi_y, x\right), \hat{h}\left(\xi_x, \xi_y, x\right)\right]^{\mathrm{tr}} \times e^{\mathrm{i}(\beta y-\omega t)}}
\end{gathered}
\end{equation}

Substituting these back into eqn.~\ref{eqn:CompBGKtilde}, we obtain the final equations for the harmonic perturbation components for the 1D normal shock:
\begin{equation}\label{eqn:Comp1DHarm}
\large
\begin{aligned}
& -\mathrm{i} \varpi \hat{g}+\xi_x \frac{\partial \hat{g}}{\partial x}+\mathrm{i} \beta \xi_y \hat{g}=\mathrm{G}_\rho \hat{\rho}+\mathrm{G}_u \hat{u}+\mathrm{G}_v \hat{v}+\mathrm{G}_E \hat{E}-\mathrm{A}_{g 1} \hat{g}, \\
&-\mathrm{i} \varpi \hat{h}+\xi_x \frac{\partial \hat{h}}{\partial x}+\mathrm{i} \beta \xi_y \hat{h}=H_\rho \hat{\rho}+H_u \hat{u}+H_v \hat{v}+H_E \hat{E}-A_{h 1} \hat{h}
\end{aligned}
\end{equation}

where $G_{\rho}(y,\boldsymbol{\xi}), G_u(y,\boldsymbol{\xi}), G_v(y,\boldsymbol{\xi}), G_E(y,\boldsymbol{\xi})$ and similarly $H_{\rho}(y,\boldsymbol{\xi}), H_u(y,\boldsymbol{\xi}), H_v(y,\boldsymbol{\xi}), H_E(y,\boldsymbol{\xi})$ are linearization coefficients that depend on the base flow quantities including: the base flow equilibrium VDFs ($g_c^e, h_c^e$), the base flow VDFs ($g_c, h_c$), base flow macroparameters ($\rho_c, u_c, T_c, P_c, E_c$), dynamic viscosity ($\mu$), reference viscosity ($\mu_{\infty}$), gas constant ($R$), and microvelocity components ($\xi_x, \xi_y$). These coefficients are constructed through combinations of partial derivatives of the equilibrium distribution functions with respect to macroparameters, weighted by pressure, viscosity, and relaxation terms; the detailed expressions are given in Zou et al.~\cite{ZouetalCompCouetteBGK2024} in equations 16a and 16b of the paper and will not be repeated in this document. Perturbation macroparameters are calculated as integrals of reduced VDFs as;

$$
\begin{gathered}
\hat \rho=\int \hat g d \xi \\
\rho \boldsymbol{\hat u}=\int \xi \hat g d \xi \\
\rho \hat E=\frac{1}{2} \int\left(\xi^2 \hat g+\hat h\right) d \xi,
\end{gathered}
$$

The final discretized equations are then assembled. For the 1D normal shock, these become:
\begin{subequations}
\begin{equation}
\label{eqn:comp_ghat1Dshock}
\begin{aligned}
& \left(\mathrm{i} \beta \xi_{y, k}+\mathrm{A}_1\left(\zeta_j\right)\right) \hat{g}_k\left(\zeta_j\right)+\xi_{x, k} \sum_{m=0}^N D_{j m} \hat{g}_k\left(\zeta_m\right) \\
& \quad -\mathrm{G}_\rho\left(\zeta_j, \boldsymbol{\xi}_k\right) \sum_{p=0}^{Q^2-1} W_p \hat{g}_p\left(\zeta_j\right)-\frac{\mathrm{G}_u\left(\zeta_j, \xi_k\right)}{\rho_c\left(\zeta_j\right)} \sum_{p=0}^{Q^2-1} W_p\left(\xi_{x, p}-u_c\left(\zeta_j\right)\right) \hat{g}_p\left(\zeta_j\right) \\
& \quad -\frac{\mathrm{G}_v\left(\zeta_j, \xi_k\right)}{\rho_c\left(\zeta_j\right)} \sum_{p=0}^{Q^2-1} W_p \xi_{y, p} \hat{g}_p\left(\zeta_j\right)-\frac{\mathrm{G}_E\left(\zeta_j, \xi_k\right)}{2 \rho_c\left(\zeta_j\right)} \sum_{p=0}^{Q^2-1} W_p\left(\xi_x^2+\xi_y^2-2 E_c\left(\zeta_j\right)\right) \hat{g}_p\left(\zeta_j\right) \\
& \quad -\frac{\mathrm{G}_E\left(\zeta_j, \xi_k\right)}{2 \rho_c\left(\zeta_j\right)} \sum_{p=0}^{Q^2-1} W_p \hat{h}_p\left(\zeta_j\right) = \mathrm{i} \varpi \hat{g}_k\left(\zeta_j\right)
\end{aligned}
\end{equation}
and for $\hat h$:
\begin{equation}
\label{eqn:comp_hhat1Dshock}
\begin{aligned}
& \left(\mathrm{i} \beta \xi_{y, k}+A_1\left(\zeta_j\right)\right) \hat{h}_k\left(\zeta_j\right)+\xi_{x, k} \sum_{m=0}^N D_{j m} \hat{h}_k\left(\zeta_m\right) \\
& \quad -H_\rho\left(\zeta_j, \xi_k\right) \sum_{p=0}^{Q^2-1} W_p \hat{g}_p\left(\zeta_j\right)-\frac{H_u\left(\zeta_j, \xi_k\right)}{\rho_c\left(\zeta_j\right)} \sum_{p=0}^{Q^2-1} W_p\left(\xi_{x, p}-u_c\left(\zeta_j\right)\right) \hat{g}_p\left(\zeta_j\right) \\
& \quad -\frac{H_v\left(\zeta_j, \xi_k\right)}{\rho_c\left(\zeta_j\right)} \sum_{p=0}^{Q^2-1} W_p \xi_{y, p} \hat{g}_p\left(\zeta_j\right)-\frac{H_E\left(\zeta_j, \xi_k\right)}{2 \rho_c\left(\zeta_j\right)} \sum_{p=0}^{Q^2-1} W_p\left(\xi_x^2+\xi_y^2-2 E_c\left(\zeta_j\right)\right) \hat{g}_p\left(\zeta_j\right) \\
& \quad -\frac{H_E\left(\zeta_j, \xi_k\right)}{2 \rho_c\left(\zeta_j\right)} \sum_{p=0}^{Q^2-1} W_p \hat{h}_p\left(\zeta_j\right) = \mathrm{i} \varpi \hat{h}_k\left(\zeta_j\right)
\end{aligned}
\end{equation}
\end{subequations}

By assuming that perturbation equilibrium distribution function $\hat{g}^e$ is a function of $g_c^e$, thus $\hat{g}=O(g_c^e, \rho, u, v, T)$ and similarly for $\hat{h}^e$, the perturbation macroparameters can be written in terms of $\hat{g}^e$ and thus are not treated as unknowns. Using the chain rule of derivation, $\hat{g}^e$ and $\hat{h}^e$ can be defined as;

\begin{equation}
\begin{aligned}
&\hat{g}^{\mathrm{e}}=C_{g\rho}(\rho,u,v,E,g_c^e) \hat{\rho}+C_{gu}(\rho,u,v,E,g_c^e) \hat{u}+C_{gv}(\rho,u,v,E,g_c^e) \hat{v}+C_{gE}(\rho,u,v,E,g_c^e) \hat{E} \\
&\hat{h}^{\mathrm{e}}=C_{h\rho}(\rho,u,v,E,h_c^e) \hat{\rho}+C_{hu}(\rho,u,v,E,h_c^e) \hat{u}+C_{hv}(\rho,u,v,E,h_c^e) \hat{v}+C_{hE}(\rho,u,v,E,h_c^e) \hat{E}
\end{aligned}
\end{equation}
The coefficients $C_{g\rho}$, $C_{gu}$, $C_{gv}$, $C_{gE}$ and similarly $C_{h\rho}$, $C_{hu}$, $C_{hv}$, $C_{hE}$ are derived from the partial derivatives of the equilibrium distribution functions ($g_c^e, h_c^e$) with respect to the macroparameters $\rho$, $u$, $v$, and $E$. Specifically, $C_{g\rho}$ depends on the ratio of $g_c^e$ to density, $C_{gu}$ and $C_{gv}$ involve the equilibrium VDF scaled by temperature and gas constant along with velocity-dependent terms and the respective microvelocity components ($\xi_x$ or $\xi_y$), and $C_{gE}$ incorporates the equilibrium VDF with temperature, gas constant, and both microvelocity components ($\xi_x, \xi_y$). The same functional dependencies apply to the $C_h$ coefficients with $h_c^e$ replacing $g_c^e$. The detailed expressions are given in equation 15 in Zou et al.~\cite{ZouetalCompCouetteBGK2024} and will not be repeated in this document.

We will consider asymptotic inlet and outlet boundary conditions, i.e. with zero perturbations for all perturbation components. Revisiting the boundary conditions as;

\begin{equation}
\begin{aligned}
&\begin{aligned}
& \hat{g}\left(\boldsymbol{r}_{\mathrm{io}}, \boldsymbol{\xi}\right)=\hat{g}^{\mathrm{e}}\left(\boldsymbol{\xi}, \hat{\rho}_{\mathrm{io}}, \hat{\boldsymbol{u}}_{\mathrm{io}}, \hat{T}_{\mathrm{io}}\right), \boldsymbol{\xi} \cdot \boldsymbol{n}>0, \\
& \hat{h}\left(\boldsymbol{r}_{\mathrm{io}}, \boldsymbol{\xi}\right)=\hat{h}^{\mathrm{e}}\left(\boldsymbol{\xi}, \hat{\rho}_{\mathrm{io}}, \hat{\boldsymbol{u}}_{\mathrm{io}}, \hat{T}_{\mathrm{io}}\right), \boldsymbol{\xi} \cdot \boldsymbol{n}>0,
\end{aligned}\\
&\text { where }\\
&\begin{aligned}
& \hat{g}^{\mathrm{e}}=C_{g\rho} \hat{\rho}+C_{gu} \hat{u}+C_{gv} \hat{v}+C_{gE} \hat{E}, \\
& \hat{h}^{\mathrm{e}}=C_{h\rho} \hat{\rho}+C_{hu} \hat{u}+C_{hv} \hat{v}+C_{hE} \hat{E},
\end{aligned}
\end{aligned}
\end{equation}
for Dirichlet boundary conditions $\hat{\rho}=\hat{u}=\hat{v}=\hat{T}=0$ for both inlet and outlet conditions where subscript io means inlet or outlet points. With the introduction of algebraic mapping of Chebyshev points, these asymptotic conditions are further strengthen as the domain can be truly large enough for the perturbations to die out naturally.

Next step is then to form the solution matrices. The solution matrices for the compressible case then becomes $A\hat{q}=i\omega \hat{q}$ and can be shown with submatrices as;

\begin{equation}\label{eqn:MatrixSystem}
\begin{bmatrix}
X & X2 \\
Y1 & Y
\end{bmatrix}
\begin{bmatrix}
\hat{g} \\ \hat{h}
\end{bmatrix}
= i\omega
\begin{bmatrix}
I & 0 \\
0 & I
\end{bmatrix}
\begin{bmatrix}
\hat{g} \\ \hat{h}
\end{bmatrix}
\end{equation}

The submatrices $X, X_2, Y_1,$ and $Y$ are constructed from the elements in Eqns.~\ref{eqn:comp_ghat1Dshock} and \ref{eqn:comp_hhat1Dshock} that multiply the vectors $\hat{g}$ and $\hat{h}$, respectively. Specifically, $X$ and $X_2$ originate from Eqn.~\ref{eqn:comp_ghat1Dshock}, while $Y_1$ and $Y$ originate from Eqn.~\ref{eqn:comp_hhat1Dshock}. Together, these submatrices constitute the global matrix $A$, acting on the state vector $\hat{q} = [\hat{g}, \hat{h}]^T$. This submatrix formulation is utilized primarily to simplify the implementation when assembling the global matrix for the eigenvalue problem.

\section{Numerical Methods for Modal Kinetic Linear Stability Theory}\label{sec:NumMethods}

This chapter outlines the discretization strategies for the 1D shock stability analysis and the corresponding base flow solutions. We employ a unified approach for both the kLST and BE-BGK solvers, utilizing Chebyshev collocation points in physical space and Gauss-Hermite quadratures in the micro-velocity domain. The technical details of these discretizations are presented first, followed by a description of the base flow solver's architecture.

\subsection{Solution of Eigenvalue Problems with Spectral Methods}\label{sec:SpectralMethodsEval} 

We will use spectral methods to discretize the domain in space as they provide spectral accuracy with $O(n \log n)$ computational complexity~\cite{gottlieb1977numerical, canutospectral, boyd2013chebyshev, MatlabSpectral}. Eigenvalue problems can have periodic boundaries or they can have specific boundary conditions such as Dirichlet or Neumann that is called a boundary value problem. The main method used to solve for the eigenvalues of the boundary value problems, i.e. no periodic boundaries, is called the pseudo-spectral methods because Fourier transforms do not work for boundaries other than the periodic. A method called Chebyshev polynomials is one of the popular ones to use for such problems. For an unknown $u$, Chebyshev polynomials are defined as ~\cite{cheb1853};
\begin{equation}
{u_m}(\cos \vartheta ) = \cos (m\vartheta ),m= 0,1,2,...,N
\end{equation}

Using trigonometric relations, the first few Chebyshev polynomials become;

\[\begin{array}{l}
{u_0}(x) = 1, {u_1}(x) = x, {u_2}(x) = 2{x^2} - 1, {u_3}(x) = 4{x^3} - 3x\\
...................
\end{array}\]
 evaluated at Chebyshev collocation points defined as;
\begin{equation}
\label{eqn:ChebPoints}
{x_i} = \cos \left( {\frac{{i\pi }}{N}} \right),i = 0,1,2,..,N
\end{equation}

Keep in mind that by using Chebyshev polynomials, we are also assuming the $u$ is a sum of several Chebyshev polynomials in the form;

\begin{equation}
u(x,t) = \sum\limits_n {{a_n}{\Phi _n}(x)} 
\end{equation}

where ${{a_n}{\Phi _n}(x)}$ are the determined from these polynomials.

When using the this collocation method to solve the eigenvalue equation originating from the Boltzmann equation, first derivatives of this formulation is needed and it is called the Chebyshev derivative operator. Rooted in the properties of Chebyshev polynomials, which form an orthogonal basis on the interval [-1,1], the Chebyshev derivative operator leverages these polynomials to accurately approximate derivatives of smooth functions with high precision and defined as follows;.

\begin{equation}\label{eqn:ChebDiag}
D_{jj} = \frac{{ - {x_j}}}{{2(1 - x_j^2)}},\quad j = 1,...,N-1
\end{equation}

\begin{equation}\label{eqn:ChebOffDiag}
\begin{array}{l}
D_{jm} = \frac{{{c_j}}}{{{c_m}}}\frac{{{{( - 1)}^{j + m}}}}{{({x_j} - {x_m})}},\quad j \ne m,\quad j,m = 0,...,N\\
{c_j} = \left\{ {\begin{array}{*{20}{c}}
{2,\quad j = 0 \text{ or } N}\\
{1,\quad \text{otherwise}}
\end{array}} \right.
\end{array}
\end{equation}

Boundary conditions are incorporated as;
\begin{equation}\label{eqn:ChebBC}
D_{00} = \frac{{2{N^2} + 1}}{6},\quad D_{N,N} =  - \frac{{2{N^2} + 1}}{6}
\end{equation}

In the above equations (Eqns.~\ref{eqn:ChebDiag}-\ref{eqn:ChebBC}), the indices $j$ and $m$ follow 0-based indexing convention, ranging from $0$ to $N$ (with $N+1$ total Chebyshev collocation points). This 0-based indexing convention is consistently used throughout all discretized equations in this chapter. The Chebyshev differentiation matrix $D$ is an $(N+1) \times (N+1)$ matrix. We will employ these Chebyshev collocation points and derivative operator in the discretized equations of Eqns.~\ref{eqn:comp_ghat1Dshock} and ~\ref{eqn:comp_hhat1Dshock} of the 1D shock stability problem. By representing a function as a sum of Chebyshev polynomials, the derivative operator, $D_{jm}$, can be efficiently applied in the spectral domain, resulting in highly accurate differentiation matrices. In the context of eigenvalue problems, the Chebyshev derivative operator facilitates the transformation of continuous differential operators into discrete matrix forms. Consequently, numerical techniques such as the Lanczos or Arnoldi algorithms can be employed to compute the eigenvalues and eigenvectors with remarkable efficiency and accuracy for those eigenvalue problems.

We next discuss the algebraic mapping for unbounded domains. Standard Chebyshev collocation points (Eqn.~\ref{eqn:ChebPoints}) are defined on the interval $y \in [-1, 1]$ and are highly clustered near the boundaries $\pm 1$. While effective for bounded problems like channel flows, this distribution is inefficient for unbounded problems such as 1D normal shocks. In shock problems, the sharpest flow gradients occur at the center of the domain (the shock wave at $x=0$), and the domain must extend to $\pm \infty$ to properly impose freestream boundary conditions.

To resolve the shock layer accurately while approximating an infinite domain, an algebraic mapping is employed \cite{boyd2013chebyshev} to transform the computational coordinate $y \in [-1, 1]$ to the physical coordinate $x \in [-s_{\infty}, s_{\infty}]$:
\begin{equation}\label{eqn:Chebmapping}
x = L \frac{y}{\sqrt{1 + s - y^2}}
\end{equation}
where $L$ is a characteristic length scale (typically chosen on the order of the shock thickness) that controls the clustering of points near the origin $x=0$. The parameter $s$ is defined as $s = (L/s_{\infty})^2$, where $s_{\infty}$ represents the far-field truncation location, i.e. the domain half-width.

To apply this mapping to the spatial discretization, the Chebyshev differentiation matrices must be scaled by the terms of the transformation, i.e.~\ref{eqn:Chebmapping}. Taking the derivative of the physical coordinate $x$ with respect to the computational coordinate $y$ and expressing the result in terms of $x$, yields the first and second-order terms:
\begin{align}
\frac{dy}{dx} &= \frac{\sqrt{1+s} \cdot L^2}{(L^2 + x^2)^{3/2}} \\
\frac{d^2y}{dx^2} &= \frac{-3 \sqrt{1+s} \cdot L^2 x}{(L^2 + x^2)^{5/2}}
\end{align}

The first derivative matrix $D_x$ and second derivative matrix $D_{xx}$ are then computed by modifying the original Chebyshev differentiation matrices $D_y$ and $D_{yy}$:
\begin{align}
(D_x)_{ij} &= \left( \frac{dy}{dx} \right)_i (D_y)_{ij} \\
(D_{xx})_{ij} &= \left( \frac{dy}{dx} \right)_i^2 (D_{yy})_{ij} + \left( \frac{d^2y}{dx^2} \right)_i (D_y)_{ij}
\end{align}
where the terms are evaluated at each mapped physical collocation point $x_i$. This mapping successfully concentrates the collocation points inside the shock layer while allowing the boundaries to be placed sufficiently far away to apply asymptotic boundary conditions without excessive computational cost.

\subsection{Integration and Node Selection for Microvelocity Nodes}\label{sec:GH} 

In this work we use discrete microvelocity nodes to solve the Boltzmann equation for both the base flows and for the kLST. This leads to the natural selection of Gauss -Hermite quadratures. A Gauss-Hermite (GH) quadrature is a specialized Gaussian quadrature method for approximating integrals over the infinite interval $(-\infty, \infty)$ with the weight function $W(x) = e^{-x^2}$. It is particularly effective for integrals of the form
\[
I = \int_{-\infty}^{\infty} f(x) e^{-x^2} \, dx,
\]
which is an excellent choice for solving the flow problems with BE-BGK since the VDFs that will be integrated to generate the macroparameters converge towards a Maxwellian form that is essentially $f(x) e^{-x^2}$. The GH quadrature approximates $I$ as
\[
I \approx \sum_{i=1}^{n} W_i f(x_i),
\]
where $x_i$ are quadrature nodes (roots of Hermite polynomials) and $W_i$ are weights. Gauss-Hermite quadrature relies on the use of Hermite polynomials $H_n(x)$, defined recursively as
\[
H_0(x) = 1, \quad H_1(x) = 2x, \quad H_{n+1}(x) = 2x H_n(x) - 2n H_{n-1}(x), \quad n \geq 1.
\]
where these polynomials are orthogonal with respect to $e^{-x^2}$, i.e.;
\[
\int_{-\infty}^{\infty} H_m(x) H_n(x) e^{-x^2} \, dx = \sqrt{\pi} \, 2^n n! \, \delta_{mn}.
\]
The nodes $x_i$ are the roots of $H_n(x_i) = 0$, and the weights are
\[
W_i = \frac{2^{n-1} n! \sqrt{\pi}}{n^2 [H_{n-1}(x_i)]^2}.
\]
This ensures exact integration for polynomials of degree up to $2n-1$.

Gauss-Hermite quadrature is a subset of Gaussian quadrature, which minimizes the approximation errors by selecting optimal nodes and weights. The nodes and weights are computed using the Golub-Welsch algorithm, which constructs a symmetric tridiagonal Jacobi matrix from the recurrence coefficients of the Hermite polynomials to solve the eigenvalue problem. The nodes and weights are computed by forming the Jacobi matrix $T_n$ for monic Hermite polynomials, with subdiagonal elements $\beta_k = \sqrt{k/2}$ for $k = 1, \dots, n-1$. The eigenvalues of $T_n$ are the nodes, and the weights are derived from the first components of the normalized eigenvectors scaled by $\sqrt{\pi}$. Gauss-Hermite (GH) quadratures provide very fast convergence with a small number of nodes for the cases where the shift in the VDFS from zero velocity is small, i.e. for low bulk velocity cases. For example for the incompressible Couette flow case six microvelocity nodes were enough to obtain grid convergence~\cite{KarpuzcukLSTScitech2025}. However for high shock speeds as well as non-equilibrium distributions it becomes challenging to obtain convergence. The GH quadratures was found to still perform well up to M=4. 


\subsection{Numerical Discretization and BE-BGK Solver Implementation}\label{sec:BEBGKsolver} 

Similar to the reasoning of the kLST methodology, we first need to reduce the Boltzmann equation to 1D2V since the computational cost for the kLST results as will be shown later in Section~\ref{sec:1DShockAnalysis} would be too great for a 1D3V approach. Thus to reduce the computational cost while retaining the necessary kinetic information, we employ the reduced distribution function approach. Defining $g(x, \xi_x, \xi_y) = \int f d\xi_z$ and $h(x, \xi_x, \xi_y) = \int \xi_z^2 f d\xi_z$, the 1D3V BGK equation is reduced to a system of two 1D2V equations:
\begin{align}\label{eqn:BoltzmannBGK_reduced}
\frac{\partial g}{\partial t} + \xi_x \frac{\partial g}{\partial x} &= \nu (g^{e} - g) \\
\frac{\partial h}{\partial t} + \xi_x \frac{\partial h}{\partial x} &= \nu (h^{e} - h)
\end{align}
where $g^{e}$ and $h^{e}$ are the corresponding moments of the Maxwellian equilibrium distribution.

The spatial derivatives are discretized using a second-order Finite Volume Method (FVM) with Monotonic Upstream-Centered Scheme for Conservation Laws (MUSCL) reconstruction~\cite{vanLeer1979MUSCL}. This approach provides higher accuracy than first-order upwind schemes while avoiding spurious oscillations near the shock through the use of a Minmod limiter. For a generic variable $\phi$ (representing $g$ or $h$) at spatial cell $i$, the value at the right interface $i+1/2$ is reconstructed based on the sign of the molecular velocity $\xi_x$:
\begin{equation}
\phi_{i+1/2} = 
\begin{cases} 
\phi_i + \frac{1}{2} \psi(r_i^+) (\phi_i - \phi_{i-1}) & \text{if } \xi_x \ge 0 \\
\phi_{i+1} - \frac{1}{2} \psi(r_{i+1}^-) (\phi_{i+2} - \phi_{i+1}) & \text{if } \xi_x < 0
\end{cases}
\end{equation}
where $\psi(r)$ is the Minmod slope limiter. In the implementation, it is evaluated as $\psi(a,b) = \frac{1}{2}(\text{sgn}(a)+\text{sgn}(b))\min(|a|,|b|)$, where $a = \phi_i - \phi_{i-1}$ and $b = \phi_{i+1} - \phi_i$ denote the backward and forward slope differences, respectively, and $r^\pm$ represents the ratio of adjacent slopes. The advection term is then computed as:
\begin{equation}
\left. \xi_x \frac{\partial \phi}{\partial x} \right|_i \approx \xi_x \frac{\phi_{i+1/2} - \phi_{i-1/2}}{\Delta x_i}
\end{equation}
The system of equations in Eqns.~\ref{eqn:BoltzmannBGK_reduced} can be rewritten in terms of the state vector $q = [g, h]^\mathrm{tr}$ as $\frac{\partial q}{\partial t} = \text{RHS}(q)$, where the right-hand side vector explicitly contains the streaming and collision contributions for each reduced distribution function:
\begin{equation}
\text{RHS}(q) = 
\begin{bmatrix} 
\text{RHS}_g \\ 
\text{RHS}_h 
\end{bmatrix} =
\begin{bmatrix} 
\nu(g^e - g) - \xi_x \frac{\partial g}{\partial x} \\ 
\nu(h^e - h) - \xi_x \frac{\partial h}{\partial x} 
\end{bmatrix}
\end{equation}

Time integration is performed using the classical fourth-order Runge-Kutta (RK4) scheme to advance the solution.

The computational domain is discretized using a mapped Chebyshev grid that clusters points near the shock location ($x=0$) to resolve the steep gradients as detailed before, while extending sufficiently far upstream and downstream to apply Dirichlet boundary conditions based on the Rankine-Hugoniot relations. In the accompanying C++ implementation of this BGK solver, the physical node locations are generated by applying the same algebraic shock mapping used in the kLST spatial operator to Chebyshev collocation nodes on a reference interval, with mapping parameters controlling the cluster width and far-field extent; the shock reference position can be aligned with the peak slope of an initial Gilbarg--Paolucci profile or a specified offset.

Since the shock drift is a known problem when solving the 1D BE-BGK equations\cite{Giddens1971}, a re-centering scheme is implemented in the code. This process might not have been necessary if the spatial grid was uniform, however since we have a grid with initial clustering of points at $x=0$, drifting of the shock from this highly spatially resolved region to a region with fewer number of grid points means that the kLST results would not be resolved. Thus, an optional \emph{adaptive shock re-centering} procedure periodically relocates the algebraic map so that the clustered grid tracks the evolving shock: the bulk-velocity profile is interpolated onto a fine probe mesh, the location of maximum $|{\mathrm d}U/{\mathrm d}x|$ is identified, and if the normalized drift of this location relative to the current map center exceeds a user threshold, the spatial grid is rebuilt with an updated map center. The reduced distribution functions $g$ and $h$ and the recovered macroscopic fields are then transferred onto the new nodes by linear interpolation in $x$, and boundary conditions are reapplied. This keeps resolution concentrated in the shock layer without increasing the nominal number of cells. Independently, optional interior monitoring of mass, momentum, and energy consistency relative to the far-field reference fluxes can be enabled for diagnostics.

Parallelism is implemented with OpenMP on the shared-memory loops that dominate cost: the finite-volume RHS evaluation (MUSCL flux differencing and BGK relaxation) and the recovery of $\rho$, $\boldsymbol{u}$, $T$, $p$, $\nu$, and local Knudsen number from moments of $g$ and $h$ at interior cells. The number of threads is set from the solver input when OpenMP is enabled at compile time; the formulation remains identical to the serial path and is validated against single-thread results. Distributed-memory (MPI) parallelism is not used in this BGK base-flow code; long runs instead rely on periodic checkpointing of the full $g$/$h$ state and mapping parameters, with restart rebuilding the spatial grid from the stored map center.

This solver generates the steady-state VDFs that are then directly used as the base flow for the kLST analysis.

\section{Parallel Generalized Eigenvalue Solver for Large Sparse Matrix Systems}\label{sec:ParallelSolver} 
In this work we solve for the eigenvalues of the systems of type 
\begin{equation}\label{eqn:MatrixProblem}
A q = \omega B q
\end{equation}
as given in Eqn.~\ref{eqn:MatrixSystem}. One of the well-known algorithms to solve for such systems is called the QZ algorithm~\cite{QZ}. In this algorithm it is postulated that there are unitary Q and Z matrices such that $QAZ$ and $QBZ$ are upper triangular and \[QAZq = \lambda QBZq\] with the equivalent eigenvalues to the original problem. The main downside to this algorithm is that it solves for all the eigenvalues of the system, i.e. total number of eigenvalues is the size of the unknown $q$. In some of the cases considered in this work the total number of unknowns exceeds well over 100,000. However in applications such as modal linear stability analysis, one only needs a certain number of eigenvalues that are close to or greater than the zero imaginary parts of the temporal problem. Thus, it is unnecessary to solve for all the eigenvalues, especially when the size of the matrices are very large. For that reason, we utilize iterative eigenvalue solvers in this study.

For the kLST problem, we pursue two complementary iterative eigensolver strategies, each suited to a different matrix size and sparsity. The first approach couples the Implicitly Restarted Arnoldi Method (IRAM)~\cite{Arnoldi1951,Lanczos1952,RadkePhD1996,Sorensen1997} with an exact shift-and-invert spectral transformation computed via direct LU factorization using the MUMPS multifrontal solver~\cite{mumps}. This strategy delivers rapid convergence in very few outer iterations because each Arnoldi step applies the exact inverse operator. However it requires storing the full LU factors globally, which can exhaust available memory for very large systems. The second approach pairs the Jacobi-Davidson (JD) algorithm~\cite{SleijpenVanDerVorst1996} with an incomplete LU (ILU) preconditioner applied block-wise across MPI ranks. Rather than an exact spectral transformation, JD solves a correction equation approximately using GMRES~\cite{SaadSchultz1986} preconditioned by block-Jacobi ILU, trading per-iteration quality for drastically reduced memory and communication costs. This makes it the only viable option when problem dimensions exceed the memory capacity required for the exact LU factorization. Both approaches are implemented within the SLEPc/PETSc framework~\cite{slepc,petsc}, which manages the Krylov basis vectors, convergence checks, and implicit subspace restarts, while interfacing with PETSc for all distributed sparse matrix structures, sparse matrix-vector products (SpMV), Krylov solvers (KSP), and preconditioners (PC). Table~\ref{tab:algo_compare} summarizes the key algorithmic differences between the two strategies. We discuss below how these two strategies are applied to the kLST problem with the requirement that we can cover a high Mach number range.

\begin{table}[H]
\centering
\caption{Detailed algorithmic comparison of the two eigensolver strategies.}
\label{tab:algo_compare}
\begin{tabular}{p{3.4cm}p{5.0cm}p{5.0cm}}
\toprule
\textbf{Aspect} & \textbf{Arnoldi + Shift-Invert} & \textbf{Jacobi--Davidson + ILU}\\
\midrule
Subspace expansion &
  $\mathbf{v} \mapsto (A-\sigma B)^{-1}B\mathbf{v}$ (one exact LU solve/step) &
  Solve correction equation with GMRES\\
Spectral transform &
  Exact: $(A-\sigma B)^{-1}$ &
  Approximate: ILU of $(A-\sigma B)$ as precond.\\
Eigenvalue extraction &
  Standard or harmonic Ritz &
  Harmonic Ritz (mandatory for interior)\\
Inner solver &
  Exact MUMPS triangular solves &
  GMRES $+$ block-Jacobi ILU\\
Outer iterations &
  Very few &
  Many \\
Cost per iteration &
  Very high (global LU solve) &
  Low (local ILU apply $+$ SpMV halo)\\
MPI communication &
  All-to-all (MUMPS tree) per step &
  Neighbour halo exchange (SpMV) per step\\
Memory: preconditioner &
  Global LU: $\rho\cdot\mathrm{nnz}(A)$ &
  Local ILU: $\rho^{\mathrm{loc}}\cdot\mathrm{nnz}(A_i)/\mathrm{rank}$\\
Optimal for &
  Moderate-size, exact convergence needed &
  Large-scale, memory/comm limited\\
Fails when &
  LU fill exceeds available RAM &
  Block too small renders ILU useless\\
\bottomrule
\end{tabular}
\end{table}

\subsection{Direct LU Factorization with Shift-Invert Arnoldi}\label{sec:IRAMArnoldi}

The Implicitly Restarted Arnoldi Method (IRAM)~\cite{RadkePhD1996,Sorensen1997} is a Krylov subspace projection technique designed for computing a small number of eigenvalues and corresponding eigenvectors of large-scale, sparse, nonsymmetric matrices. It addresses the limitations of classical iterative methods like the QZ method and the basic Arnoldi process by combining efficient restarting strategies with spectral transformations, enabling robust solutions to problems where only matrix-vector products are feasible. This method is particularly suited for the kLST problem since the matrices become very large for high-speed flows since a large number of micro-velocity nodes are needed to achieve micro-velocity grid convergence. The approach builds on the Arnoldi factorization and draws inspiration from the implicitly shifted QR algorithm for dense eigenproblems, but truncates it for scalability~\cite{RadkePhD1996, Sorensen1997}.

IRAM operates within Krylov subspaces generated by the Arnoldi process. Starting from an initial vector $v_1$, the $m$-step Arnoldi factorization yields:
\[
AV_m = V_m H_m + f_m e_m^T,
\]
where $V_m$ is an $n \times m$ orthonormal matrix ($V_m^T V_m = I_m$), $H_m$ is an $m \times m$ upper Hessenberg matrix, $f_m = A v_m - V_m h_{m,m} e_m^T$ is the residual vector (with $\|f_m\| = h_{m+1,m}$), and $e_m$ is the $m$-th standard basis vector. The eigenvalues of $H_m$ (Ritz values $\theta(H_m)$) approximate those of $A$, and the corresponding eigenvectors (Ritz vectors) are obtained via $V_m y$, where $H_m y = \theta y$.

\begin{figure}[H]
\centering
\includegraphics[width=\linewidth]{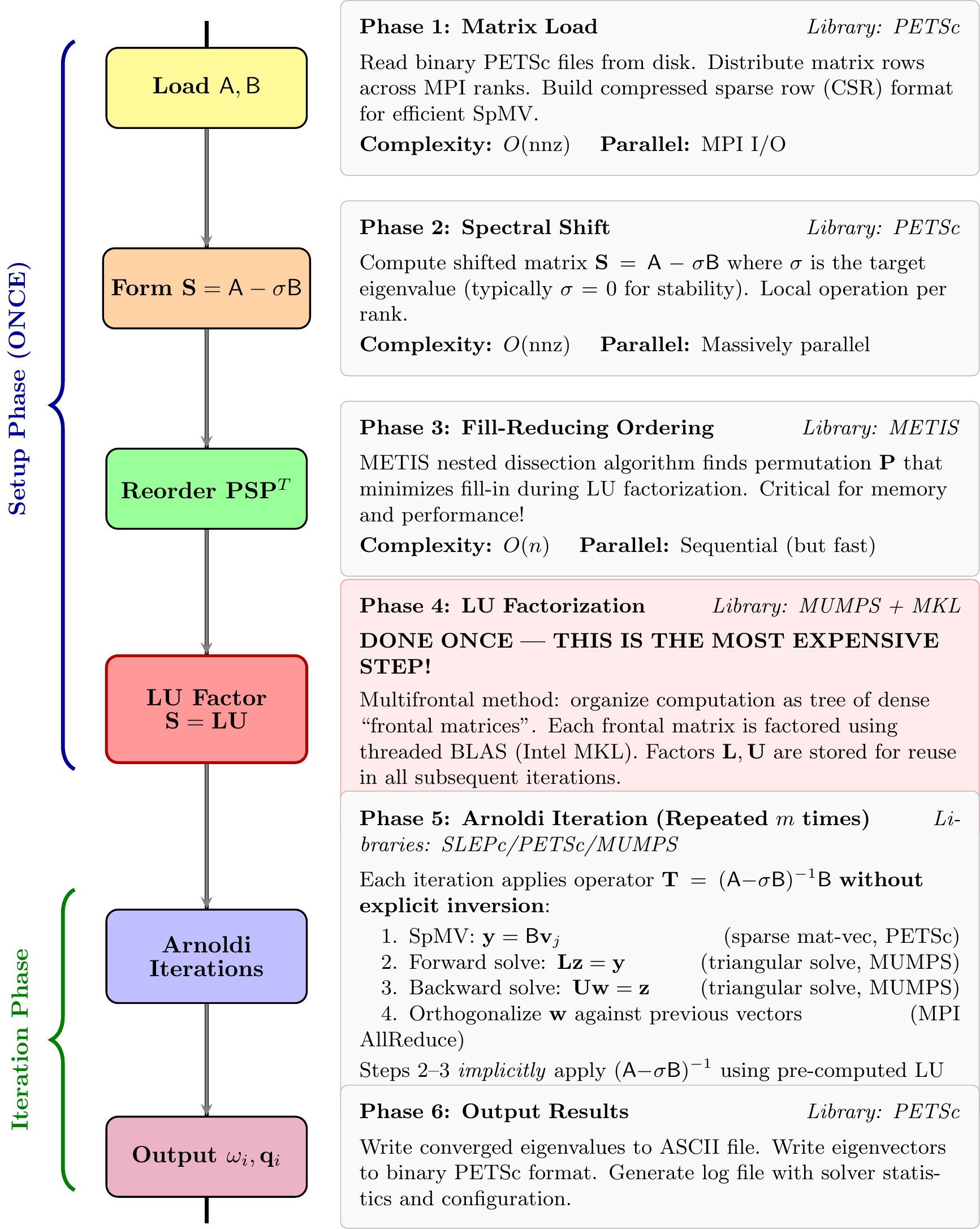}
\caption{\textbf{Complete execution timeline}. The Setup Phase runs once; the Iteration Phase repeats. The LU factorization (Phase 4) dominates setup time but enables fast triangular solves in all subsequent Arnoldi iterations.}
\label{fig:timeline}
\end{figure}

The method's restarting mechanism filters the subspace to focus on desired eigenvalues (e.g., those near a target value or in a spectral interval) while discarding unwanted components. This is achieved implicitly through a polynomial filter $\phi(\lambda)$ of degree $p = m - k$ (where $k$ is the number of desired eigenpairs), such that the restarted subspace approximates:
\[
\mathcal{K}_m(A, v_1) \approx \operatorname{Span}\{\phi(A) z_1, \phi(A) z_2, \dots, \phi(A) z_k\},
\]
with $\{z_j\}$ being approximate eigenvectors. Spectral transformations (e.g., shift-invert $(A - \sigma I)^{-1}$) can enhance convergence for interior or clustered eigenvalues. The detailed computational steps for the Arnoldi shift-invert pipeline are given in Fig.~\ref{fig:timeline}.

\begin{figure}[H]
\centering
\includegraphics[width=0.45\linewidth]{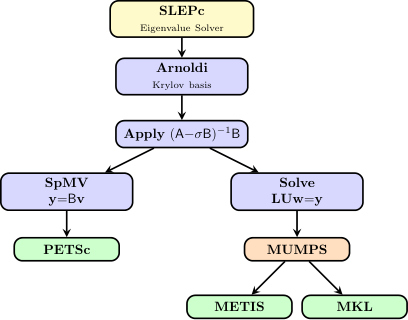}
\caption{Software stack for the parallel eigenvalue solver: SLEPc $\rightarrow$ PETSc/MUMPS $\rightarrow$ METIS/MKL.}
\label{fig:solverStack}
\end{figure}

The parallel solver implementation builds upon a hierarchical stack of numerical libraries, each providing specific capabilities essential for efficient large-scale eigenvalue computations. Fig.~\ref{fig:solverStack} shows how all software components connect, from the high-level eigenvalue solver down to the low-level linear algebra libraries.

To compute eigenvalues near the origin (least stable modes), we employ the shift-and-invert transformation. Instead of solving $\mathsf{A}\mathbf{q} = \omega \mathsf{B}\mathbf{q}$ directly, we solve:
\begin{equation}
    \underbrace{(\mathsf{A} - \sigma \mathsf{B})^{-1} \mathsf{B}}_{\mathbf{T}} \mathbf{q} = \mu \mathbf{q}, \quad \mu = \frac{1}{\omega - \sigma}
    \label{eqn:parShiftInvert}
\end{equation}
where $\sigma$ is the shift parameter (target eigenvalue). Eigenvalues $\omega$ near $\sigma$ correspond to large $|\mu|$, which Arnoldi finds first.

Crucially, rather than relying on iterative techniques for the demanding high-condition shifted matrix operations, PETSc passes the system to MUMPS~\cite{mumps}, an external package dedicated to distributed direct LU factorization. In the setup phase, the shifted matrix is decomposed via Gaussian elimination into a product of a lower triangular matrix $\mathbf{L}$ and an upper triangular matrix $\mathbf{U}$, such that $\mathsf{A}-\sigma\mathsf{B} = \mathbf{L}\mathbf{U}$. Because the kLST matrices are exceedingly large and sparse, computing this factorization can rapidly exhaust available memory if zero-valued elements become non-zero during elimination (a phenomenon known as ``fill-in''). To mitigate this, before MUMPS begins any factorization it invokes METIS~\cite{metis} to perform a fill-reducing analysis using a nested dissection algorithm. METIS determines a critical permutation matrix $\mathbf{P}$ guaranteeing that the resultant reordered matrix $\mathbf{P}\mathbf{S}\mathbf{P}^T$ incurs minimal structural fill-in during the LU decomposition, effectively reducing the necessary memory footprint and computational factorization time by several orders of magnitude.

MUMPS computes the LU factorization of the shifted matrix:
\begin{equation}
    \mathbf{P}(\mathsf{A} - \sigma\mathsf{B})\mathbf{Q} = \mathbf{L}\mathbf{U}
    \label{eqn:parLUfactor}
\end{equation}
where $\mathbf{P}, \mathbf{Q}$ are permutation matrices from METIS ordering.

By framing the shift-and-invert transformation explicitly through the multifrontal method, the formidable $O(n^3)$ complexity of dense matrix inversion is entirely bypassed. Given an optimal nested dissection ordering provided by METIS, the initial LU factorization cost is reduced to approximately $O(n^{1.5})$ for physically-mapped problems. Furthermore, for exceptionally large discretized domain requirements where classical exact LU bounds approach cluster limitations, MUMPS provides Block Low-Rank (BLR) compression options that analytically discard less consequential block data to further shrink the structural memory footprint by factors of 2 to 10.

It is important to emphasize that MUMPS is responsible exclusively for the LU factorization; it does not perform the Arnoldi iteration itself. Once the $\mathbf{L}$ and $\mathbf{U}$ factors are computed and stored, control returns to SLEPc, which drives the iterative Arnoldi process (Phase~5 in Fig.~\ref{fig:timeline}). At each Arnoldi step, SLEPc implicitly applies the inverse operator by first computing a sparse matrix-vector product $\mathbf{y} = \mathsf{B}\mathbf{v}_j$ via PETSc, and then solving the triangular systems $\mathbf{L}\mathbf{z} = \mathbf{y}$ (forward substitution) and $\mathbf{U}\mathbf{w} = \mathbf{z}$ (backward substitution) using the pre-computed MUMPS factors. This strategy avoids any explicit matrix inversion while applying the exact mathematical inverse operator, rendering each subsequent Arnoldi step computationally fast. Following the triangular solves, the new vector $\mathbf{w}$ is orthogonalized against all previously computed basis vectors using distributed dot products (requiring MPI AllReduce operations) and local AXPY updates managed by PETSc. Because the LU factorization is performed exactly once and all successive Arnoldi iterations seamlessly reuse the stored factors, the individual iterative substitution solve cost is reduced to $O(n \log n)$. This efficient scaling strictly hinges on the quality of the METIS ordering; poor permutations invariably lead to rapid fill-in explosions, causing memory allocations and execution times to surge uncontrollably.

The parallel characteristics of each operation in the solver pipeline, corresponding to the phases shown in Fig.~\ref{fig:timeline}, are summarized in Table~\ref{tab:parallelOps}.

\begin{table}[h!]
\caption{Parallel characteristics of each operation in the solver pipeline.}
\label{tab:parallelOps}
\centering
\begin{tabular}{p{3.5cm}p{3cm}p{4cm}p{3.5cm}}
\toprule
\textbf{Operation} & \textbf{Library} & \textbf{Parallel Model} & \textbf{Scaling} \\
\midrule
Matrix load & PETSc & MPI (parallel I/O) & Excellent \\
Form $\mathsf{A}-\sigma\mathsf{B}$ & PETSc & MPI (local ops) & Excellent \\
METIS ordering & METIS & Sequential & $O(n)$ \\
LU factorization & MUMPS & MPI + OpenMP & Good (tree-limited) \\
SpMV $\mathsf{B}\mathbf{v}$ & PETSc & MPI (row-distributed) & Excellent \\
Triangular solve & MUMPS & MPI + OpenMP & Moderate \\
Dot products & PETSc & MPI AllReduce & Poor at high $P$ \\
AXPY, norms & PETSc & Local + AllReduce & Moderate \\
\bottomrule
\end{tabular}

\end{table}



\subsection{Incomplete LU Factorization with Jacobi-Davidson}\label{sec:ILUJD}
When exact LU factorization becomes unfeasible due to memory or computational cost constraints such as at M=4 as occurs for the highest-dimensional kLST problems, the Jacobi, Davidson (JD) algorithm~\cite{SleijpenVanDerVorst1996} paired with an incomplete LU (ILU) preconditioner provides a viable alternative. An incomplete LU factorization ILU($p$) gives approximate factors $\tilde{L}$, $\tilde{U}$ by dropping fill entries whose level exceeds a prescribed threshold $p$:
\begin{equation}
  \tilde{L}\,\tilde{U} = C - R,
  \label{eq:ilu}
\end{equation}
where $R$ is the remainder matrix containing the dropped fill. Because $\tilde{L}\tilde{U} \neq C$ in general, these factors cannot serve as a direct solver; instead they are used as a preconditioner for an iterative solver such as GMRES. The fill level $p$ controls the trade-off between approximation quality and memory cost. At level zero, ILU(0) permits no fill beyond the original sparsity pattern of $C$, so the incomplete factors have the same sparsity structure as the original matrix. At level one, ILU(1) allows fill in positions that arise from a single elimination step: entry $(i,j)$ is retained if there exists an index $k$ such that both $(i,k)$ and $(k,j)$ belong to the original sparsity pattern $\mathcal{S}$, which significantly enriches the factor pattern and improves approximation quality. 

In distributed memory, each MPI rank $i$ owns a contiguous row-block $C_i$ of the global matrix. The block-Jacobi preconditioner applies ILU($p$) independently to each local block:
\begin{equation}
  M^{-1}_{\mathrm{BJ}} = \mathrm{diag}\!\left(
    (\tilde{L}_1\tilde{U}_1)^{-1},\,
    (\tilde{L}_2\tilde{U}_2)^{-1},\ldots,
    (\tilde{L}_p\tilde{U}_p)^{-1}\right).
  \label{eq:bjilu}
\end{equation}
Application of $M^{-1}_{\mathrm{BJ}}$ requires only local forward and backward triangular solves on $\tilde{L}_i$ and $\tilde{U}_i$, with no MPI communication. The trade-off is that couplings between different ranks' blocks are ignored, so preconditioner quality degrades as the block size shrinks with increasing rank count. Table~\ref{tab:lu_ilu_compare} summarises the key mathematical and algorithmic differences between exact LU and block-Jacobi ILU.

\begin{table}[H]
\centering
\caption{Mathematical and algorithmic comparison of ILU($p$) and exact LU.}
\label{tab:lu_ilu_compare}
\begin{tabular}{p{3.2cm}p{5.2cm}p{5.2cm}}
\toprule
\textbf{Aspect} & \textbf{Exact LU (MUMPS)} & \textbf{ILU($p$) block-Jacobi}\\
\midrule
Factorisation identity &
  $PAQ = LU$ exactly &
  $\tilde L\tilde U = C - R$, $R\neq0$\\
Fill control &
  Reduced only by reordering (AMD/METIS) &
  Hard cutoff: fill level $>p$ dropped\\
Solve quality &
  Exact (to machine eps) &
  Approximate; used as preconditioner for GMRES\\
Communication &
  All-to-all (multifrontal tree) &
  None (local triangular solves only)\\
Memory (global) &
  $O(\rho_{\mathrm{fill}}\cdot\mathrm{nnz}(C))$ &
  $O(\rho^{\mathrm{loc}}_{\mathrm{fill}}\cdot\mathrm{nnz}(C_i))$ per rank\\
Setup cost &
  $O(m^\alpha)$, $\alpha\approx1.5$--$2$ &
  $O\!\left((1+p)\cdot\mathrm{nnz}(C_i)\right)$ per rank\\
Quality vs.\ parallelism &
  Fixed (independent of rank count) &
  Degrades as block size shrinks\\
Scalability &
  Memory-limited &
  Good to 512+ ranks (this work)\\
\bottomrule
\end{tabular}
\end{table}

The Jacobi--Davidson (JD) method~\cite{SleijpenVanDerVorst1996} is a subspace expansion eigensolver that differs fundamentally from Arnoldi in how it grows the search space. Rather than applying a fixed linear operator to generate Krylov vectors, JD computes an approximate eigenvector (a Ritz vector) from the current search space and then expands the space by solving a \emph{correction equation} for a direction $\mathbf{t} \perp \tilde{\mathbf{x}}$ that improves the current approximation $\tilde{\mathbf{x}}$. Because the kLST eigenvalues of interest lie in the interior of the spectrum near the target shift $\sigma$, standard Ritz extraction is unreliable; instead, \emph{harmonic Ritz} pairs are employed, which satisfy
\begin{equation}
  (AV_m - BV_m\tilde\lambda)^H(AV_m - BV_m\tilde\lambda) \mathbf{y} = \mathbf{0}.
\end{equation}
This projects the residual onto the search space, providing robust eigenvalue estimates near the shift.

The JD iteration proceeds as follows. During setup, the block-Jacobi ILU(1) preconditioner is computed for the shifted matrix $(A - \sigma B)$ and an initial search vector is set. At each outer iteration, the small projected eigenproblem $M_m = V_m^H A V_m$, $N_m = V_m^H B V_m$ is formed and solved to extract a harmonic Ritz pair $(\tilde\lambda, \tilde{\mathbf{x}})$ and its residual $\mathbf{r} = A\tilde{\mathbf{x}} - \tilde\lambda B\tilde{\mathbf{x}}$. If the residual norm satisfies $\|\mathbf{r}\| < \texttt{tol} \cdot |\tilde\lambda|$, the eigenpair is accepted and deflated, and extraction continues for the next pair. Otherwise, the correction equation is solved approximately using GMRES(30) preconditioned by block-Jacobi ILU(1), the resulting vector is orthogonalised against the existing basis and appended to $V_m$. When the search space reaches ncv$=200$ vectors, SLEPc performs a restart retaining $k_{\min}=6$ directions, and the process repeats. Table~\ref{tab:parallelOpsJD} summarises the parallel characteristics of each operation in the JD solver pipeline.

\begin{table}[h!]
\caption{Parallel characteristics of each operation in the JD solver pipeline.}
\label{tab:parallelOpsJD}
\centering
\begin{tabular}{p{3.5cm}p{3cm}p{4cm}p{3.5cm}}
\toprule
\textbf{Operation} & \textbf{Library} & \textbf{Parallel Model} & \textbf{Scaling} \\
\midrule
Matrix load & PETSc & MPI (parallel I/O) & Excellent \\
Form $\mathsf{A}-\sigma\mathsf{B}$ & PETSc & MPI (local ops) & Excellent \\
Block-Jacobi ILU(1) setup & PETSc & Local per rank & Excellent \\
SpMV $(\mathsf{A}-\tilde\lambda\mathsf{B})\mathbf{v}$ & PETSc & MPI halo exchange & Excellent \\
ILU apply (precond.) & PETSc & Local per rank & Excellent \\
GMRES dot products & PETSc & MPI AllReduce & Moderate \\
JD orthogonalisation & SLEPc & MPI AllReduce & Poor at high $P$ \\
Projected eigenproblem & SLEPc & Replicated (small $m$) & N/A \\
\bottomrule
\end{tabular}
\end{table}

\begin{figure}[H]
\centering
\includegraphics[width=\linewidth]{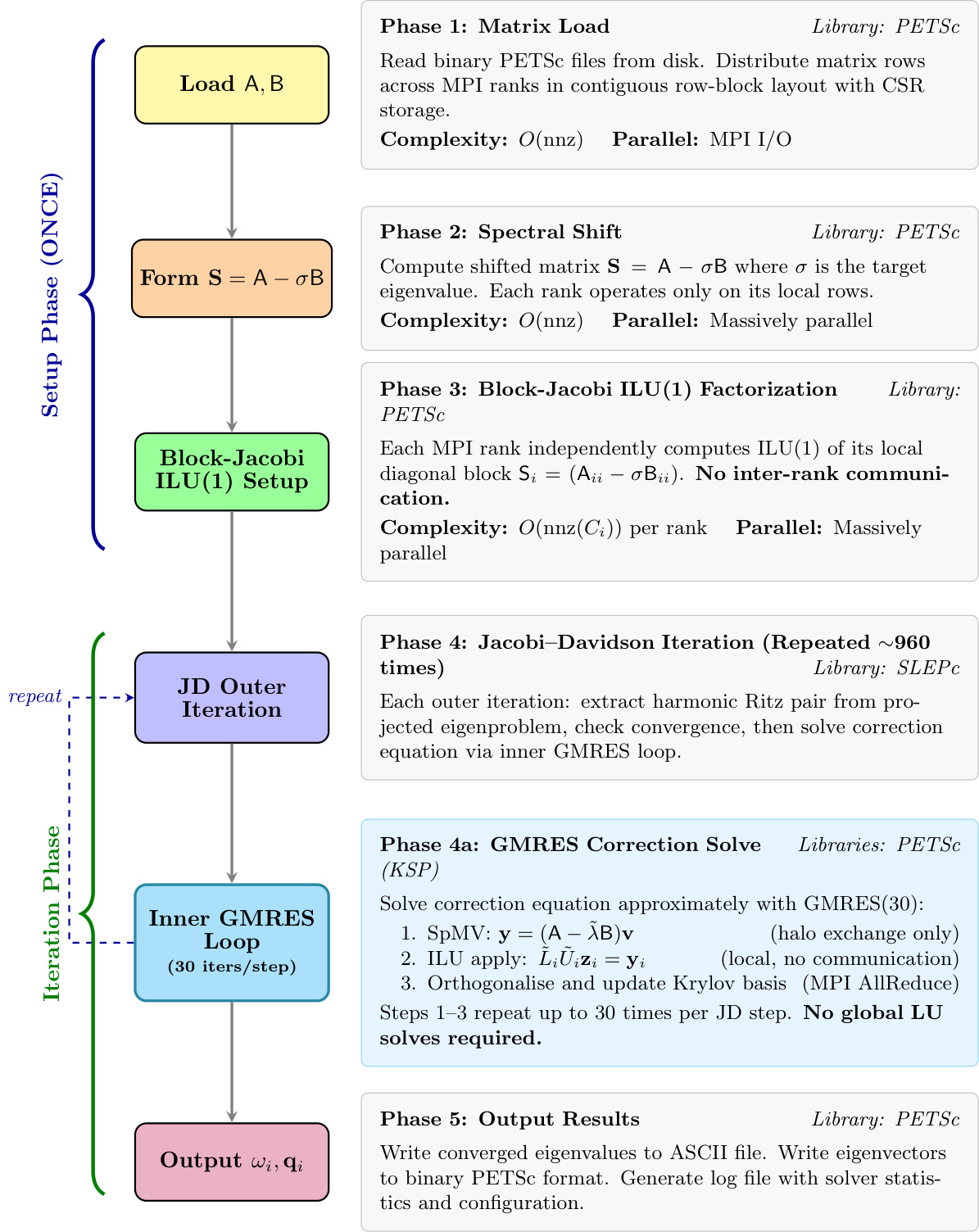}
\caption{\textbf{Complete execution timeline for the Jacobi--Davidson solver}. The Setup Phase replaces the expensive global LU factorization with massively parallel block-Jacobi ILU(1). The Iteration Phase features a nested loop: each JD outer step solves the correction equation via an inner GMRES(30) loop that requires only local ILU applies and neighbour halo exchanges.}
\label{fig:timelineJD}
\end{figure}

The complete execution timeline for the JD solver is shown in Fig.~\ref{fig:timelineJD}, which could be compared with the Arnoldi timeline in Fig.~\ref{fig:timeline}. The most notable difference is that the expensive global LU factorization (Phase~4 in Fig.~\ref{fig:timeline}) is replaced by a cheap, massively parallel block-Jacobi ILU setup, while the single-step Arnoldi operator application is replaced by a multi-step inner GMRES loop that iterates until the correction equation is solved to sufficient accuracy.

The JD eigensolver utilizes the same SLEPc/PETSc infrastructure and CSR-based matrix distribution as the Arnoldi solver but replaces the exact shift-invert transformation with an approximate preconditioned approach. Each MPI rank independently computes a local block-Jacobi ILU preconditioner, which is then applied within an inner GMRES loop to solve the correction equation. Although this approximate solve necessitates significantly more outer iterations than the exact Arnoldi approach, the JD steps are vastly cheaper and more scalable. By replacing global multifrontal LU factorization with a massively parallel ILU setup and limiting communication to nearest-neighbour halo exchanges and minimal all-reduces, the JD strategy remains competitive in wall-clock time for the largest kLST systems.

Before moving to the next section to delve into the linear stability analysis of the 1D shocks, it should be noted that the numerical framework was verified using classical compressible Couette flow extensively and compared against the results of Zou et al.~\cite{ZouetalCompCouetteBGK2024}; the details are given in Appendix~A, see Fig.~\ref{fig:CompCouetteEvals} and Table~\ref{LeastStableCompCouette}.

\subsection{High-Performance Computing Benchmarks}\label{sec:HPCBenchmarks}

To extend the kinetic linear stability theory framework to high Mach number flows, the solver must handle the extreme computational demands imposed by dense microscopic velocity discretizations and fine spatial grids. These requirements result in massive system matrices that exceed the memory and processing capacity of standard workstations. As described in Sections~\ref{sec:IRAMArnoldi} and \ref{sec:ILUJD}, the parallel eigenvalue solver addresses these challenges by combining SLEPc iterative methods with two distinct approaches, the first one is the direct LU factorization with Arnoldi solver, which will be referred to as Method-1 and the second one is incomplete LU factorization with Jacobi-Davidson solver, which will be referred to as Method-2.

For the performance benchmarks presented in this section, we evaluate the $M_\infty=1.2$ case at a spanwise wavenumber of $\beta=16$ and look for $k=50$ eigenvalues that are closest to the origin $0+0i$; the corresponding free-stream conditions are detailed in Table~\ref{tab:1DShockConditions}. The complete SLEPc/PETSc solver settings for all benchmark runs are provided in Appendix~C. Since one of the earliest applications of the Implicitly Restarted Arnoldi Method (IRAM)~\cite{Sorensen1997} was programmed specifically for MATLAB~\cite{RadkePhD1996} and is now embedded natively in its \texttt{eigs()} function, our baseline performance metric is derived from serial MATLAB solution times. The MATLAB baseline runs used MATLAB R2023b Update~6 (version 23.2.0.2485118, 64-bit \texttt{glnxa64}, released December~28,~2023). Throughout these benchmark studies, the Texas Advanced Computing Center's (TACC) Frontera supercomputer~\cite{Frontera} was utilized, specifically leveraging compute nodes equipped with dual-socket Intel Xeon Platinum 8280 (``Cascade Lake'') processors (56 cores total) and 192~GB of DDR4 memory. Throughout the benchmark discussion, we use the standard hybrid MPI+OpenMP terminology summarized in Table~\ref{tab:parallel_terminology}. An MPI rank is a distributed-memory process participating in the MPI communicator, while OpenMP threads are shared-memory worker threads spawned within each rank. 

\begin{table}[htbp]
  \centering
  \caption{Parallel-resource terminology used in the benchmark tables. The total core count assumes one OpenMP thread is bound to one physical CPU core and no oversubscription is used.}
  \label{tab:parallel_terminology}
  \begin{tabular}{lll}
    \toprule
    Symbol / column & Meaning & Relation \\
    \midrule
    Nodes & Number of Frontera compute nodes & $N_{\mathrm{node}}$ \\
    Ranks/node & MPI ranks placed on each node & $r_{\mathrm{node}}=n_r/N_{\mathrm{node}}$ \\
    MPI ranks & Total MPI ranks in the run & $n_r=N_{\mathrm{node}}r_{\mathrm{node}}$ \\
    OMP threads/rank & OpenMP threads spawned by each MPI rank & $t$ \\
    Total cores & Nominal CPU cores used by the hybrid run & $N_{\mathrm{core}}=n_r t$ \\
    Local rows & Matrix rows assigned to each MPI rank & $n/n_r$ \\
    \bottomrule
  \end{tabular}
\end{table}

Three velocity-space resolutions ($Q=20$, 24, and 32) are benchmarked on the fixed $N=80$ spatial grid, producing generalized eigenproblems with $n=2(N+1)Q^2$ complex unknowns and $\mathsf{A}\in\mathbb{C}^{n\times n}$. Table~\ref{tab:benchmark_cases} summarizes the matrix dimensions, sparsity, and storage requirements for each case. A dense treatment would allocate all $n^2$ complex entries regardless of the underlying BE-BGK structure. For the smallest benchmark ($Q=20$, $n=64{,}800$), this corresponds to more than $4.2\times10^9$ entries and roughly 63\,GB for $\mathsf{A}$ alone; for the finest case ($Q=32$, $n=165{,}888$), the count exceeds $2.7\times10^{10}$ entries and 410\,GB. These footprints do not include the mass matrix $\mathsf{B}$, Krylov basis vectors, or the much larger memory demands of an exact LU factorization, and they are far beyond the capacity of a typical workstation. Dense direct eigensolvers such as the QZ algorithm are therefore not a viable option for kLST at these resolutions. In practice, however, the kLST discretization is sparse: only about $1.3\%$ of the entries in $\mathsf{A}$ are nonzero. Storing $\mathsf{A}$ in compressed sparse row (CSR) form reduces the matrix size to approximately 1.3--7.8\,GB for the cases in Table~\ref{tab:benchmark_cases}, a reduction of roughly two orders of magnitude relative to dense storage. This sparse representation is essential for the parallel workflow developed in Sections~\ref{sec:IRAMArnoldi} and~\ref{sec:ILUJD}. Both solver strategies rely on repeated sparse matrix--vector products, and Method~1 additionally requires a sparse direct LU factorization whose cost scales with fill-in rather than with $n^2$. Neither operation is feasible if $\mathsf{A}$ is assembled densely. The benchmarks below therefore load $\mathsf{A}$ and $\mathsf{B}$ as sparse PETSc matrices and solve the eigenproblem with the iterative Arnoldi and Jacobi--Davidson solvers described above.

\begin{table}[htbp]
  \centering
  \caption{Summary of the benchmark cases used in this section for the $M_\infty=1.2$ normal shock with $N=80$ Chebyshev collocation points. The number of unknowns is $n=2(N{+}1)Q^2$. ``Dense entries'' counts all $n^2$ complex positions; dense storage assumes one \texttt{PetscScalar} ($a+\mathrm{i}b$, 16 bytes) per position. ``PETSc binary storage'' estimates the file size as $24\,\mathrm{nnz}+8(n{+}1)$ bytes (CSR values $+$ 64-bit column indices $+$ row pointers); the $Q=20$ entry matches the measured export file.}
  \label{tab:benchmark_cases}
  \begingroup
  \footnotesize
  \setlength{\tabcolsep}{2.5pt}
  \begin{tabular}{ccccccccc}
    \toprule
    $Q$ & $n$ & $\mathsf{A}$ dimensions & Dense entries ($n^2$) & $\mathrm{nnz}(\mathsf{A})$ & Sparsity & Dense storage & PETSc binary storage \\
    \midrule
    20 & \num{64800}  & $\num{64800}\times\num{64800}$   & \num{4.20e9}  & \num{56320800}   & 1.34\,\% & 62.6\,GB & 1.26\,GB \\
    24 & \num{93312}  & $\num{93312}\times\num{93312}$   & \num{8.71e9}  & \num{113542272} & 1.30\,\% & 129.7\,GB & 2.54\,GB \\
    32 & \num{165888} & $\num{165888}\times\num{165888}$ & \num{2.75e10} & \num{348653568}  & 1.27\,\% & 410.1\,GB & 7.79\,GB \\
    \bottomrule
  \end{tabular}
  \endgroup
\end{table}

\begin{figure}[htbp]
\centering
\subfigure[Continuum NS LSA ($n=387$)]{\label{fig:sparsity_NS}\includegraphics[trim=10 10 10 10,clip,width=0.45\linewidth]{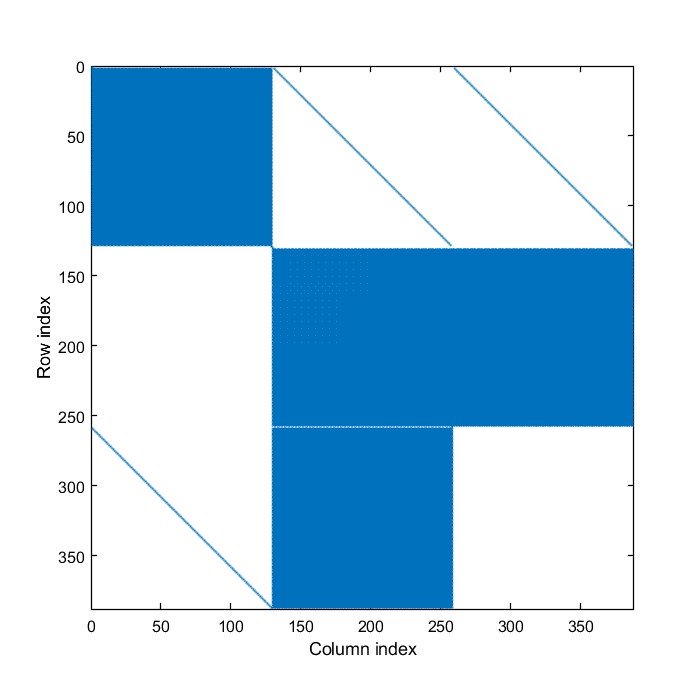}}
\subfigure[kLST extract ($350\times350$ submatrix)]{\label{fig:sparsity_kLST}\includegraphics[trim=10 10 10 10,clip,width=0.45\linewidth]{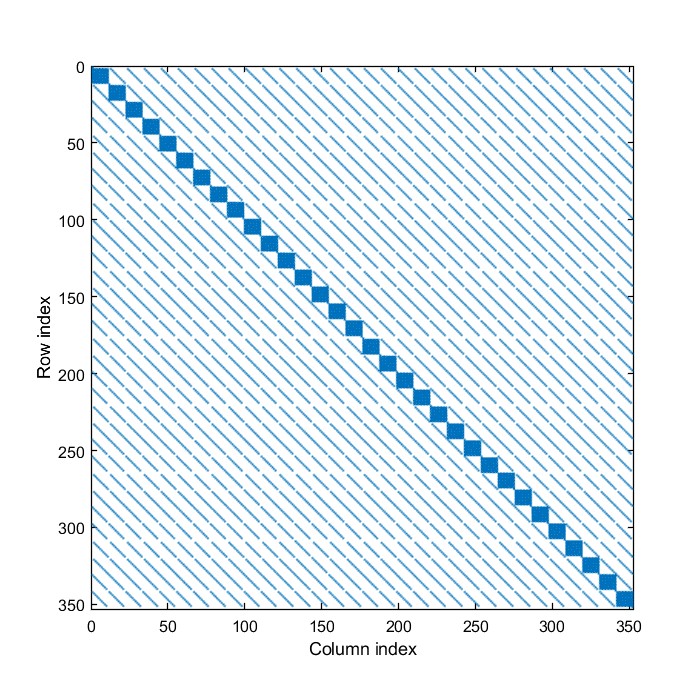}}
\caption{\textbf{Qualitative comparison of sparsity patterns} for continuum Navier--Stokes linear stability analysis (NS LSA) and kinetic linear stability theory (kLST). (a)~Incompressible NS operator with three macroscopic unknowns per spatial node ($\hat{u}$, $\hat{v}$, $\hat{p}$). (b)~Representative extract from the kLST stability matrix for $M_\infty=1.2$, $N=80$, $Q=20$. The panels differ in dimension and physics and are shown only for qualitative structural comparison.}
\label{fig:sparsity_patterns}
\end{figure}

Figure~\ref{fig:sparsity_patterns} compares the qualitative sparsity structure of a continuum Navier--Stokes linear stability operator with that of a kLST operator. The panels are not equivalent in size or physics; they illustrate why the kinetic formulation poses different demands on sparse solvers than classical macroscopic LSA~\cite{TheofilisLecture2014}. The NS operator in Fig.~\ref{fig:sparsity_NS} ($n=387$, assembled following the 1D LNSE framework of~\cite{TheofilisLecture2014}) is organized into three compact macroscopic blocks. Within each block, nonzeros lie in low-width bands set by Chebyshev collocation derivatives, with limited cross-field coupling. By contrast, the kLST extract in Fig.~\ref{fig:sparsity_kLST} displays many closely spaced diagonal bands, reflecting Gauss--Hermite velocity coupling and spatial streaming; the full $M_\infty=1.2$ operator ($N=80$, $Q=20$) has $n=64{,}800$ unknowns and the $2\times2$ block structure of Eqn.~\eqref{eqn:MatrixSystem}. Although Table~\ref{tab:benchmark_cases} shows that only $\approx1.3\%$ of the global kLST entries are nonzero, the effective bandwidth and fill-producing couplings are far larger than in the NS case.

These structural differences explain why kLST is substantially more demanding for Method~1 (direct multifrontal LU $+$ shift-and-invert Arnoldi) in the sections that follow.  First, the LU fill-in ratio grows with both $\mathrm{nnz}(\mathsf{A})$ and the topological width of the sparsity pattern; the multi-band kinetic coupling visible in Fig.~\ref{fig:sparsity_kLST} produces much heavier fill than the compact NS bands in Fig.~\ref{fig:sparsity_NS}, even when the baseline $N=81$, $Q=20$ case already stores $\approx2.9$B factor nonzeros (Table~\ref{tab:m12_nq_comparison}).  Second, the global kLST dimension ($n\ge 10^4$--$10^5$) forces the factorization into shared-memory or out-of-core regimes.  Third, the $N/Q$ discretization balance changes the \emph{shape} of the pattern, not merely its size, thereby altering both fill-in and the spectral distribution seen by shift-and-invert Arnoldi, this sensitivity is absent in fixed macroscopic NS discretizations.  Method~2 (block-Jacobi ILU $+$ Jacobi--Davidson) is less sensitive to exact fill-in but still pays a per-iteration cost proportional to the wider kLST bandwidth.  The benchmark results below should therefore be interpreted in light of Fig.~\ref{fig:sparsity_patterns}, kLST is harder not only because $n$ is larger, but because its sparsity pattern carries stronger velocity-space coupling that directly drives LU fill-in, memory footprint, and Arnoldi iteration counts.

Before presenting the runtime results, we note two constraints that guided the benchmark design. Hybrid MPI+OpenMP configurations scaled poorly or failed outright in both single-node and multi-node tests, so the runs below use either pure MPI (Method~2) or a single MPI rank with OpenMP threading within the rank (Method~1). Likewise, increasing the node count beyond the minimum needed to accommodate the LU factorization in aggregate memory degraded performance rather than improving it. The multi-node scaling studies therefore report results at the smallest node allocations that provide sufficient memory, with rank-thread sweeps carried out at those fixed node counts.

\subsubsection{Single Node Performance Benchmarks}\label{sec:SNBenchmarks}

First we consider the performance utilization of the two methods on single node applications, starting with the baseline problem size of $N=81, Q=20$, followed by an investigation into the effect of the sparsity pattern for similar sized problems but with different $N$ and $Q$ combinations. For the strong scaling study, Method~1 used one MPI rank with varying OpenMP threads per rank, whereas Method~2 used pure MPI with one OpenMP thread per rank and varying total rank count. The corresponding SLEPc/PETSc options are given in Table~\ref{tab:single_node_solver_settings} of Appendix~C.

\begin{figure}[h!]
\center
\subfigure[Run times (s)]{\label{fig:SingleNodeScalingM12runtimes}\includegraphics[trim=10 10 10 10,clip,width=0.45\linewidth]{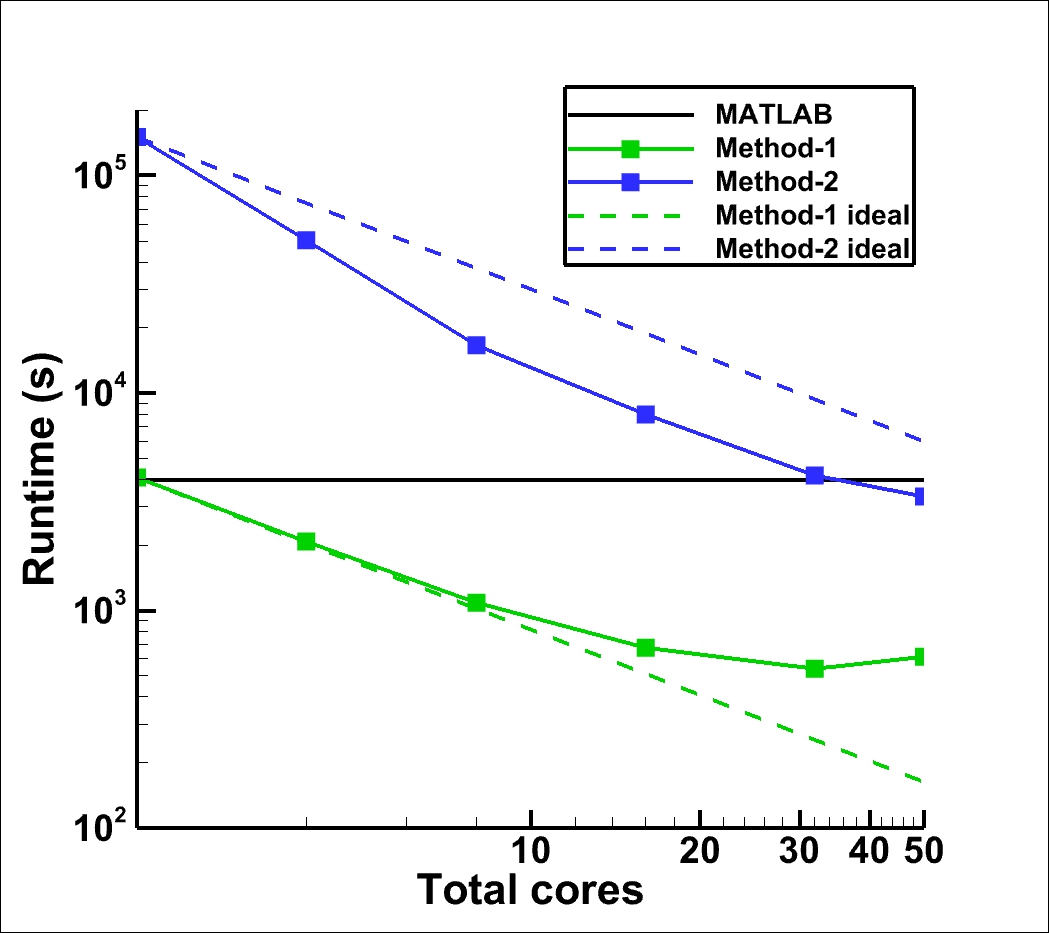}}
\subfigure[Speed-up compared to MATLAB]{\label{fig:SingleNodeScalingM12speed-up}\includegraphics[trim=10 10 10 10,clip,width=0.45\linewidth]{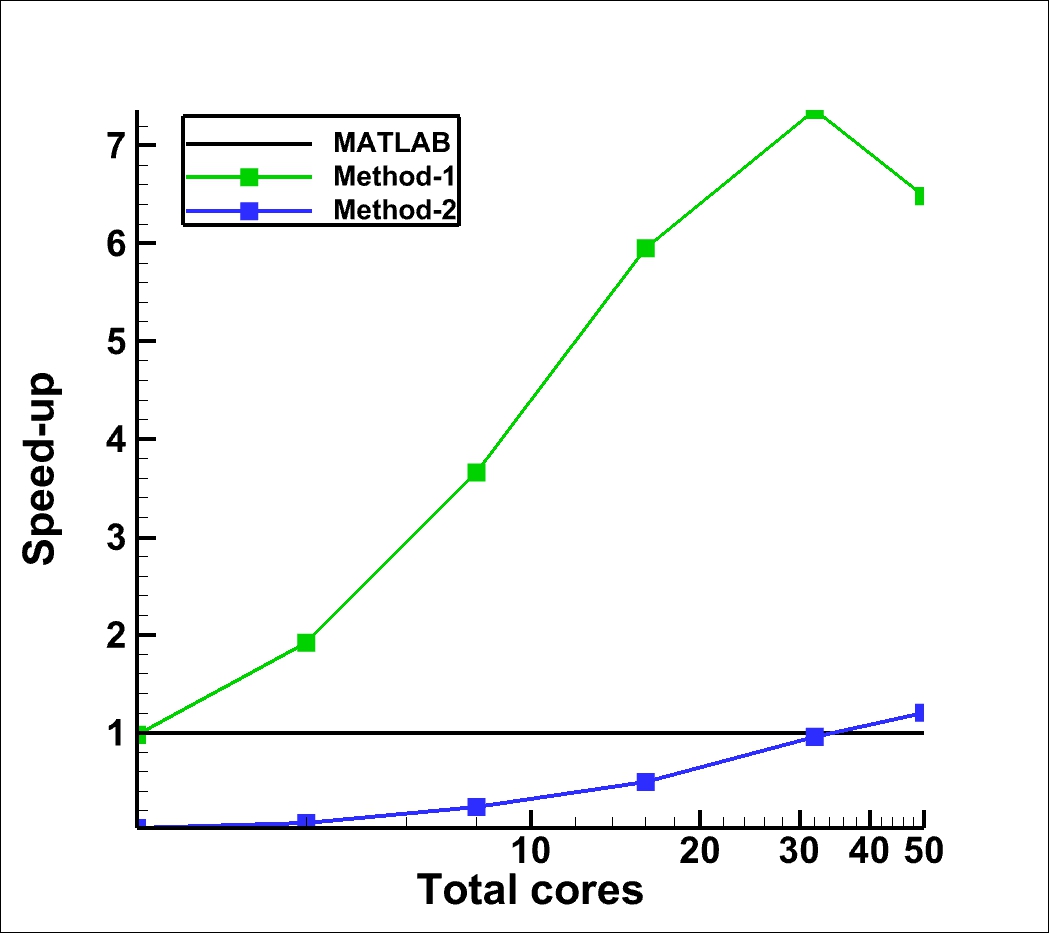}}
\caption{Single-node performance benchmarks for the $M_\infty=1.2$ normal shock stability case ($N=80, Q=20$) on the Frontera supercomputer. (a) Absolute wall-clock run times for Method 1 (LU + Arnoldi) and Method 2 (ILU + JD) as a function of the active scaling coordinate: OpenMP threads per rank for Method~1 and total MPI ranks for Method~2. (b) Parallel speed-up factors evaluated against the serial MATLAB baseline, plotted alongside the theoretical ideal strong scaling limits.}
\label{fig:SingleNodeScalingM12}
\end{figure}

Next, we present the results of the strong scaling study executed on a single compute node. The serial MATLAB execution (runtime: \num{4000}\,s) is used as the baseline against which Method~1 and Method~2 are evaluated where Table~\ref{tab:q20_single_node} lists the fixed single-node sweeps for the baseline $Q=20$ problem ($n=64{,}800$).  It must be explicitly noted that the MATLAB solver is inherently restricted in this context because we cannot directly distribute its workload across the varying MPI ranks or OpenMP threads shown on the abscissa of Fig.~\ref{fig:SingleNodeScalingM12}. Consequently, the MATLAB performance is plotted as a constant horizontal line across all process counts purely for visual representation and comparative baseline purposes, representing its fixed execution time on the allocated node. Fig.~\ref{fig:SingleNodeScalingM12runtimes} reports the absolute wall-clock run times for both parallel methods as a function of the number of active processes, plotted alongside the theoretical ideal strong scaling trajectories while Fig.~\ref{fig:SingleNodeScalingM12speed-up} illustrates the corresponding parallel speed-up relative to this fixed MATLAB baseline. It is immediately clear that Method 1 drastically outperforms both Method 2 and the MATLAB baseline, achieving a peak $7.3\times$ speed-up. Conversely, Method 2 exhibits a super-linear scaling behavior; this phenomenon occurs because as the problem is distributed over more MPI ranks, the local matrix sub-blocks shrink, leading to a disproportionately large reduction in the ILU factorization memory fill-in and processing time. Despite this favorable scaling trend, the heavy iterative nature of Method 2 dictates that its absolute performance only surpasses the MATLAB baseline when utilizing the maximum available cores on the single node. Nevertheless, for a fixed node-hour allocation on Frontera, both parallel approaches represent a tangible improvement over the MATLAB serial execution, with Method 1 definitively establishing itself as the superior choice for moderately sized problems that fit within a single node's shared memory.

\begin{table}[htbp]
  \centering
  \caption{Method~1 (LU-AR) OpenMP-thread sweep and Method~2 (ILU-JD) MPI-rank sweep on one fixed node for the $Q=20$ problem ($n=64{,}800$), compared with the serial MATLAB baseline (runtime: \num{4000}\,s).}
  \label{tab:q20_single_node}
  \begingroup
  \footnotesize
  \setlength{\tabcolsep}{2.5pt}
  \begin{tabular}{ccccccccc}
    \toprule
    Solver Method & Nodes & Ranks/node & MPI ranks & OMP threads/rank & Total cores & Local rows/rank & Runtime (s) & Speedup vs. MATLAB \\
    \midrule
    Method~1 (LU-AR) & 1 & 1 & 1 & 2  & 2  & 64{,}800 & 4{,}083 & 0.98$\times$ \\
                     & 1 & 1 & 1 & 4  & 4  & 64{,}800 & 2{,}083 & 1.92$\times$ \\
                     & 1 & 1 & 1 & 8  & 8  & 64{,}800 & 1{,}092 & 3.66$\times$ \\
                     & 1 & 1 & 1 & 16 & 16 & 64{,}800 & 672   & 5.95$\times$ \\
                     & 1 & 1 & 1 & 32 & 32 & 64{,}800 & \textbf{543}   & \textbf{7.36}$\times$ \\
                     & 1 & 1 & 1 & 50 & 50 & 64{,}800 & 617   & 6.48$\times$ \\
    \midrule
    Method~2 (ILU-JD) & 1 & 2  & 2  & 1 & 2  & 32{,}400 & 150{,}045 & 0.03$\times$ \\
                      & 1 & 4  & 4  & 1 & 4  & 16{,}200 & 50{,}182  & 0.08$\times$ \\
                      & 1 & 8  & 8  & 1 & 8  & 8{,}100  & 16{,}524  & 0.24$\times$ \\
                      & 1 & 16 & 16 & 1 & 16 & 4{,}050  & 8{,}004   & 0.50$\times$ \\
                      & 1 & 32 & 32 & 1 & 32 & 2{,}025  & 4{,}171   & 0.96$\times$ \\
                      & 1 & 50 & 50 & 1 & 50 & 1{,}296  & \textbf{3{,}332}  & \textbf{1.20}$\times$ \\
    \bottomrule
  \end{tabular}
  \endgroup
\end{table}

One important distinction of kLST matrices compared to their counterparts in continuum NS equations is their unique sparsity patterns, which heavily depend on the balance between spatial and micro-velocity discretizations (Fig.~\ref{fig:sparsity_patterns}). To further investigate the computational performance of the HPC solver, we constructed cases for $M_\infty=1.2$ with similar total unknown counts ($n \approx 65{,}000$) resulting from different combinations of Chebyshev collocation points ($N$) and Gauss-Hermite quadrature nodes ($Q$). The computational breakdown for Method 1 is detailed in Table~\ref{tab:sparsity_performance}. The sparsity sweep reveals a solver-specific sensitivity to the $N/Q$ balance that has no analogue in the compact NS pattern of Fig.~\ref{fig:sparsity_NS}. For Method 1, the $N=61, Q=24$ case is nearly as fast as the $N=81, Q=20$ baseline despite having more unknowns and more nonzeros.  In contrast, the $N=101, Q=18$ and $N=121, Q=16$ cases are much slower and require 39--44 Arnoldi iterations instead of 5--6.  This indicates that the performance change is not simply due to matrix size or raw sparsity fraction.  Increasing $N$ at fixed $n$ thins the spatial bands but widens the velocity-space coupling footprint within each spatial block, reshaping the pattern sketched in Fig.~\ref{fig:sparsity_kLST} and increasing LU fill-in (Table~\ref{tab:m12_nq_comparison}) while simultaneously stiffening the shift-and-invert spectrum; the lower-$Q$ cases therefore require many more Krylov steps. For Method 2, the three alternate sparsity cases remain in a narrower runtime band of roughly \SIrange{3935.8}{4712.0}{s}, but the iteration count still changes substantially.  The $N=121, Q=16$ case is the slowest Method 2 run and also has the largest JD iteration count, while the $N=61, Q=24$ case has the fewest JD iterations but is not the fastest because its higher \texttt{nnz(A)} raises the per-iteration matrix and preconditioner cost.

\begin{table}[h!]
\centering
\caption{Sparsity-pattern performance results. Method~1 uses one MPI rank with 32 OpenMP threads per rank, while Method~2 uses 50 MPI ranks with one OpenMP thread per rank.}
\label{tab:sparsity_performance}
\begin{tabular}{lS[table-format=4.1]S[table-format=2.0]S[table-format=4.1]S[table-format=4.0]}
\toprule
\multirow{2}{*}{\textbf{Case}} &
\multicolumn{2}{c}{\textbf{Method 1: Arnoldi + MUMPS}} &
\multicolumn{2}{c}{\textbf{Method 2: JD + ILU}}\\
\cmidrule(lr){2-3}\cmidrule(lr){4-5}
& {\textbf{Time (s)}} & {\textbf{Iters}} &
{\textbf{Time (s)}} & {\textbf{Iters}}\\
\midrule
$N=81, Q=20$  & 543.4  & 6  & 3331.5 & 1700\\
$N=61, Q=24$  & 561.0  & 5  & 4309.4 & 1566\\
$N=101, Q=18$ & 2223.2 & 39 & 3935.8 & 1882\\
$N=121, Q=16$ & 2522.7 & 44 & 4712.0 & 2428\\
\bottomrule
\end{tabular}
\end{table}

We next investigate the details of the computations for Method~1, which showed the largest performance variation between different sparsity patterns. A comparison of the configurations in Table~\ref{tab:m12_nq_comparison} demonstrates that while higher spatial resolutions generally lead to sparser matrices (e.g., $N=121, Q=16$ yields an $\mathrm{nnz}(A)$ of 38.8M compared to 83.8M for $N=61, Q=24$), they simultaneously cause severe degradation in solver convergence. As shown in Table~\ref{tab:m12_nq_comparison}, the initial LU factorization time scales favorably with the lower non-zero count, dropping from \SI{245.9}{s} for the $Q=24$ grid down to \SI{168.5}{s} for the $Q=16$ grid. However, the subsequent Arnoldi iteration phase exhibits dramatic sensitivity to this discretization balance. Configurations with high spatial resolution but lower micro-velocity resolution suffer from extremely poor spectral convergence. For instance, the $N=121, Q=16$ configuration requires 44 outer Arnoldi iterations and nearly 5,600 triangular solves, resulting in over \SI{2520}{s} of total execution time. This performance degradation occurs because an under-resolved velocity space fails to accurately represent the complex distribution function dynamics inside the shock layer, altering the eigenvalue distribution seen by the shift-invert Arnoldi process and introducing spectral stiffness that drastically hinders the extraction of the least stable modes. In contrast, increasing the velocity resolution to $Q=20$ ($N=81$) restores rapid convergence, requiring only 6 Arnoldi iterations and solving the entire system in \SI{543.4}{s}, a speedup of approximately $4.6\times$ over the $N=121, Q=16$ case despite the heavier LU factorization cost. These results underscore that for kinetic stability problems, the efficiency of Method~1 is dictated as much by the \emph{shape} of the sparsity pattern (Fig.~\ref{fig:sparsity_patterns}) and the resulting LU fill-in as it is by the raw nonzero count or the physical fidelity of the velocity grid alone. Maintaining sufficient velocity resolution is thus critical for achieving efficient parallel execution with direct solvers.

\begin{table}[h!]
\centering
\caption{Detailed computational breakdown for Method 1 (direct LU + Arnoldi) utilizing one MPI rank and 32 OpenMP threads per rank. The $M_\infty=1.2$ sparsity-pattern cases compare different spatial ($N$) and velocity ($Q$) resolutions while maintaining a nearly constant matrix size $n \approx 65{,}000$.}
\label{tab:m12_nq_comparison}
\begingroup
\small
\setlength{\tabcolsep}{3.5pt}
\begin{tabular}{lcccc}
\toprule
\textbf{Metric} & \textbf{$N{=}81$, $Q{=}20$} & \textbf{$N{=}61$, $Q{=}24$} &
\textbf{$N{=}101$, $Q{=}18$} & \textbf{$N{=}121$, $Q{=}16$} \\
\midrule
Matrix size $n$ & 64,800 & 70,272 & 65,448 & 61,952 \\
$\mathrm{nnz}(A)$ & 56.3M & 83.8M & 48.5M & 38.8M \\
$\mathrm{nnz}(\mathrm{LU})$ & 2.898B & 3.387B & 2.961B & 2.662B \\
Fill-in ratio & 51.46 & 40.43 & 61.09 & 68.55 \\
\midrule
LU factorization & \SI{195.8}{s} & \SI{245.9}{s} & \SI{197.5}{s} & \SI{168.5}{s} \\
Arnoldi iterations (outer) & 6 & 5 & 39 & 44 \\
PETSc \texttt{KSPSolve} calls & 758 & 611 & 4714 & 5599 \\
Tri.\ solves (total) & \SI{334.6}{s} & \SI{305.1}{s} & \SI{1983.1}{s} & \SI{2310.1}{s} \\
\midrule
Total time & \SI{543.4}{s} & \SI{561.0}{s} & \SI{2223.2}{s} & \SI{2522.7}{s} \\
Memory (MB) & 47,516 & 55,905 & 48,379 & 43,396 \\
\bottomrule
\end{tabular}
\endgroup
\end{table}

\subsubsection{Multi-Node Benchmarks}\label{sec:MNBenchmarks}

Having established the performance behavior for moderately sized single-node problems, we now extend our analysis to multi-node distributed environments necessary for larger system matrices. For that purpose we performed a multi-node scaling study with fixed-node rank--thread sweeps for the $Q=24$ and $Q=32$ cases. The corresponding multi-node SLEPc/PETSc options are given in Table~\ref{tab:multi_node_solver_settings} of Appendix~C.

Table~\ref{tab:q24_fixed_node} shows the fixed-node sweeps for both methods on the $Q=24$ problem, compared against the serial MATLAB baseline solver (runtime: \num{14290}\,s). For Method~1 on 4 nodes, switching to METIS ordering and sweeping OpenMP threads per rank from 5 to 50 results in a monotonic runtime reduction from 2{,}468\,s to 1{,}192\,s, achieving a 12.0$\times$ speedup over MATLAB. For Method~2 on 2 nodes, sweeping total MPI ranks from 10 to 100 shows that using 10 ranks is slower than MATLAB (28{,}667\,s vs. \num{14290}\,s), but increasing the rank count to 50 yields the optimal runtime of 4{,}952\,s (a 2.89$\times$ speedup over MATLAB). Fully populating the two nodes with 100 ranks degrades the runtime to 5{,}696\,s. This confirms that keeping a larger local block size ($\approx 1{,}866$ rows per rank) to maintain preconditioner quality is more efficient than maximizing raw CPU core utilization.

\begin{table}[htbp]
  \centering
  \caption{Method~1 (LU-AR) OpenMP-thread sweep on 4 fixed nodes and Method~2 (ILU-JD) MPI-rank sweep on 2 fixed nodes for the $Q=24$ problem (unknown-vector dimension $n=93{,}312$), compared with the serial MATLAB baseline (runtime: \num{14290}\,s).}
  \label{tab:q24_fixed_node}
  \begingroup
  \footnotesize
  \setlength{\tabcolsep}{2.5pt}
  \begin{tabular}{ccccccccc}
    \toprule
    Solver Method & Nodes & Ranks/node & MPI ranks & OMP threads/rank & Total cores & Local rows/rank & Runtime (s) & Speedup vs. MATLAB \\
    \midrule
    Method~1 (LU-AR) & 4 & 1  & 4 & 5  & 20  & 23{,}328 & 2{,}468  & 5.79$\times$ \\
                     & 4 & 1  & 4 & 10 & 40  & 23{,}328 & 1{,}664  & 8.59$\times$ \\
                     & 4 & 1  & 4 & 25 & 100 & 23{,}328 & 1{,}347  & 10.61$\times$ \\
                     & 4 & 1  & 4 & 50 & 200 & 23{,}328 & 1{,}192  & 11.99$\times$ \\
    \midrule
    Method~2 (ILU-JD) & 2 & 5  & 10  & 1 & 10  & 9{,}331 & 28{,}667 & 0.50$\times$ \\
                      & 2 & 10 & 20  & 1 & 20  & 4{,}665 & 12{,}925 & 1.11$\times$ \\
                      & 2 & 25 & 50  & 1 & 50  & 1{,}866 & \textbf{4{,}952}  & \textbf{2.89}$\times$ \\
                      & 2 & 50 & 100 & 1 & 100 & 933   & 5{,}696  & 2.51$\times$ \\
    \bottomrule
  \end{tabular}
  \endgroup
\end{table}

For the larger $Q=32$ problem (unknown-vector dimension $n=165{,}888$), Method~1 required a 64-node allocation to provide sufficient aggregate memory for the direct LU factorization. Table~\ref{tab:q32_fixed_node} shows the fixed 64-node OpenMP sweep for Method~1 and the fixed 4-node MPI-rank sweep for Method~2 on the $Q=32$ problem, compared against the serial MATLAB baseline solver (runtime: \num{119953}\,s). Method~1 obtains its best runtime at 25 OpenMP threads per rank, where the 64-node run completes in 1{,}232\,s and converges 41 eigenvalues in one Arnoldi iteration, achieving a 97.4$\times$ speedup over MATLAB. Method~2 runtimes scale positively from 41{,}024\,s at 20 ranks (5 ranks per node) to an optimal runtime of 11{,}273\,s at 100 ranks (25 ranks per node), yielding a 10.6$\times$ speedup over MATLAB, before degrading to 13{,}969\,s at 200 ranks (50 ranks per node).

\begin{table}[htbp]
  \centering
  \caption{Method~1 (LU-AR) OpenMP-thread sweep on 64 fixed nodes and Method~2 (ILU-JD) MPI-rank sweep on 4 fixed nodes for the $Q=32$ problem (unknown-vector dimension $n=165{,}888$), compared with the serial MATLAB baseline (runtime: \num{119953}\,s).}
  \label{tab:q32_fixed_node}
  \begingroup
  \footnotesize
  \setlength{\tabcolsep}{2.5pt}
  \begin{tabular}{ccccccccc}
    \toprule
    Solver Method & Nodes & Ranks/node & MPI ranks & OMP threads/rank & Total cores & Local rows/rank & Runtime (s) & Speedup vs. MATLAB \\
    \midrule
    Method~1 (LU-AR) & 64 & 1 & 64 & 5  & 320   & 2{,}592 & 2{,}062          & 58.2$\times$ \\
                     & 64 & 1 & 64 & 10 & 640   & 2{,}592 & 1{,}410          & 85.1$\times$ \\
                     & 64 & 1 & 64 & 25 & 1{,}600 & 2{,}592 & \textbf{1{,}232} & \textbf{97.4}$\times$ \\
                     & 64 & 1 & 64 & 50 & 3{,}200 & 2{,}592 & 1{,}303          & 92.0$\times$ \\
    \midrule
    Method~2 (ILU-JD) & 4 & 5  & 20  & 1 & 20  & 8{,}294 & 41{,}024          & 2.92$\times$ \\
                      & 4 & 10 & 40  & 1 & 40  & 4{,}147 & 17{,}953          & 6.68$\times$ \\
                      & 4 & 25 & 100 & 1 & 100 & 1{,}659 & \textbf{11{,}273} & \textbf{10.6}$\times$ \\
                      & 4 & 50 & 200 & 1 & 200 & 829     & 13{,}969          & 8.59$\times$ \\
    \bottomrule
  \end{tabular}
  \endgroup
\end{table}

The scaling performance of Method~2 is governed by a per-rank saturation window of 900 to 1700 local rows per MPI rank. This empirical window represents the optimal balance between the local block-Jacobi ILU(1) preconditioning quality and parallel communication overhead. Ranks with fewer rows suffer from collapsed preconditioner fill levels and high Krylov synchronization latency, whereas ranks with too many rows are dominated by local solve costs. This saturation window provides a robust guideline for estimating the optimal rank allocation $n_r^{\star} \approx n / n_{\mathrm{loc}}^{\star}$ for large-scale stability analyses.

\subsubsection{Extension to Larger Problems}\label{sec:LargeProblemExtension}

The scaling benchmarks above characterize solver performance at $M_\infty=1.2$, but the highest-speed shock cases considered in this study ($M_\infty=3.0$ and $4.0$) require substantially finer velocity-space resolution and therefore much larger discrete systems. Extending the kLST implementation to these regimes introduces practical challenges that the moderate-size benchmarks do not fully expose. For Method~1, the first and most consequential of these is matrix ordering: while Approximate Minimum Degree (AMD) ordering is inexpensive to compute, it produces excessive fill-in during direct LU factorization on the multi-band, velocity-space-coupled structures inherent to kinetic formulations (Fig.~\ref{fig:sparsity_kLST}), which bear little resemblance to the narrow-band NS pattern of Fig.~\ref{fig:sparsity_NS}. METIS nested dissection is therefore strongly preferred, as it exploits the topological structure of the spatial and velocity grids to reduce both peak memory usage and factorization time. For instance, in a large-scale setup with $261{,}792$ unknowns, METIS significantly mitigates peak memory usage and factorization time compared to AMD. By contrast, Method~2 solver settings transfer with little modification across problem sizes, relying consistently on a local block-Jacobi ILU(1) preconditioner and an inner GMRES loop to solve the correction equation without an exact global factorization.

These ordering and preconditioning considerations, together with the global memory cost of exact LU factorization, ultimately govern solver selection for the most demanding high-speed cases. While Method~1 (Arnoldi + MUMPS) is highly robust and performs exceptionally well for moderately sized kLST problems whose sparsity pattern remains sufficiently ``velocity-resolved'' (cf.\ Fig.~\ref{fig:sparsity_patterns} and Table~\ref{tab:sparsity_performance}), its global memory footprint scales severely with the non-zeros of the exact LU factorization. For the $M_\infty=4.0$ normal shock stability analysis requiring an ultra-fine grid of $N=61, Q=48$ (resulting in $281{,}088$ unknowns), Method~1 suffers from an intractable memory bottleneck during the direct factorization phase. In this massive computational regime, Method~2 (Jacobi-Davidson + ILU) becomes the only viable strategy. Its local ILU preconditioning maintains a modest and distributed memory footprint, allowing it to scale across multi-node allocations. As shown in Table~\ref{tab:solver_comparison}, Method~2 successfully solves the massive $M_\infty=4.0$ system in under three hours, whereas Method~1 is restricted to solving the moderately smaller $M_\infty=3.0$ case ($N=81, Q=32$, $165{,}888$ unknowns) and cannot scale up to the $M_\infty=4.0$ requirements.

\begin{table}[h!]
\centering
\caption{Comparison between Arnoldi with direct LU ($M_\infty=3.0$ case with N=81 and Q=32) and Jacobi-Davidson with iterative ILU ($M_\infty=4.0$ case with N=61 and Q=48).}
\label{tab:solver_comparison}
\begin{tabular}{lccc}
\toprule
Metric & Arnoldi + MUMPS & JD + ILU (S8) \\
\midrule
Matrix Size & 165,888 & 281,088 \\
Total Wall Time & 4.71 h & 2.89 h \\
Memory Footprint & 1.83 TB (Global) & $\approx$ 30 GB (Local) \\
Outer Iterations & 58 & 958 \\
Cost per Iteration & 291.9 s & 10.8 s \\
Scalability & Limited by memory & Good to 512 ranks \\
\bottomrule
\end{tabular}
\end{table}

The suitability of these solvers is ultimately determined by the structure of the eigenvalue spectrum. The 1D normal shock problem contains a continuous spectrum that appears as dense clusters of discrete modes in the complex plane. This clustering enables the JD solver to function effectively even with a coarse ILU preconditioner because the correction equation has many approximate solutions that guide the solver toward the target branch. Conversely, the discrete spectrum of compressible Couette flow consists of well-separated, isolated modes. This structure requires an exact or highly accurate inner solve to identify the specific direction of each mode, causing the JD and ILU strategy to fail where the exact shift-invert Arnoldi method succeeds. The dense modal clusters of the shock problem are thus a key factor that enables the use of memory-efficient iterative solvers for high-speed stability problems.

\section{Kinetic Linear Stability Analysis of 1D Shocks}\label{sec:1DShockAnalysis}

In this section, we apply the kinetic linear stability framework to evaluate the modal stability of one-dimensional argon shocks across a range of Mach numbers. The primary objective is to investigate the effect of non-Maxwellian distributions (referred to as translational non-equilibrium) on the resulting eigenvalue spectra and disturbance modes. For that purpose, three different Mach numbers are chosen which results in varying non-equilibrium levels in the VDFs due to the bi-modal distribution of VDFs in shock layers. The base flows for these analysis are obtained from the BE-BGK solver and directly used in the linear stability analyses with both using the same spatial and micro-velocity grid.  The free-stream conditions and argon gas properties used for all the one-dimensional shock flow analyses at various Mach numbers are summarized in Table~\ref{tab:1DShockConditions}. The Prandtl number is exactly 1.0 due to the use of the BGK collision operator and to obtain the G\&P base flow it is also set as Pr=1 to have one-to-one comparison.

\begin{table}[htbp]
\centering
\caption{Free-stream conditions and argon molecular properties for 1D normal shock cases.}
\begin{tabular}{llcc}
\toprule
\textbf{Parameter} & \textbf{Symbol} & \textbf{Value} & \textbf{Unit} \\
\midrule
\multicolumn{4}{l}{\textit{Free-stream Conditions}} \\
Mach Number & $M_{\infty}$ & 1.2, 3.0, 4.0 & - \\
Temperature & $T_{\infty}$ & 300.0 & K \\
Pressure & $P_{\infty}$ & 41.4 & Pa \\
\midrule
\multicolumn{4}{l}{\textit{Argon Gas Properties}} \\
Specific Heat Ratio & $\gamma$ & 5/3 & - \\
Specific Gas Constant & $R$ & 208 & J/(kg$\cdot$K) \\
Atomic Mass & $m_{Ar}$ & $6.63 \times 10^{-26}$ & kg \\
Collision Diameter & $d_{Ar}$ & $4.17 \times 10^{-10}$ & m \\
Reference Viscosity (at $300$ K) & $\mu_\infty$ & $2.688 \times 10^{-5}$ & Pa$\cdot$s \\
Viscosity Exponent & $s$ & 0.5 & - \\
Prandtl Number (BGK) & $Pr$ & 1.0 & - \\
\bottomrule
\end{tabular}
\label{tab:1DShockConditions}
\end{table}

\begin{table}[htbp]
\centering
\caption{Calculated shock thickness ($\Delta$), Reynolds number ($Re$), and Knudsen number ($Kn$) for the 1D normal shock cases based on the free-stream conditions.}
\begin{tabular}{lccc}
\toprule
\textbf{\textbf{$M_\infty$}} & \textbf{$\Delta$ (mm)} & \textbf{$Re_\Delta$} & \textbf{$Kn_\Delta$} \\
\midrule
1.2 & 2.548 & 24.34 & 0.081 \\
3.0 & 0.405 & 9.68 & 0.511 \\
4.0 & 0.346 & 11.03 & 0.598 \\
\bottomrule
\end{tabular}
\label{tab:1DShockThickness}
\end{table}

Table~\ref{tab:1DShockThickness} summarizes the resulting shock thickness ($\Delta$), Reynolds number ($Re_\Delta$), and Knudsen number ($Kn_\Delta$) for each of the three Mach number cases. We utilize shock thicknesses from G\&P results to define the length scale, as these can be calculated a priori to the BGK solutions. This approach avoids the complexity of BGK solvers, offering a more accessible method for readers to verify results using a simpler G\&P solver. The corresponding Reynolds and Knudsen numbers are formally defined as:
\begin{equation}
Re_\Delta = \frac{\rho_\infty u_\infty \Delta}{\mu_\infty}, \quad Kn_\Delta = \frac{\lambda_\infty}{\Delta}
\end{equation}
where the upstream mean free path for the given argon free-stream conditions evaluates to $\lambda_\infty \approx 0.207$ mm. As the Mach number increases from 1.2 to 4.0, the shock thickness decreases about an order of magnitude. This sharp steepening of the shock layer causes a corresponding drop in the Reynolds number and a significant rise in the Knudsen number. The elevated Knudsen numbers at higher speeds highlight the strongly rarefied and translationally non-equilibrium nature of the internal shock structure.

\begin{figure}[H]
\center
\subfigure[$M_\infty=1.2$ base flow]{\label{fig:1DShockProfilesM12base}\includegraphics[trim=10 10 10 10,clip,width=0.40\linewidth]{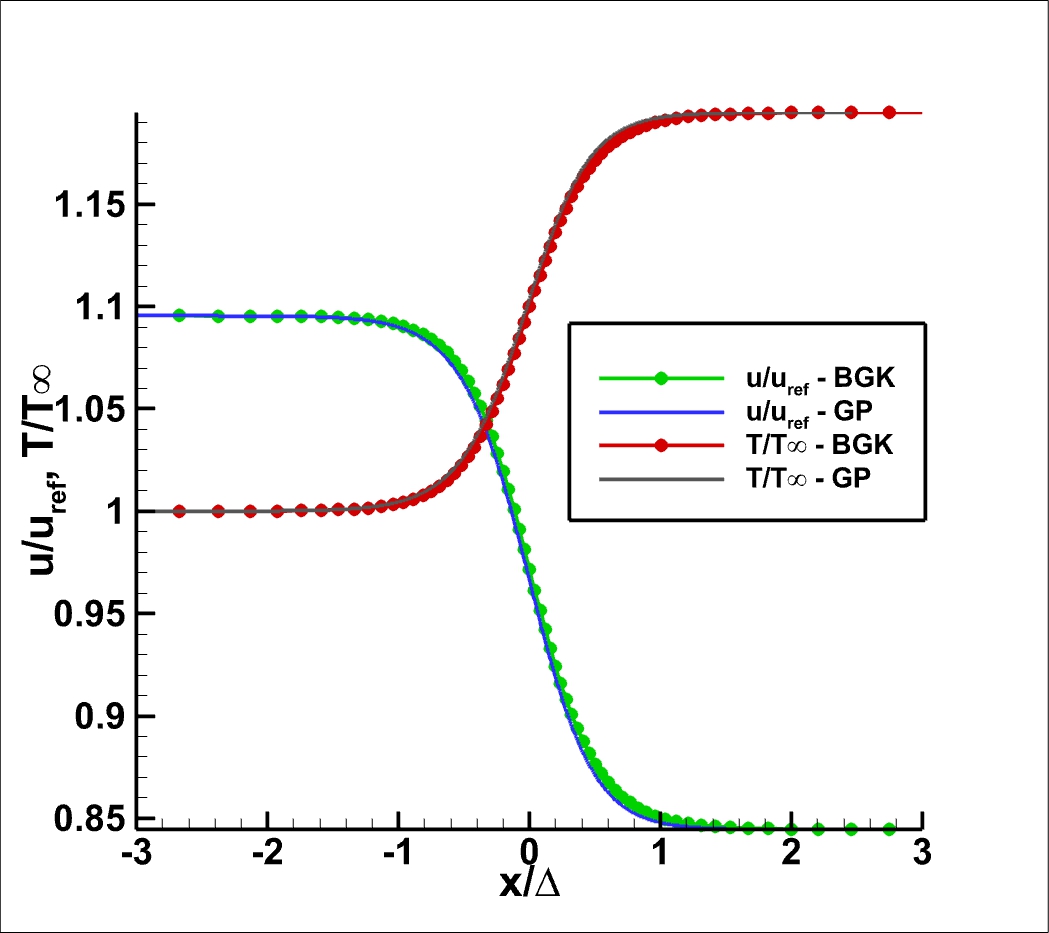}}
\subfigure[$M_\infty=1.2$ non-equilibrium]{\label{fig:1DShockProfilesM12nonequi}\includegraphics[trim=10 10 10 10,clip,width=0.40\linewidth]{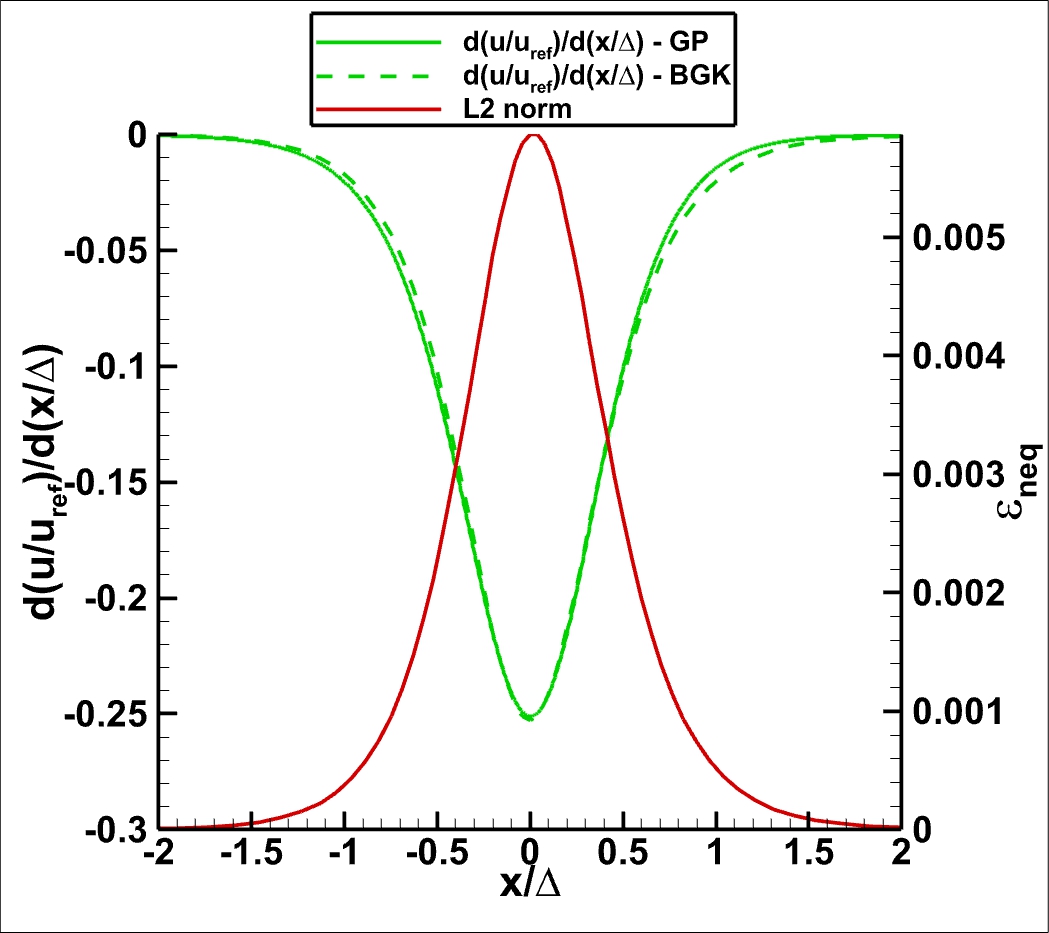}}
\subfigure[$M_\infty=3.0$ base flow]{\label{fig:1DShockProfilesM3Macro}\includegraphics[trim=10 10 10 10,clip,width=0.40\linewidth]{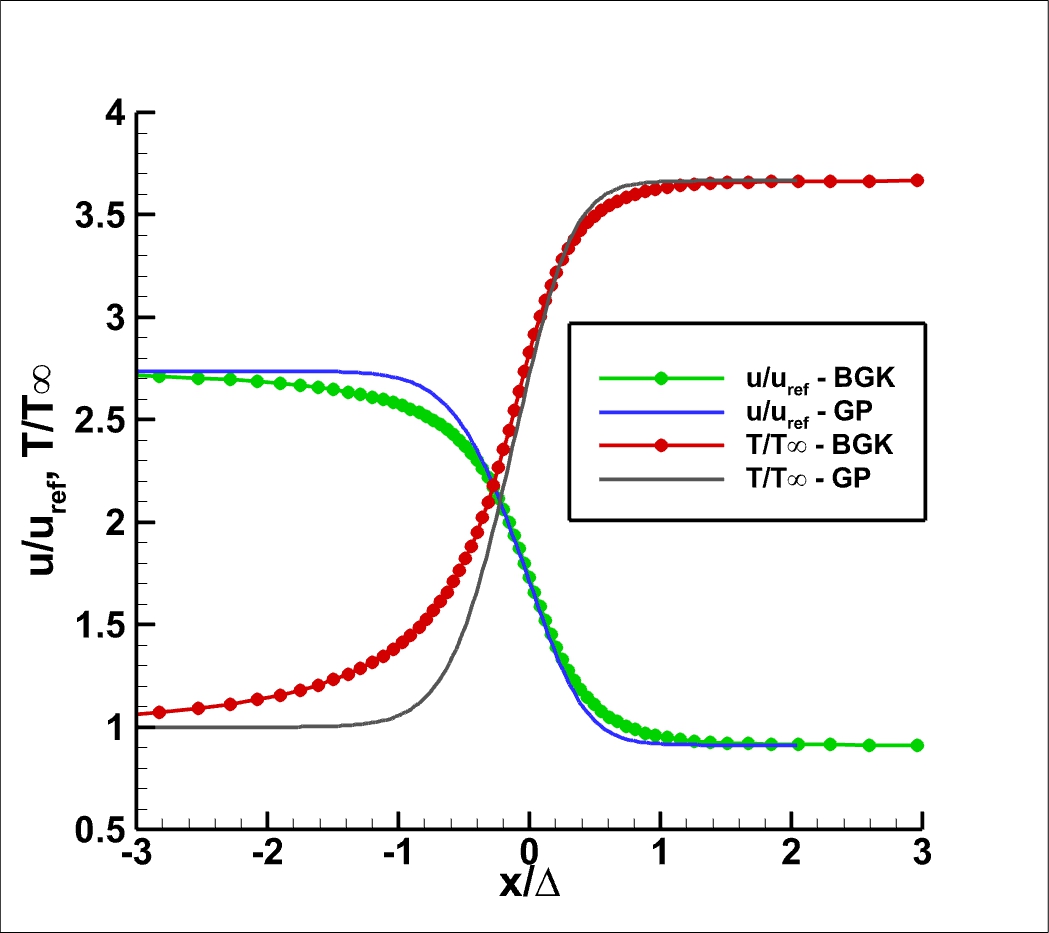}}
\subfigure[$M_\infty=3.0$ non-equilibrium]{\label{fig:1DShockNONEqLevelM3}\includegraphics[trim=10 10 10 10,clip,width=0.40\linewidth]{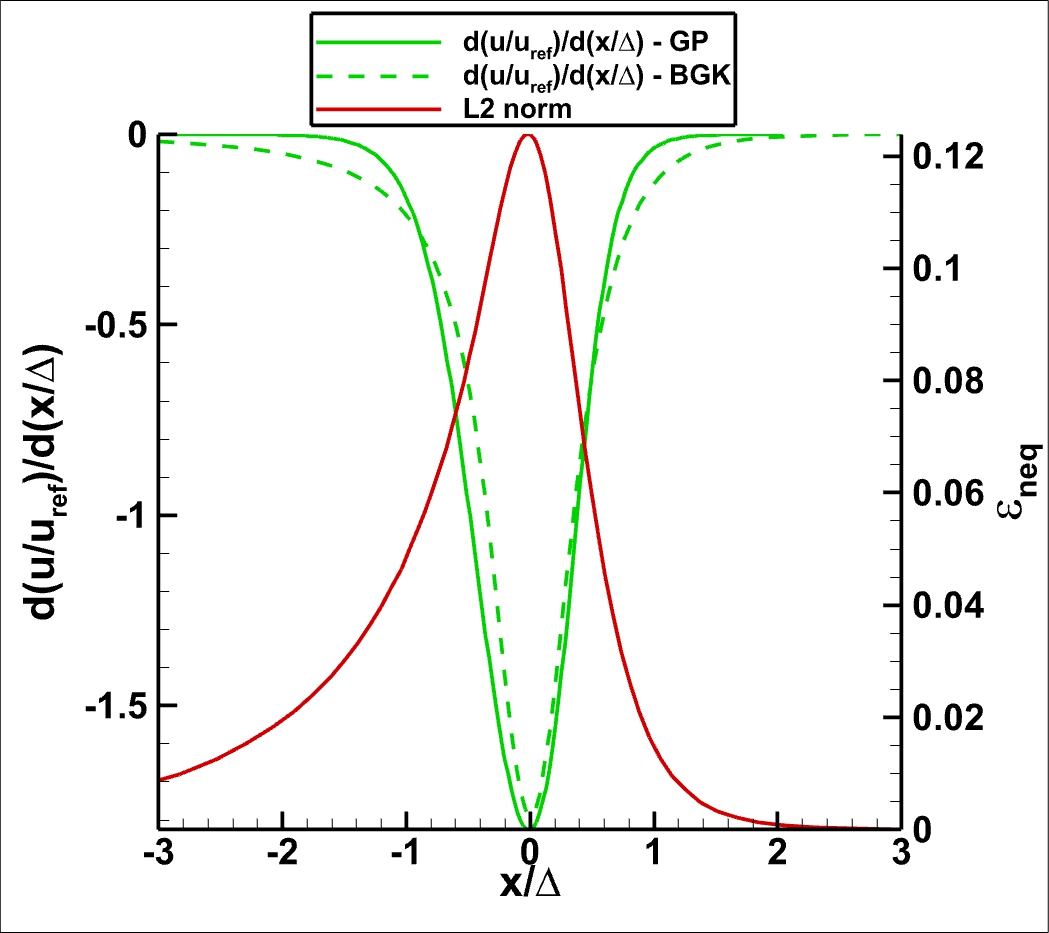}}
\subfigure[$M_\infty=4.0$ base flow]{\label{fig:1DShockProfilesM4Macro}\includegraphics[trim=10 10 10 10,clip,width=0.40\linewidth]{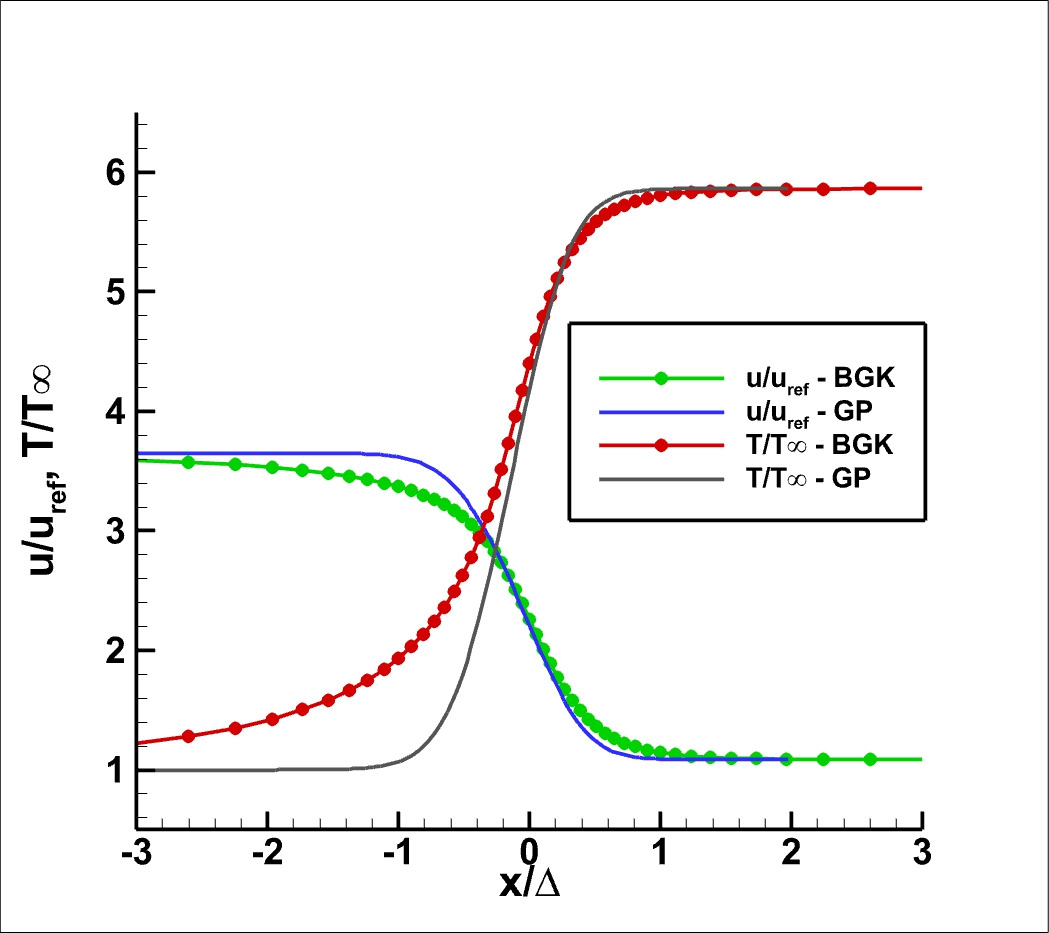}}
\subfigure[$M_\infty=4.0$ non-equilibrium]{\label{fig:1DShockNONEqLevelM4}\includegraphics[trim=10 10 10 10,clip,width=0.40\linewidth]{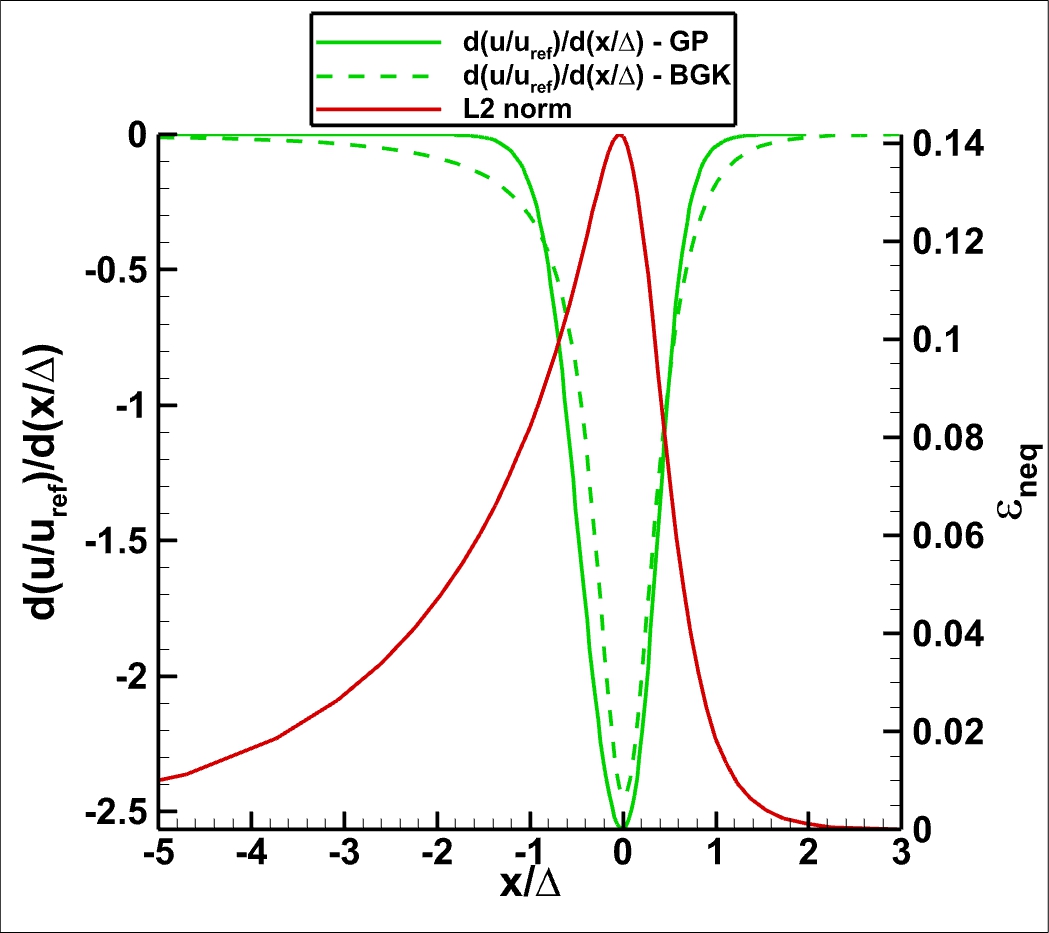}}
\caption{Comparative base-flow and translational non-equilibrium structure for 1D normal shocks at $M_\infty=1.2$, 3.0, and 4.0. Base-flow panels compare velocity and temperature profiles from the kinetic BE-BGK and continuum G\&P formulations. Non-equilibrium panels compare velocity gradients with the $L_2$ norm defined in Eqn.~\ref{eq:L2norm}. The reference length is the shock thickness $L=\Delta$ given by Eqn.~\ref{eqn:ShockThickness}.}
\label{fig:ShockBaseFlowComparison}
\end{figure}

We first compare the base-flow structure across the three Mach numbers in order to isolate the increasing non-equilibrium levels before discussing the eigenspectra. To establish a baseline for evaluating non-equilibrium conditions, we extract structural metrics directly from the kinetic BE-BGK solver. The spatial profile used below traces the $L_2$ norm (cumulative weighted RMS) of the local absolute deviation between the true VDF and the instantaneous Maxwellian equilibrium distribution function ($\Delta f = f - f^e$). This discrete deviation is integrated over the two-dimensional velocity space domain utilizing the Gauss-Hermite quadrature weights $W_{k_x}$ and $W_{k_y}$:
\begin{equation}\label{eq:L2norm}
\mathcal{E}_{\text{neq}}(x/\Delta) = \sqrt{\sum_{k_x=1}^{Q} \sum_{k_y=1}^{Q} W_{k_x} W_{k_y} \left( f(x, \xi_{k_x}, \xi_{k_y}) - f^e(x, \xi_{k_x}, \xi_{k_y}) \right)^2}
\end{equation}
Consequently, it establishes a reliable macroscopic measure of the global non-equilibrium intensity, clearly demonstrating a peak corresponding to the steepest velocity-gradient region inside the shock layer. The base-flow comparison is summarized in Fig.~\ref{fig:ShockBaseFlowComparison}. At $M_\infty=1.2$ (Figs.~\ref{fig:1DShockProfilesM12base} and \ref{fig:1DShockProfilesM12nonequi}), where very weak non-equilibrium effects are expected, the G\&P and BGK solutions result in nearly identical shock profiles and velocity gradients. For this low Mach number the $L_2$ norm $\mathcal{E}_{\text{neq}}(x/\Delta)$ is quite small and its effect is expected to be small on the eigenspectra as well. As the flow Mach number increases to $M_\infty=3.0$ (Figs.~\ref{fig:1DShockProfilesM3Macro} and \ref{fig:1DShockNONEqLevelM3}), the shock wave becomes significantly stronger and the continuum assumptions inherent in the Navier-Stokes equations begin to break down precipitously. At this Mach number, the profiles diverge within the high-gradient interior of the shock structure, highlighting a domain of translational non-equilibrium where the continuum theory fails to accurately capture the flow physics. At $M_\infty=4.0$ (Figs.~\ref{fig:1DShockProfilesM4Macro} and \ref{fig:1DShockNONEqLevelM4}), the shock is even stronger and the shock layer is dominated by translational non-equilibrium. The two formulations diverge through the shock layer more severely than at $M_\infty=3.0$, reflecting the stronger gradients and more pronounced VDF deviation from equilibrium at this Mach number. Note that since we set Pr$=1$ for the G\&P solution as well, the differences truly lie only in the kinetic treatment of the shock.

\begin{figure}[H]
\center
\subfigure[$M_\infty=3.0$]{\label{fig:M3VDFs}\includegraphics[trim=10 10 10 10,clip,width=0.45\linewidth]{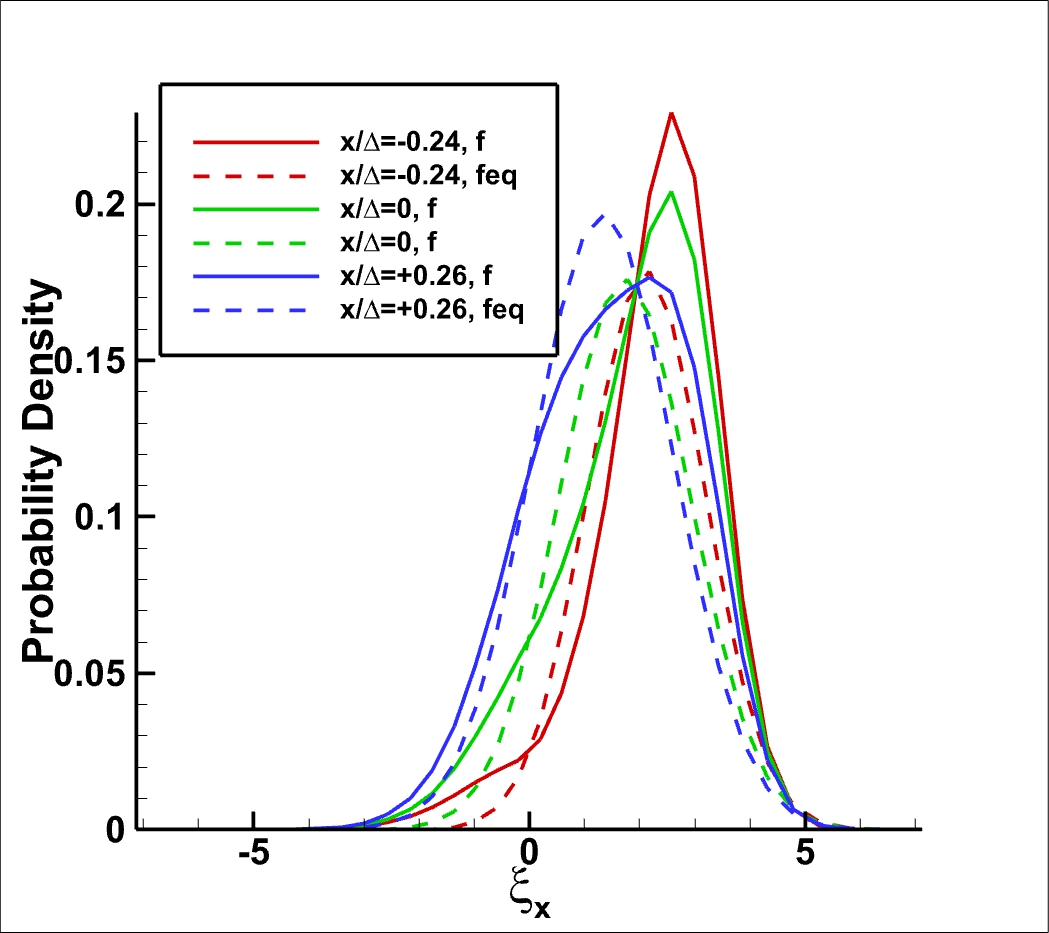}}
\subfigure[$M_\infty=4.0$]{\label{fig:M4VDFs}\includegraphics[trim=10 10 10 10,clip,width=0.45\linewidth]{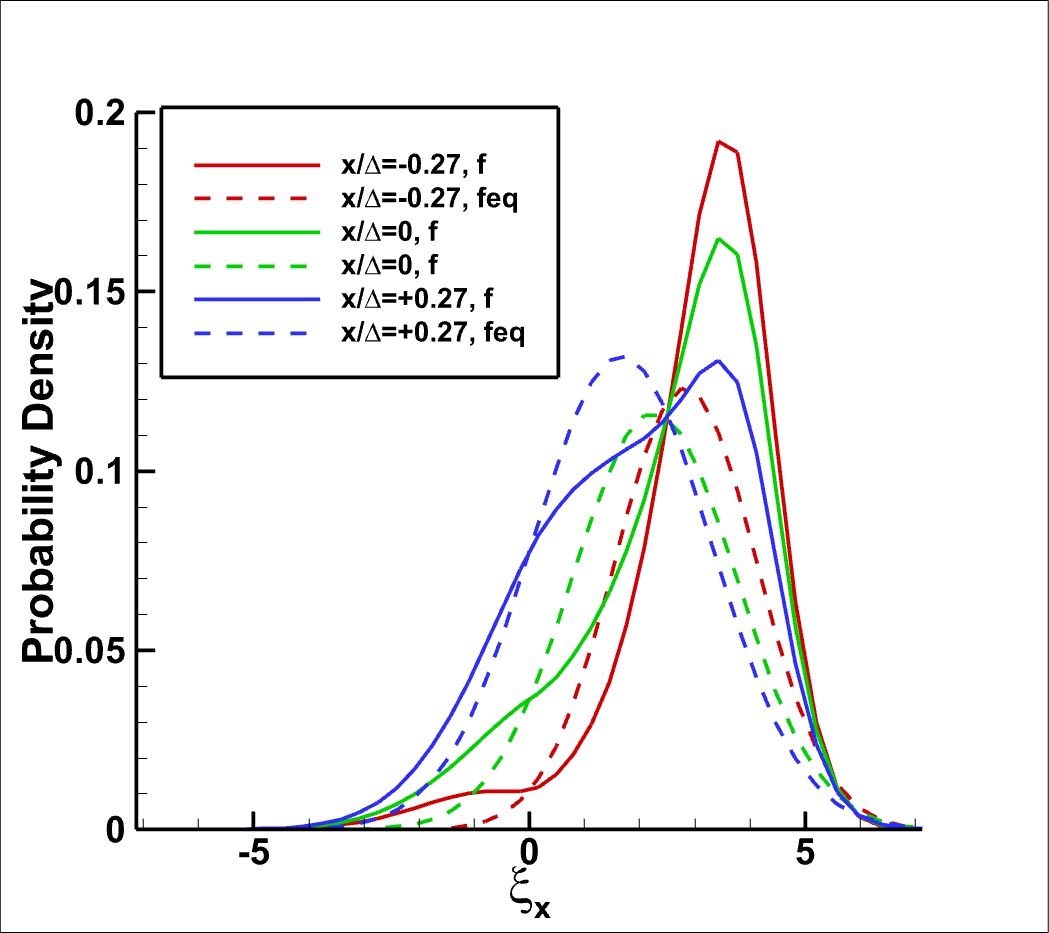}}
\caption{Evolution of the velocity distribution functions (VDFs) across the shock layer for (a) $M_\infty=3.0$ and (b) $M_\infty=4.0$, highlighting the increasingly bimodal distribution in the high-gradient region.}
\label{fig:ShockVDFComparison}
\end{figure}

The corresponding evolution of the velocity distribution functions (VDFs) through the shock layer is shown in Fig.~\ref{fig:ShockVDFComparison} for the $M_\infty=3.0$ and 4.0 cases. For $M_\infty=3.0$ (Fig.~\ref{fig:M3VDFs}), the deviation of the VDFs from the Maxwellian distribution is clear both upstream and downstream of the shock location. For $M_\infty=4.0$ (Fig.~\ref{fig:M4VDFs}), the VDF deviation from Maxwellian equilibrium is much clearer, consistent with the larger $\mathcal{E}_{\text{neq}}(x/\Delta)$ values in Fig.~\ref{fig:1DShockNONEqLevelM4}. This comparison emphasizes that the degree of VDF bimodality grows with Mach number and motivates the equilibrium versus non-equilibrium stability comparisons below.

\begin{figure}[H]
\center
\subfigure[Eigenspectra for $\beta=16$]{\label{fig:1DShockEvalM12spectra}\includegraphics[trim=10 10 10 10,clip,width=0.45\linewidth]{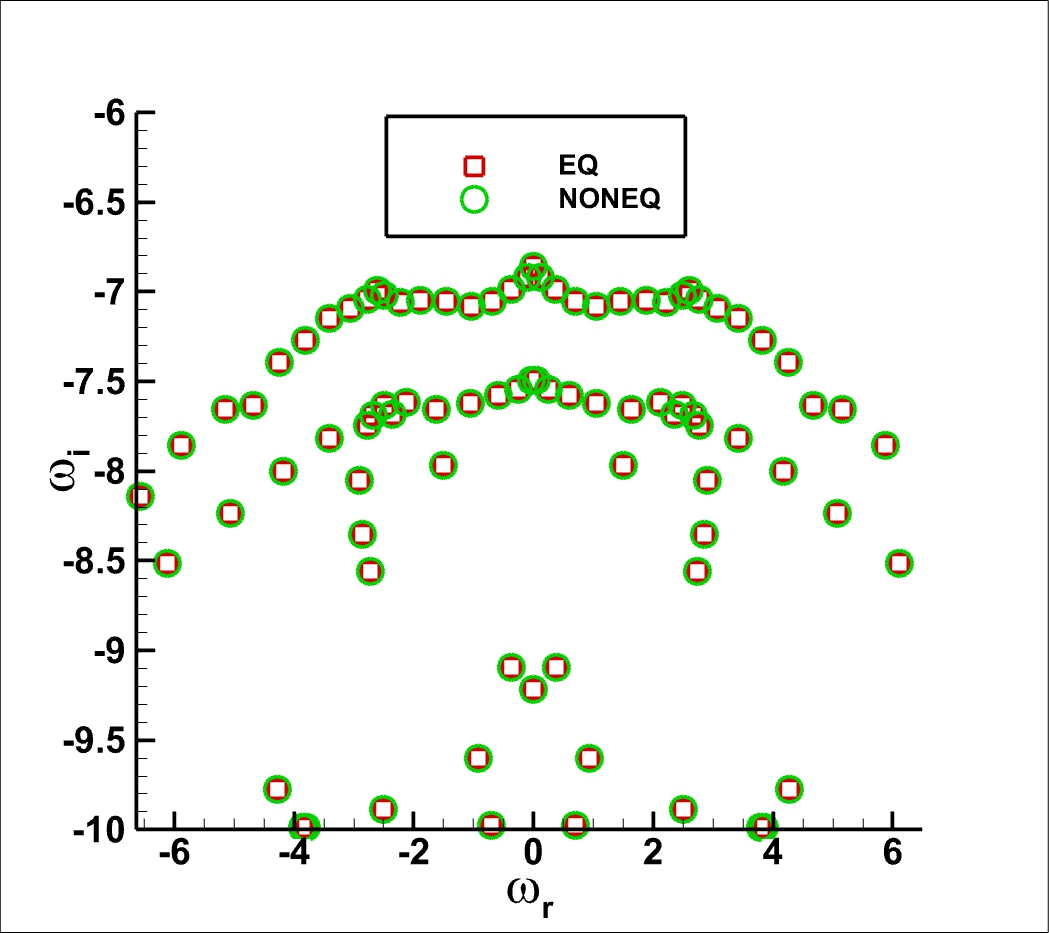}}
\subfigure[Real and imaginary parts of $\hat{u}$]{\label{fig:1DShockEvalM12evec}\includegraphics[trim=10 10 10 10,clip,width=0.45\linewidth]{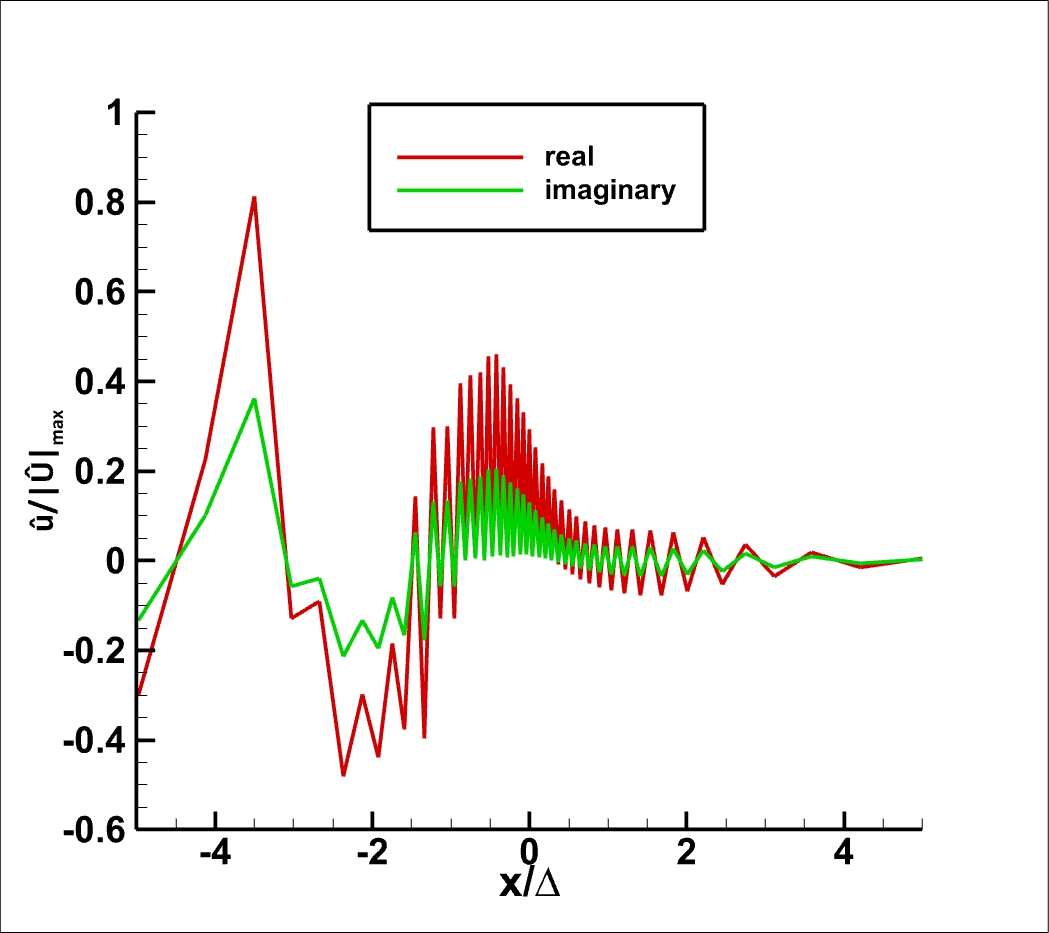}}
\subfigure[$\beta$ sweep]{\label{fig:1DShockEvalM12betasweep}\includegraphics[trim=10 10 10 10,clip,width=0.45\linewidth]{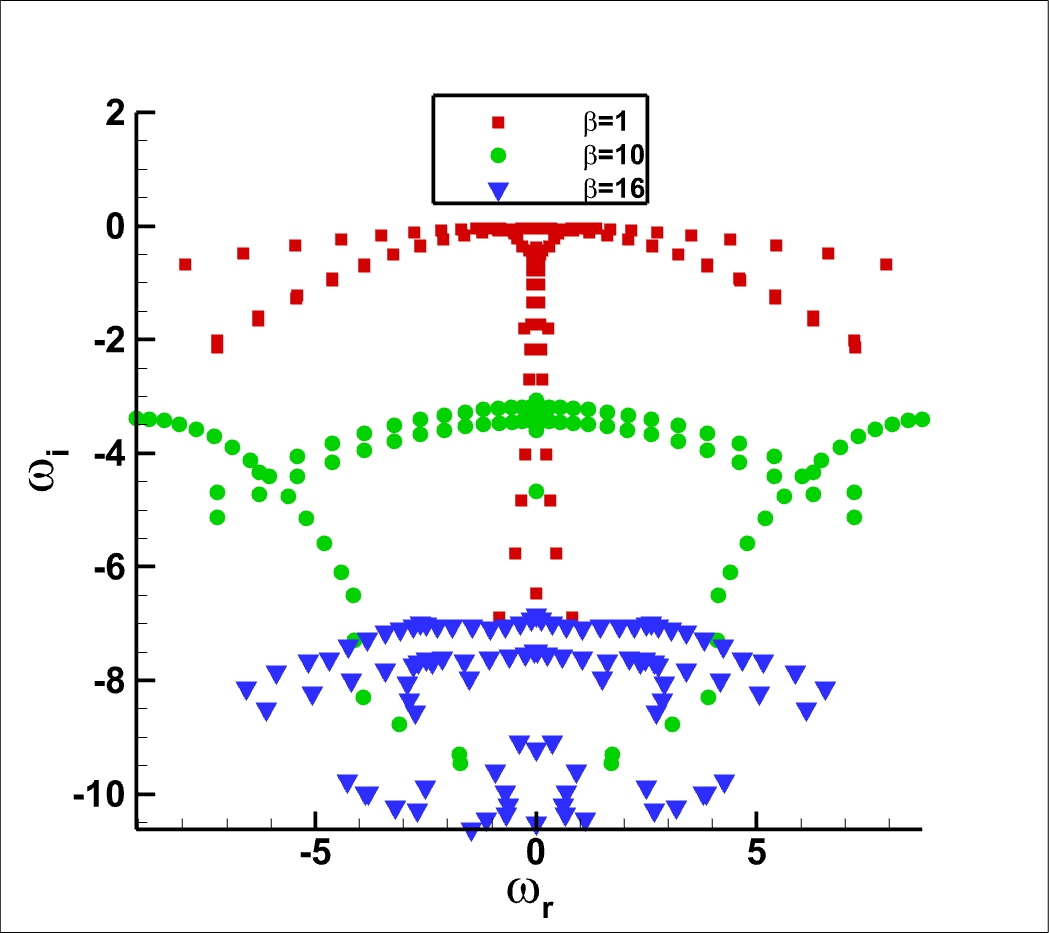}}
\caption{Results for a 1D normal shock at $M_\infty=1.2$, $Re_\Delta=24.34$ on the $N=81$, $Q=20$ grid. (a) Eigenvalue spectra at $\beta=16$ comparing equilibrium and non-equilibrium VDF solutions computed from the BE-BGK base flow. (b) Real and imaginary parts of the streamwise perturbation velocity eigenvector for the least stable mode at $0-6.85i$. (c) Non-equilibrium eigenspectra for $\beta=1$, 10, and 16.}
\label{fig:1DShockEvalM12}
\end{figure}

Using the base flows generated directly from the kinetic BE-BGK solver, eigenspectra are generated for the $M_\infty=1.2$, $Re_\Delta=24.34$, and $\beta=16$ case. Figure~\ref{fig:1DShockEvalM12} summarizes the results: the $\beta=16$ spectra are shown in Fig.~\ref{fig:1DShockEvalM12spectra}, computed with 200 Krylov subspaces to extract the 100 least stable eigenvalues using the computational framework detailed in Section~\ref{sec:ParallelSolver} with N=81 chebyshev collocation points and Q=20x20 microvelocity nodes for which we utilized the LU decomposition + Arnoldi method. The results labeled EQ means that we force $g_c=g_c^e$ and $h_c=h_c^e$ in the kLST solution whereas NONEQ means that we use the BGK solver output directly for this case and in the subsequent Mach number cases. Looking at the eigenspectra, the characteristic continuous branches of the 1D shock stability spectra are clearly identified~\cite{DuckBalakumarNormalShock,sawantPhDThesis}. Because $M_\infty=1.2$ represents a weak shock with minimal structural deformations, comparing the true non-equilibrium VDF solution against an enforced Maxwellian equilibrium yields nearly identical spectral branches, establishing this low-Mach configuration as a reliable continuum baseline before the severe non-equilibrium effects emerge at higher speeds. Fig.~\ref{fig:1DShockEvalM12betasweep} confirms that the same near-coincidence of EQ and NONEQ spectra persists across $\beta=1$, 10, and 16, with the continuous branches shifting to more negative growth rates as $\beta$ increases. Complementing this spectral data, Fig.~\ref{fig:1DShockEvalM12evec} displays the real and imaginary parts of the streamwise perturbation velocity eigenvector normalized by the maximum absolute magnitude of the total velocity vector ($\hat{u}/|\hat{U}|_{max}$), corresponding to the least stable mode at $0-6.85i$. All the eigenvectors corresponding to the velocity perturbations are normalized using;
\begin{equation}\label{eq:Unormpert}
|\hat{U}| = \sqrt{|\hat{u}|^2 + |\hat{v}|^2}
\end{equation}
As expected, the modal perturbation amplitudes are strongly localized within the high-gradient interior of the shock layer and decay to zero in the free stream.

\begin{figure}[H]
\center
\subfigure[$\beta=1$]{\label{fig:1DShockEvalM3b1}\includegraphics[trim=10 10 10 10,clip,width=0.45\linewidth]{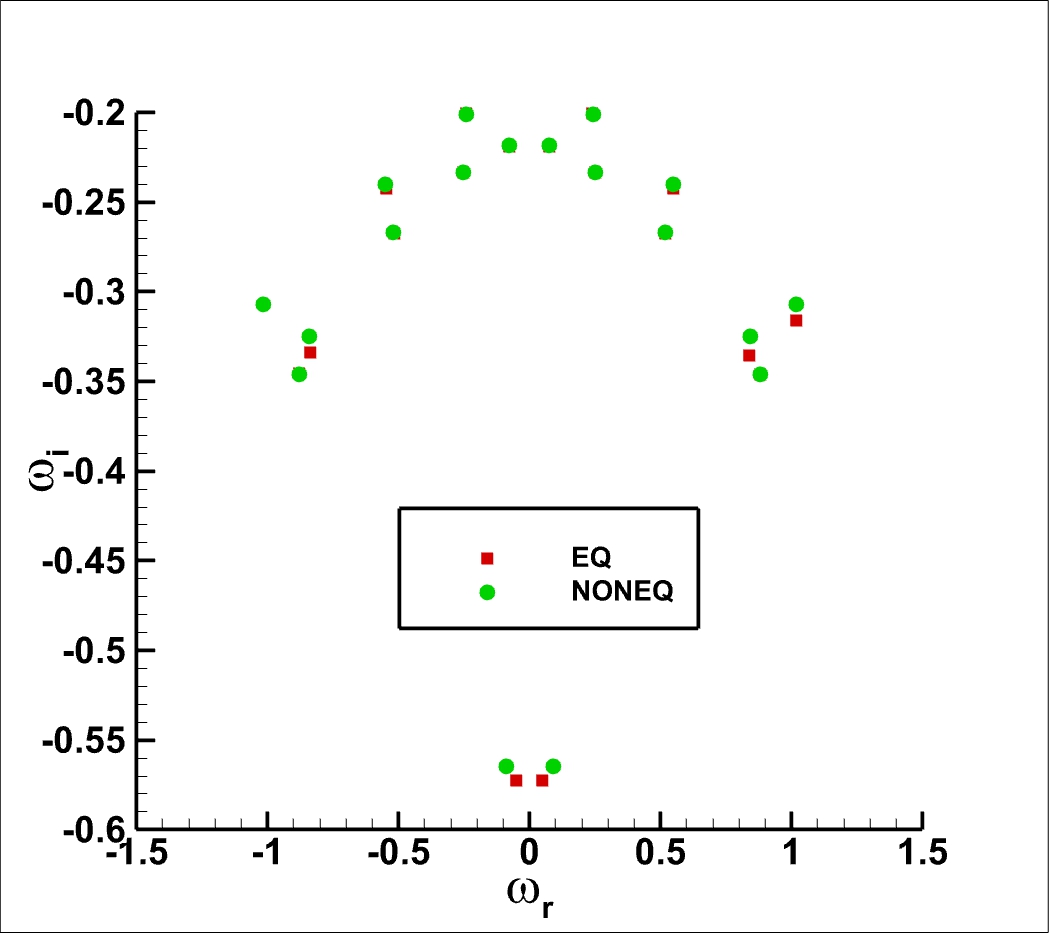}}
\subfigure[$\beta=10$]{\label{fig:1DShockEvalM3b10}\includegraphics[trim=10 10 10 10,clip,width=0.45\linewidth]{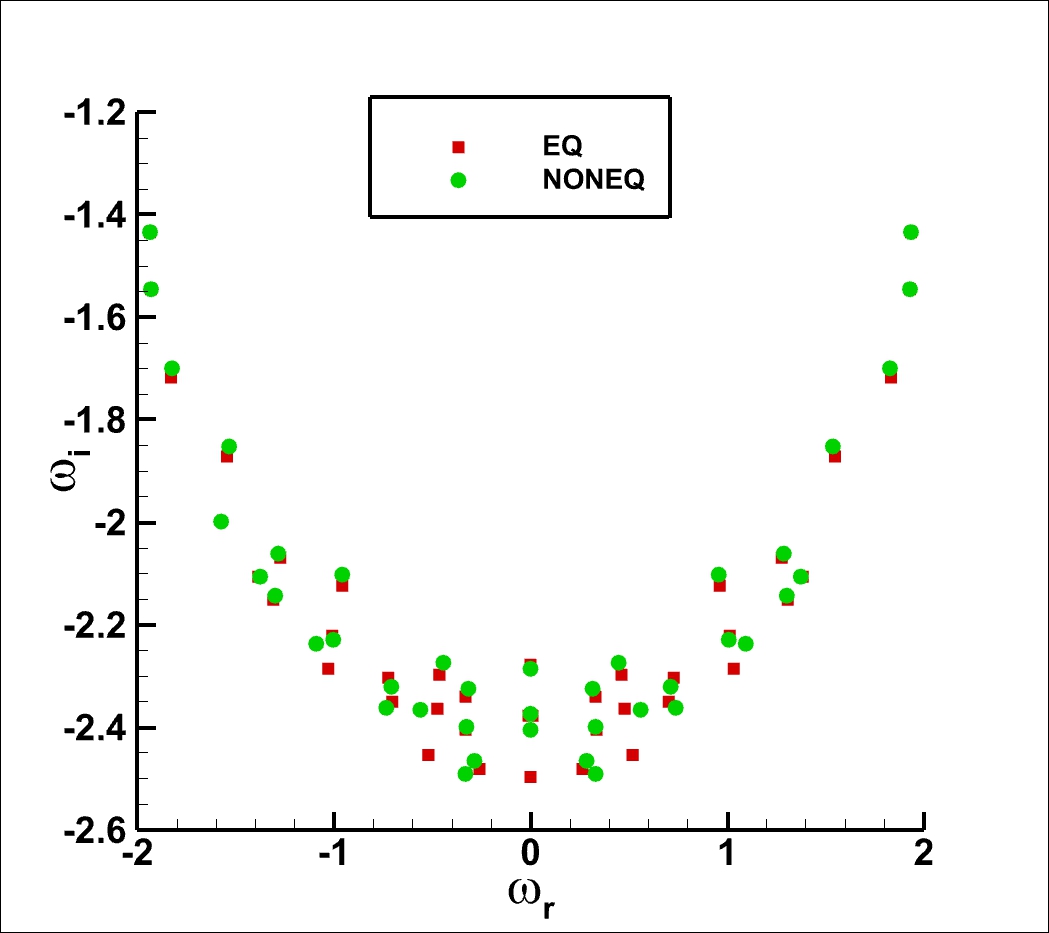}}
\subfigure[$\beta=16$]{\label{fig:1DShockEvalM3b16}\includegraphics[trim=10 10 10 10,clip,width=0.45\linewidth]{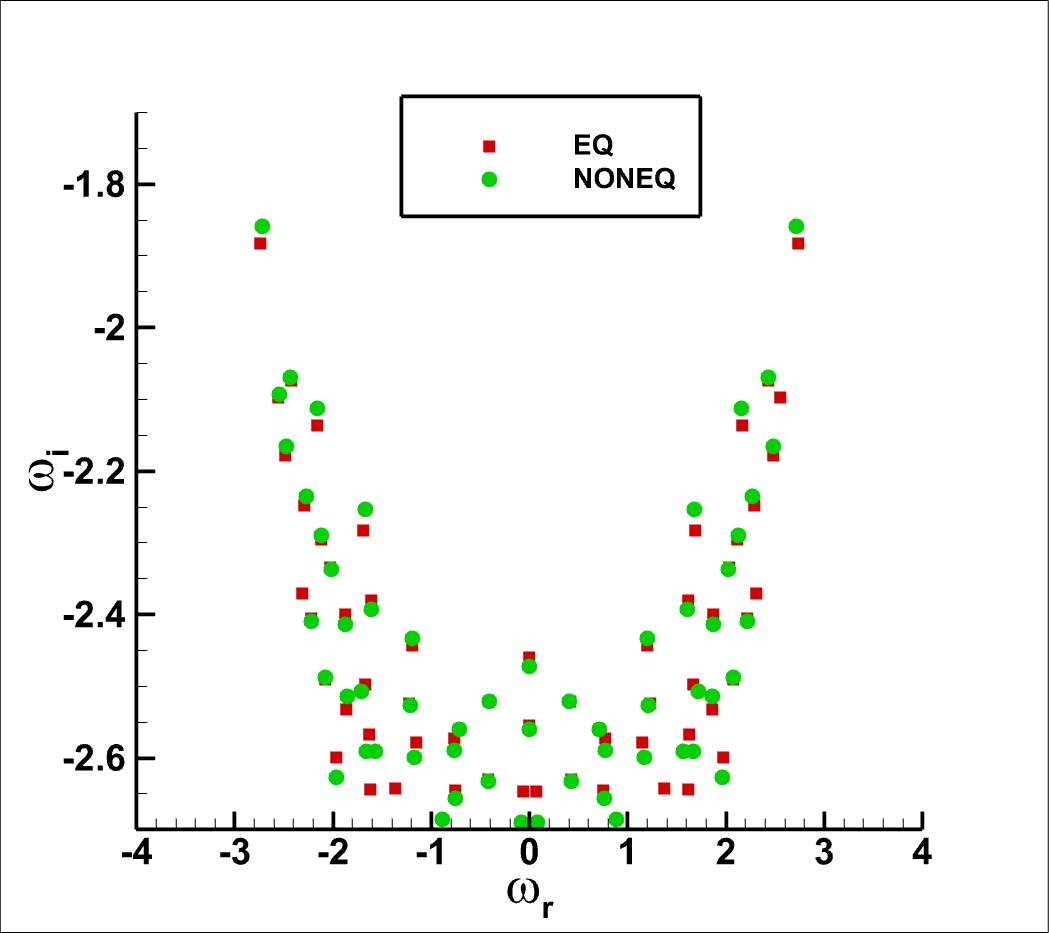}}
\caption{Eigenvalue spectra for a 1D normal shock at $M_\infty=3.0$, establishing the differences between using equilibrium versus non-equilibrium VDF formulations for (a) $\beta=1$, (b) $\beta=10$, and (c) $\beta=16$ on the $N=81$, $Q=32$ grid.}
\label{fig:1DShockEvalM3}
\end{figure}

The stability analysis for the $M_\infty=3.0$ case is severely complicated by the demanding micro-velocity resolution required. To ensure convergence at this Mach number, a velocity space discretization of at least $Q=32$ Gauss-Hermite points was necessary, paired with $N=81$ spatial nodes. This generates a massive generalized eigenvalue problem whose feasibility relies entirely upon the parallel SLEPc/PETSc HPC framework developed in this work, again utilizing the direct LU + Arnoldi approach as detailed in Section~\ref{sec:HPCBenchmarks}. The resulting eigenspectra are plotted in Fig.~\ref{fig:1DShockEvalM3} for $\beta=1$, 10, and 16 (Figs.~\ref{fig:1DShockEvalM3b1}, \ref{fig:1DShockEvalM3b10}, and \ref{fig:1DShockEvalM3b16}, respectively). It is immediately apparent that incorporating the true non-equilibrium velocity distribution functions profoundly shifts the continuous branches compared to assuming equilibrium distributions. Across all three spanwise wavenumbers, the non-equilibrium spectra are shifted toward a less stable state relative to the equilibrium spectra, with the least stable branches moving closer to the imaginary axis.

\begin{figure}[H]
\center
\includegraphics[trim=10 10 10 10,clip,width=0.65\linewidth]{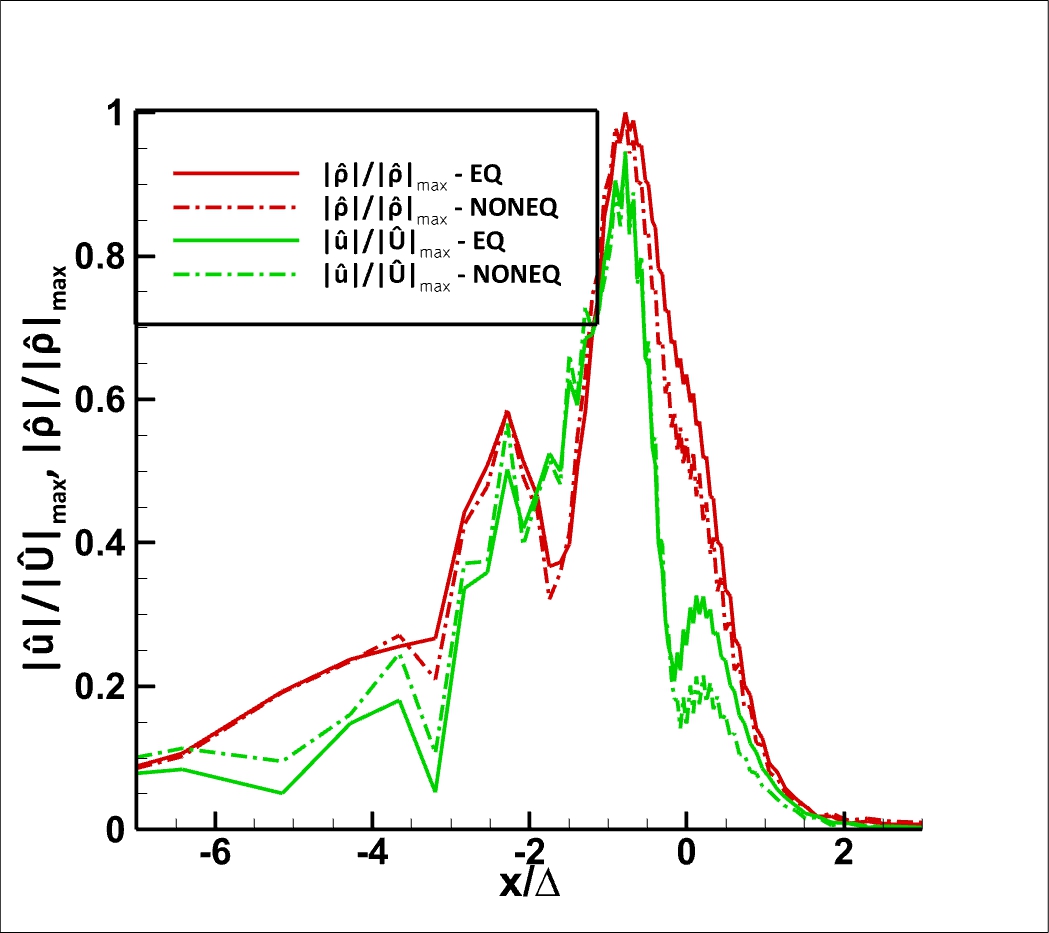}
\caption{Comparison of the absolute magnitude of the density ($\hat{\rho}$) and streamwise velocity ($\hat{u}$) eigenvectors between equilibrium (solid) and non-equilibrium (dashed) VDF formulations corresponding to the equilibrium eigenvalue at $2.4815-2.1779i$ and non-equilibrium eigenvalue at $2.477-2.166i$ for $M_\infty=3.0$ ($N=81$, $Q=32$, $\beta=16$).}
\label{fig:M3EvecEQNONEQ}
\end{figure}

To further examine the effect of the non-equilibrium VDFs on the modal structure at $\beta=16$, Fig.~\ref{fig:M3EvecEQNONEQ} compares the absolute magnitude of the density ($\hat{\rho}$) and streamwise velocity ($\hat{u}$) eigenvectors between the equilibrium and non-equilibrium formulations for the least stable mode at $M_\infty=3.0$. The density perturbation eigenvectors (red curves) peak within the shock layer and are broader in the upstream region for the non-equilibrium case, indicating that the true kinetic VDFs redistribute the perturbation energy over a wider spatial extent compared to the equilibrium assumption. The velocity perturbation eigenvectors (green curves) exhibit a similar trend, with the non-equilibrium modes showing a more diffuse structure upstream of the shock center. These differences in eigenvector spatial distributions directly reflect the bimodal character of the upstream VDF in Fig.~\ref{fig:M3VDFs}, which cannot be captured by a single Maxwellian.

\begin{figure}[H]
\center
\subfigure[$\beta=1$]{\label{fig:1DShockEvalM4b1}\includegraphics[trim=10 10 10 10,clip,width=0.45\linewidth]{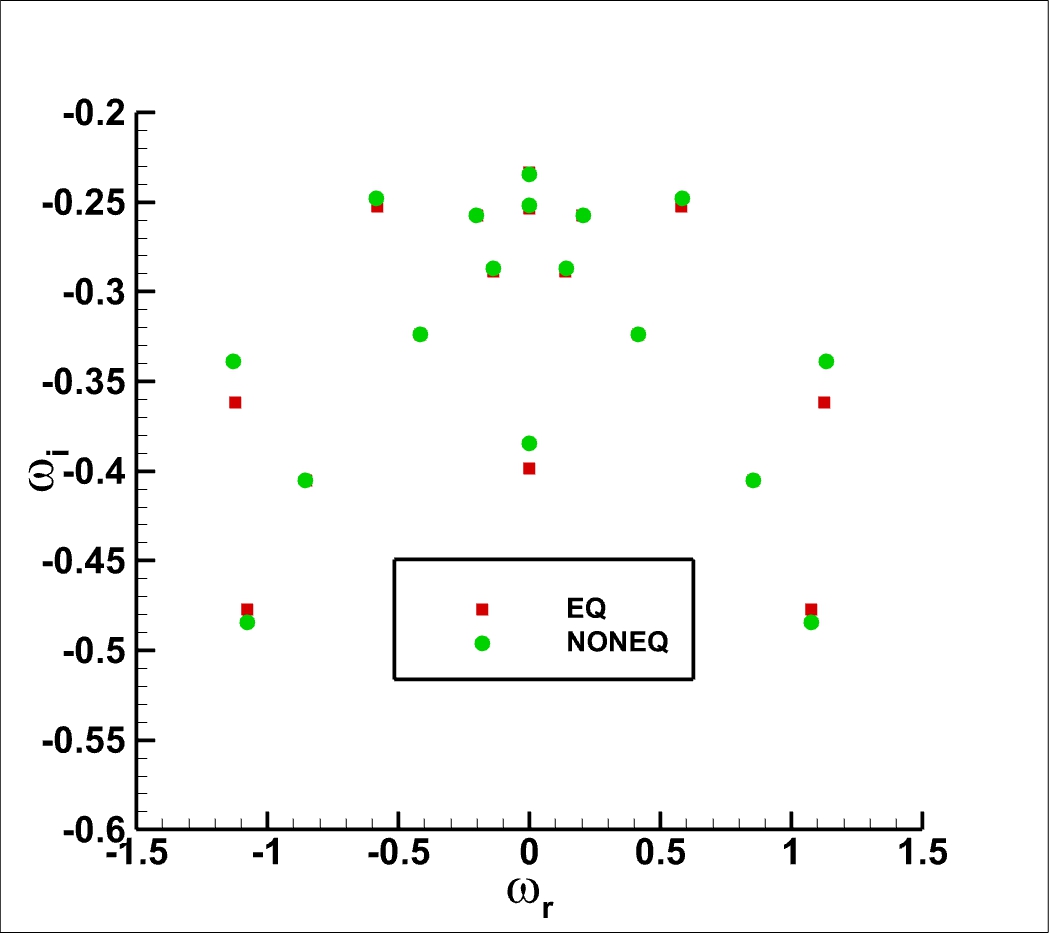}}
\subfigure[$\beta=10$]{\label{fig:1DShockEvalM4b10}\includegraphics[trim=10 10 10 10,clip,width=0.45\linewidth]{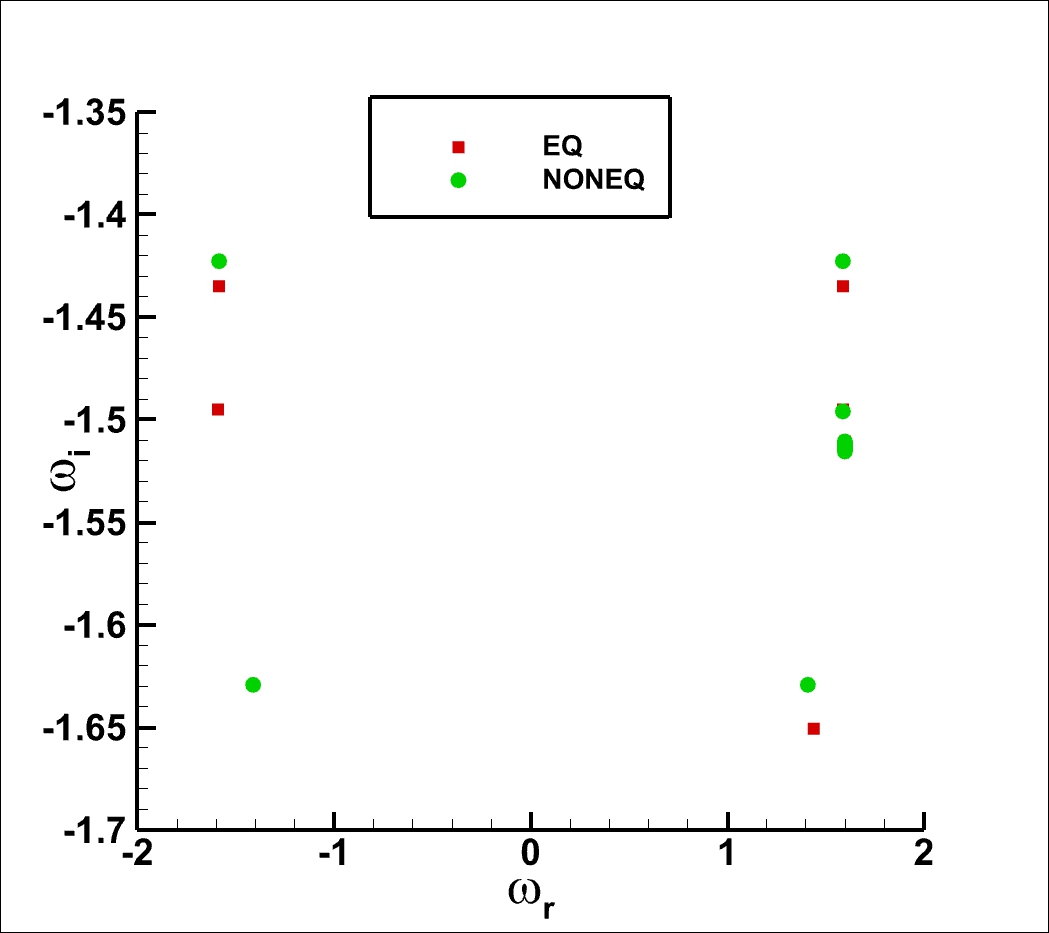}}
\subfigure[$\beta=16$]{\label{fig:1DShockEvalM4b16}\includegraphics[trim=10 10 10 10,clip,width=0.45\linewidth]{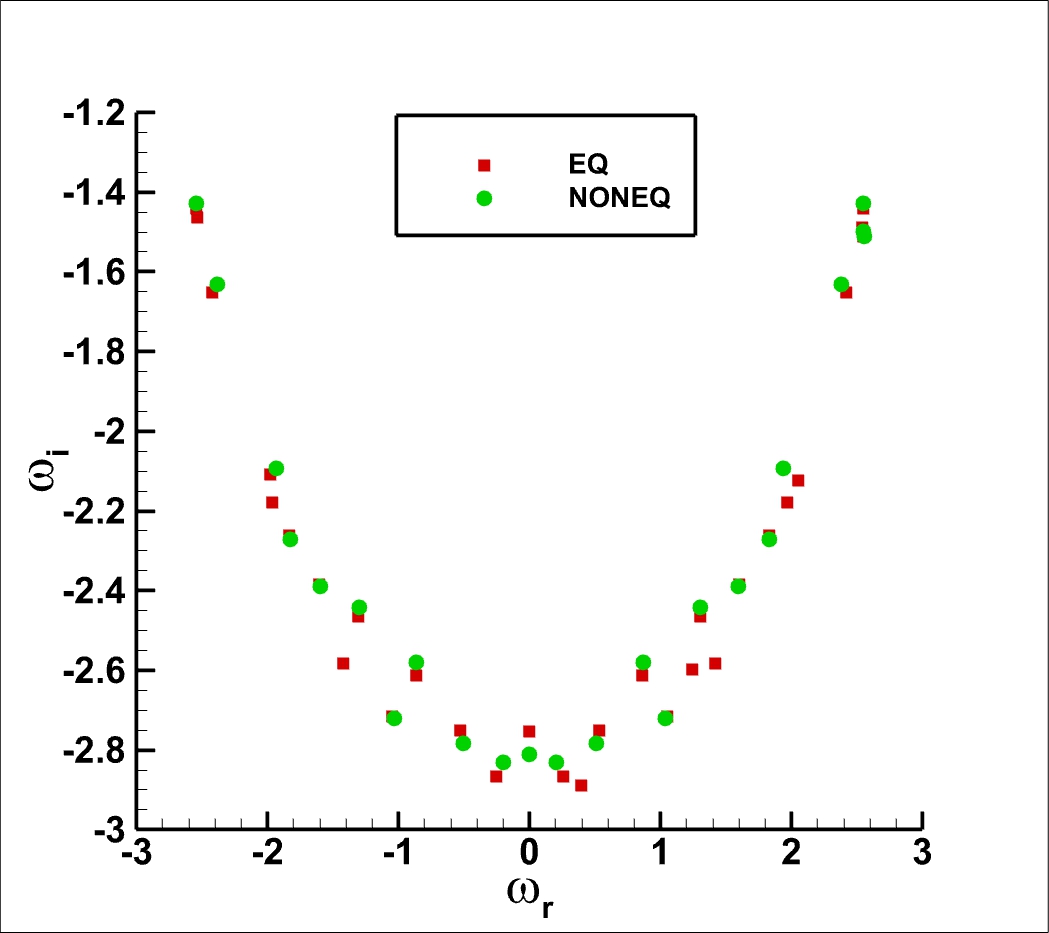}}
\caption{Eigenvalue spectra for a 1D normal shock at $M_\infty=4.0$, comparing equilibrium-based and non-equilibrium VDF formulations for (a) $\beta=1$, (b) $\beta=10$, and (c) $\beta=16$ on the N=61 and Q=48 grid.}
\label{fig:1DShockEvalM4}
\end{figure}

The linear stability eigenvalue problem at $M_\infty=4.0$ is solved with N=61 and Q=48 together with the parallel SLEPc/PETSc framework of Section~\ref{sec:ParallelSolver}. However for this case we needed to utilize the ILU + Jacobi-Davidson method as the distribution of the matrices for the LU factorization became a memory bottleneck. While it may seem counter-intuitive, the Jacobi-Davidson solverâ€™s reliance on a precise initial guess and search direction proved manageable in this instance. Because the spectrum is continuous, meaning that the eigenvalues are closely packed, making it significantly easier to converge on target values compared to cases with sparsely spaced, discrete eigenvalues as detailed in Section~\ref{sec:HPCBenchmarks}. Fig.~\ref{fig:1DShockEvalM4} reports the resulting eigenspectra (equilibrium versus non-equilibrium VDF linearizations) for $\beta=1$, 10, and 16 (Figs.~\ref{fig:1DShockEvalM4b1}, \ref{fig:1DShockEvalM4b10}, and \ref{fig:1DShockEvalM4b16}, respectively). For all three spanwise wavenumbers, the eigenspectra move toward a less stable state when the non-equilibrium distributions are considered. This non-equilibrium destabilization is more pronounced than in the $M_\infty=3.0$ case (Fig.~\ref{fig:1DShockEvalM3}), consistent with the stronger VDF bimodality in Fig.~\ref{fig:M4VDFs} and the larger non-equilibrium levels in Figs.~\ref{fig:1DShockProfilesM4Macro} and \ref{fig:1DShockNONEqLevelM4}. At these higher Mach numbers, the equilibrium-based and kinetic non-equilibrium spectra differ systematically across the continuous branches, underscoring that stability conclusions at this Mach number must be drawn from the full kinetic formulation when translational non-equilibrium is significant.

\begin{figure}[H]
\center
\subfigure[Overall profiles]{%
  \label{fig:M4EvecEQNONEQfull}%
  \includegraphics[trim=10 10 10 10,clip,width=0.45\linewidth]{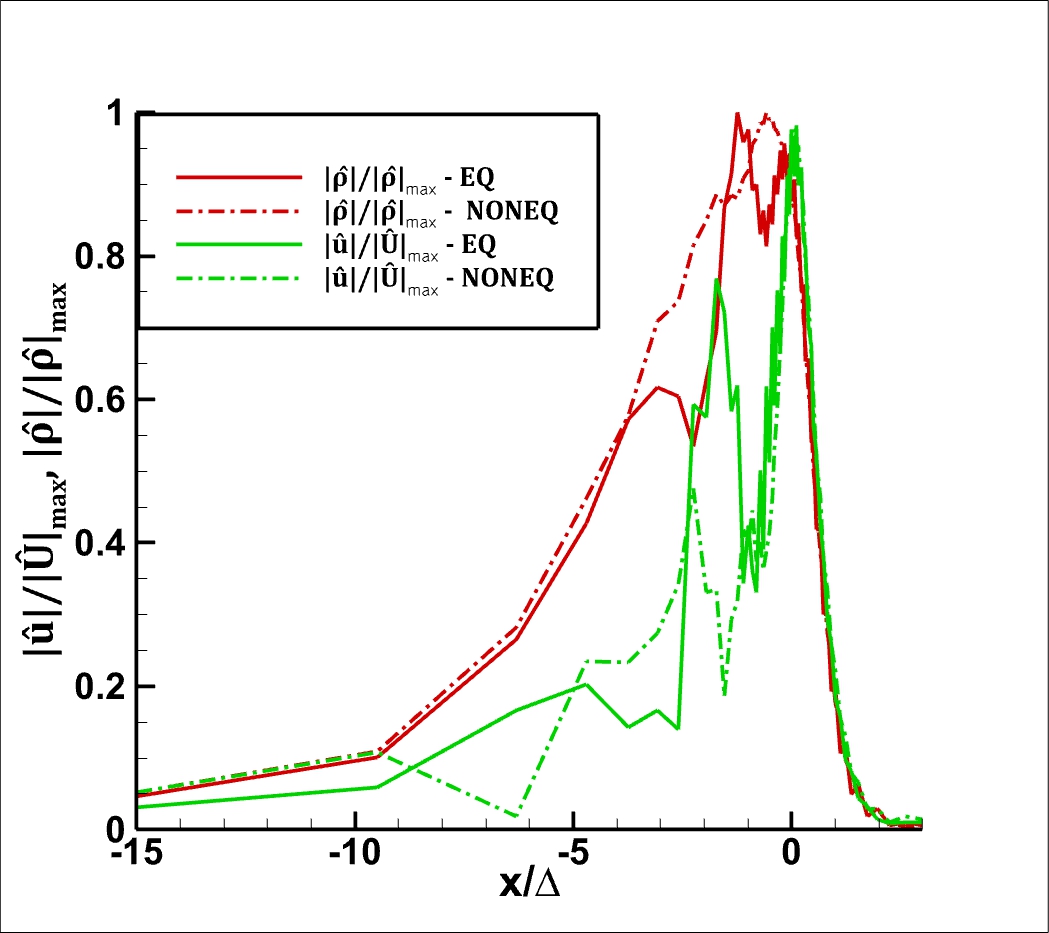}}
\subfigure[Zoomed near-shock region]{%
  \label{fig:M4EvecEQNONEQzoom}%
  \includegraphics[trim=10 10 10 10,clip,width=0.45\linewidth]{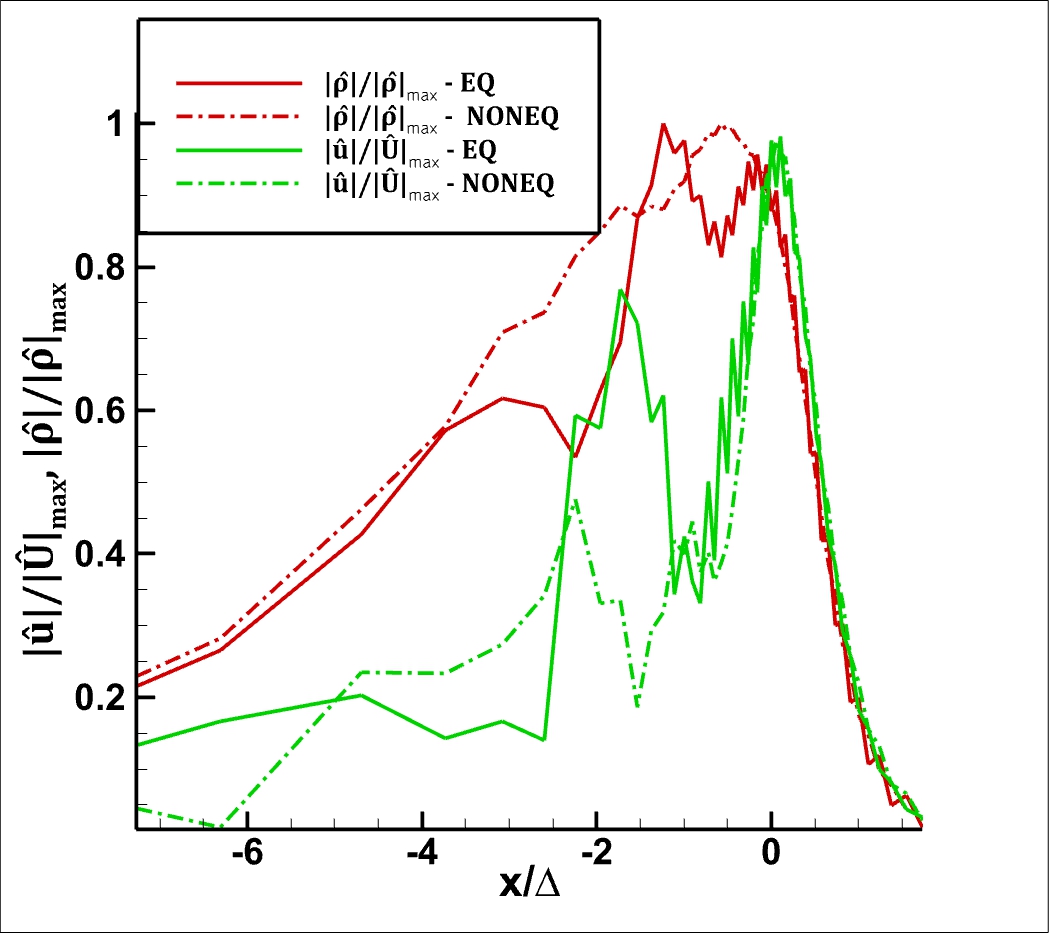}}
\caption{Comparison of the absolute magnitude of the density ($\hat{\rho}$) and streamwise velocity ($\hat{u}$) eigenvectors between equilibrium (solid) and non-equilibrium (dashed) VDF formulations for the equilibrium eigenvalue at $2.0554-2.1245i$ and non-equilibrium eigenvalue at $1.9360-2.0923i$ at $M_\infty=4.0$. (a) Overall profiles showing the extended upstream support of the non-equilibrium modes. (b) Zoomed view of the near-shock region ($-7 \lesssim x/\Delta \lesssim 3$) highlighting the localised non-equilibrium velocity perturbation peak at the shock centre.}
\label{fig:M4EvecEQNONEQ}
\end{figure}

Figure~\ref{fig:M4EvecEQNONEQ} presents the analogous eigenvector comparison for the $M_\infty=4.0$ case at $\beta=16$ for the equilibrium eigenvalue at $2.0554-2.1245i$ and non-equilibrium eigenvalue at $1.9360-2.0923i$. The full-domain view (Fig.~\ref{fig:M4EvecEQNONEQfull}) reveals that the density eigenvectors for both formulations peak near the shock centre and are nearly indistinguishable in the downstream region, but diverge markedly upstream: the non-equilibrium eigenvector maintains a higher amplitude over a significantly wider upstream extent ($x/\Delta \lesssim -15$), consistent with the increasingly bimodal VDF in Fig.~\ref{fig:M4VDFs}. The velocity eigenvector differences are even more pronounced: the non-equilibrium $\hat{u}$ mode is effectively zero upstream of the shock for the equilibrium formulation, yet exhibits substantial amplitude extending far upstream in the non-equilibrium case. This behavior directly reflects the upstream VDF bimodality in Fig.~\ref{fig:M4VDFs}, where a secondary population of fast particles penetrates ahead of the shock front and introduces perturbation structure that the single-Maxwellian assumption entirely misses.

The zoomed near-shock view (Fig.~\ref{fig:M4EvecEQNONEQzoom}) focuses on $-7 \lesssim x/\Delta \lesssim 3$. Over $-6 \lesssim x/\Delta \lesssim -1$, the non-equilibrium density eigenvector lies systematically above the equilibrium curve and reaches its maximum slightly downstream of the equilibrium peak ($x/\Delta \approx -0.5$ versus $\approx -1.2$); both profiles then decay sharply past the shock centre. The velocity eigenvectors show stronger local structure in this window: the equilibrium $\hat{u}$ mode exhibits a distinct pre-shock peak near $x/\Delta \approx -2$ before rising to its global maximum just downstream of the centre, whereas the non-equilibrium mode remains smaller over much of the upstream region but grows rapidly for $x/\Delta \gtrsim -6$ with pronounced oscillations before attaining a comparable near-centre maximum. These upstream differences in density support and velocity shape reflect the VDF bimodality that the equilibrium Maxwellian linearization cannot capture.

The results at $M_\infty=3.0$ and $M_\infty=4.0$ (Figs.~\ref{fig:1DShockEvalM3} and \ref{fig:1DShockEvalM4}) clearly show that the micro-velocity states of the flowfield affect the linear stability of the macro-states of the 1D shock layers due to the non-equilibrium distributions of VDFs. It is important to emphasize that ``equilibrium'' in this context does not imply a continuum or Navier-Stokes solution, but rather a kinetic solution constructed purely from macroscopic variables assuming a local Maxwellian equilibrium. For instance, while methods like Direct Simulation Monte Carlo (DSMC) correctly predict the macroscopic shock profiles where continuum solvers fail, using only those macroscopic parameters to construct an equilibrium state for stability analysis would still completely miss the stability implications of translational non-equilibrium; the velocity distribution functions (VDFs) must be considered directly, as illustrated by the eigenvector comparisons in Figs.~\ref{fig:M3EvecEQNONEQ} and \ref{fig:M4EvecEQNONEQ}. Ultimately, comparing the kinetic equilibrium and non-equilibrium spectra would be like comparing for example an infrared emission spectrum derived by assuming a Boltzmann distribution characterized by a single vibrational temperature against one derived using a state-to-state collisional-radiative model with true non-Boltzmann populations~\cite{ThiranietalJTHT2025}. The comparison discussed in this work thus definitively isolates and illustrates the profound effect that microscopic states exert on the stability of the macroscopic flow.

\section{Conclusions and Future Work}\label{sec:Conclusions}

In this work, we formulated and implemented kinetic linear stability theory (kLST) for finite-thickness one-dimensional normal shocks by linearizing the Boltzmann--BGK equation about BE-BGK base flows, thereby retaining translational non-equilibrium in both the mean state and the perturbations. The discrete problem was built with Gauss--Hermite quadrature in micro-velocity space and Chebyshev collocation on a mapped spatial grid that concentrates resolution in the shock layer, and the assembled operators were verified against compressible Couette-flow benchmarks (Appendix~A) and mesh-convergence studies for the shock cases (Appendix~B). To make high-Mach analyses tractable, we developed a massively parallel SLEPc/PETSc eigensolver infrastructure combining shift-invert Arnoldi with direct LU factorization for moderate systems and Jacobi--Davidson with local ILU preconditioning for the largest problems. Benchmarks on the Frontera HPC quantified the effects of sparsity patterns, solver settings, and scaling limits, providing a guide for transitioning between these strategies.

Applying the kLST framework to normal shock solutions at $M=1.2$, we obtained an eigenvalue spectra that showed the expected continuous modal branches. Extending the analysis to kinetic base flows for higher Mach numbers of 3 and 4 shocks revealed the significant impact of translational non-equilibrium on flow stability. The results demonstrated that non-equilibrium velocity distribution functions (VDFs) systematically shift the eigenvalue spectrum toward less stable regions. Even when using kinetic methods for the base flow calculations, this indicates that relying solely on an approach where macroparameters used in the LST is insufficient, as perturbations about the VDFs are required. Consequently, kinetic effects must be accounted for both in the base flows and in the micro-velocity space to fully trust stability predictions in high-speed shock-dominated flows.

Finally, we addressed the significant computational demands of high-resolution kinetic stability analysis through a parallel generalized eigenvalue solver integrated with the SLEPc and PETSc frameworks. Extensive performance benchmarks conducted on the TACC Frontera cluster demonstrated the scalability of the framework. We established that while exact shift-invert Arnoldi methods are robust for moderate problem sizes, the iterative Jacobi-Davidson (JD) approach with local ILU preconditioning is the only viable strategy for the massive system matrices generated by high-resolution velocity discretizations. The JD solver demonstrated super-linear speedup and efficient hybrid parallelism, reducing runtimes for problems with over 280,000 unknowns. This implementation of the kLST workflow unlocks the capability to explore high-speed flow physics previously deemed computationally intractable. With this computational framework, an analysis of the Mach 4 flowfields is made possible which is the highest Mach number achieved to date with a kinetic linear stability method.

Nevertheless, the present study should be viewed as a first step rather than a complete treatment of kinetic shock stability. The BGK collision model adopted here fixes the Prandtl number at unity and therefore does not yet represent the correct Prandtl-number dependence of real gases. Thus, extending the formulation to more faithful collision operators or variable-Prandtl BGK variants is an important direction for future work. Likewise, the HPC benchmarks identified practical limits of the current solver strategies, and further gains in scalability will require continued development of preconditioning, ordering, and parallel-decomposition options tailored to the wide-band kLST sparsity pattern. By establishing the first kinetic linear stability analysis of finite-thickness normal shocks together with a verified numerical and parallel framework, this work is intended primarily to pave the way for broader applications of kLST to multidimensional, reacting, and strongly non-equilibrium flows in future studies.\pagebreak

\section*{Appendix A: Verification of the kLST code with Couette Flow}\label{sec:kLSTVerification}

Compressible Couette flow serves as a controlled benchmark for the kinetic linear stability formulation summarized in Sections~\ref{sec:kLSTformulation} and~\ref{sec:NumMethods}, because its one-dimensional mean state admits sharp comparisons between predictions and reference behavior while still exercising compressibility, heat transfer at the walls, and streamwise wavenumber-dependent perturbations.

\begin{figure}[H]
\centering
\includegraphics{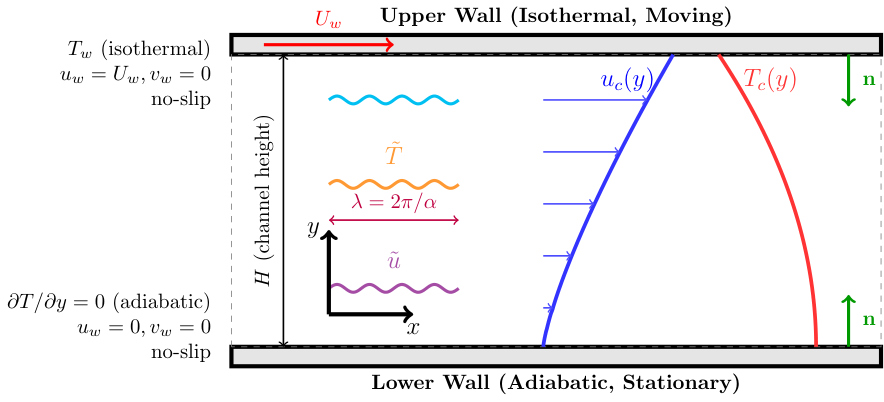}
\caption{Schematic of compressible Couette flow with linear stability perturbations. The upper wall moves with velocity $U_w$ and is isothermal, while the lower wall is stationary and adiabatic. Base flow profiles $u_c(y)$ and $T_c(y)$ exhibit compressibility effects. Wave structures in the flow field represent perturbations $\tilde{u}$ and $\tilde{T}$ with wavelength $\lambda=2\pi/\alpha$ in the streamwise direction. Wall normals $\mathbf{n}$ point toward the fluid domain.}
\label{fig:couetteSetup}
\end{figure}

The flow setup is given in Fig.~\ref{fig:couetteSetup}, consisting of Argon gas confined between two infinitely long walls. The upper isothermal wall moves at a high speed, while the adiabatic lower wall is stationary. The flow is assumed to be fully developed with no variations in the streamwise direction, i.e. any profile along the wall-normal direction is identical at any point in the streamwise direction, making the base flow essentially one-dimensional. As seen in the figure, both velocity and temperature profiles are non-linear while pressure remains constant across the channel height. As mentioned before argon is the working fluid with $\gamma=5/3$ for a hard sphere approximation for the temperature dependence of the viscosity since we use the BGK approximation for $Pr=1$. Perturbations to these mean flow profiles in the streamwise direction are characterized by wavenumbers $\alpha$. Wall normals are also shown in Fig.~\ref{fig:couetteSetup} to define the boundary conditions in the derivation.

Using the kLST code developed, we have conducted a validation study for the compressible Couette cases and compared with the NS based solution. In this section cLST means the continuum based, i.e. using Navier-Stokes-Fourier equations, linear stability theory analysis. The linear stability analysis for the compressible Couette flow is conducted for M=1, Kn=0.0001 base flow. The eigenvalue spectra is presented in phase velocity space, i.e. $c=\omega/\alpha$ where $\omega=i\omega_i+\omega_r$ is the complex temporal wavenumber and $\alpha$ is the spatial wavenumber. Since we are performing a temporal linear stability analysis we seek solutions for $\omega$ for a given $\alpha$. For this validation case there is no non-equilibriunm in the flow as the Kn numbers is small, so both the VDFs and equilibrium VDFs are generated using the macroparameters and Maxwellian distribution.

\begin{figure}[H]
\center
\subfigure[$\alpha=0.5$]{\label{fig:PCFKLSAEvala05}\includegraphics[trim=10 10 10 10,clip,width=0.45\linewidth]{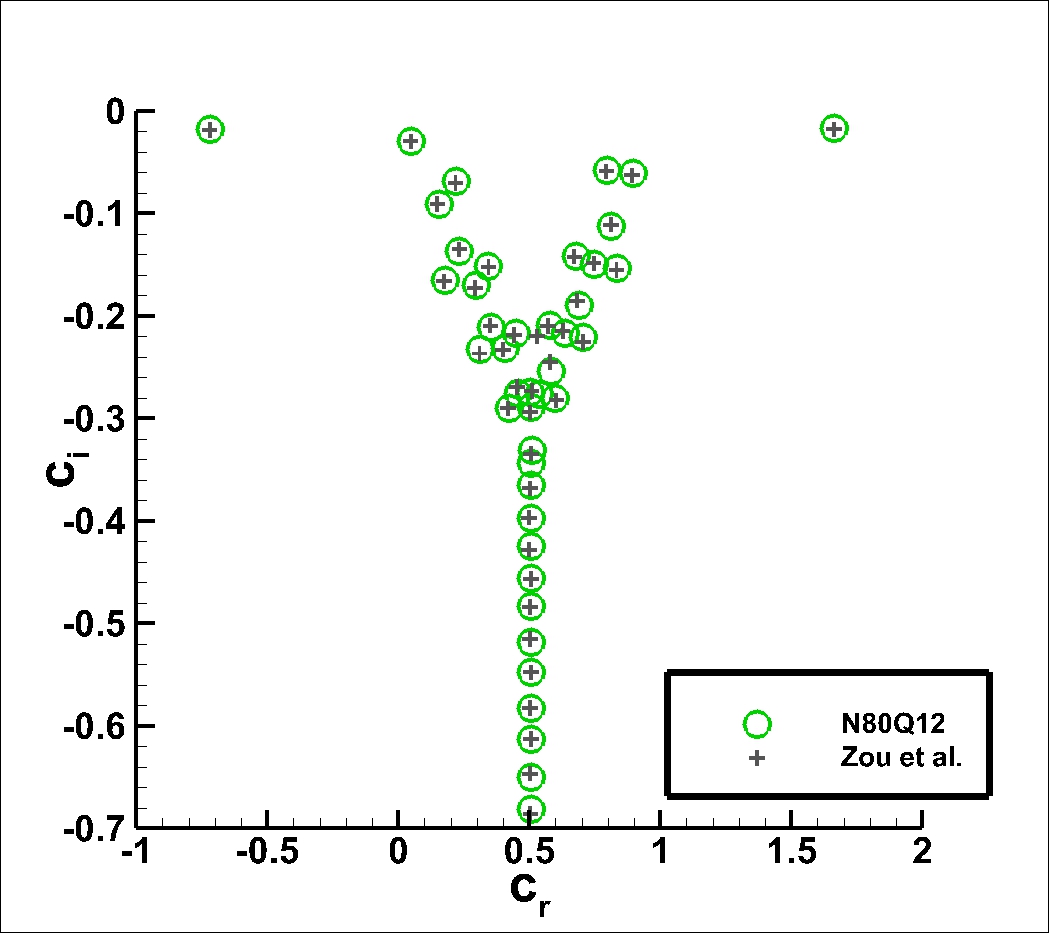}}
\subfigure[$\alpha=6.0$]{\label{fig:PCFKLSAEvala6}\includegraphics[trim=10 10 10 10,clip,width=0.45\linewidth]{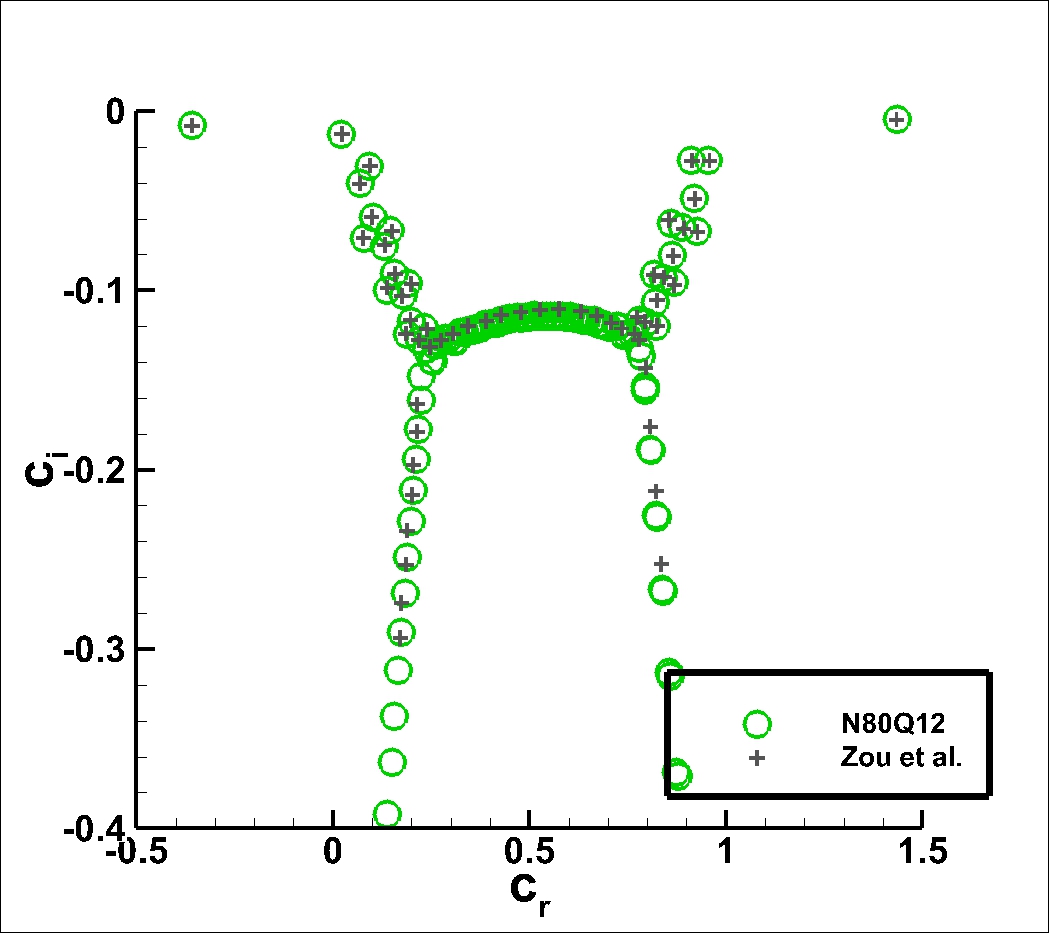}}
\caption{Comparison of eigenvalues with the reference solution for M=1, Kn=0.0001 with $\alpha=0.5$ and $\alpha=6.0$.}
\label{fig:CompCouetteEvals}
\end{figure}

Comparing our results with those from Zou et al.~\cite{ZouetalCompCouetteBGK2024} in Fig. \ref{fig:CompCouetteEvals} for $\alpha=0.5$ and $6$ with N=80 Chebyshev points and Q=12 micro-velocity nodes, it is clear that the results agree very well. For both wavenumbers least stable modes are captured properly and the shape of the spectrra is the same with the reference solution. The results for the figures are digitized from the manuscript of Zou et al.~\cite{ZouetalCompCouetteBGK2024} to have visual comparison. Furthermore, to have quantitative comparison with the reported values in Zou et al.~\cite{ZouetalCompCouetteBGK2024}, a comparison is made  in Table~\ref{LeastStableCompCouette} using the two modes that has the smallest imaginary part. For both modes current results are within 1\% of the reported results and the remaining differences can be attributed to the Arnoldi solver convergence tolerances or the small differences in the handling of base flows as they were not explicitly reported in Zou et al.~\cite{ZouetalCompCouetteBGK2024}.

\begin{table}[htbp]
\centering
\caption{Comparison of eigenvalues for the least stable mode with $\alpha=0.5\&6.0$, Kn=0.0001 and M=1.0}
\begin{tabular}{|c|c|c|c|c|}
\hline
\multirow{2}{*}{} & \multicolumn{2}{c|}{$\alpha = 0.5$} & \multicolumn{2}{c|}{$\alpha = 6.0$} \\
\cline{2-5}
                 & Current & Zou et al.~\cite{ZouetalCompCouetteBGK2024} & Current & Zou et al.~\cite{ZouetalCompCouetteBGK2024}\\
\hline
Mode 1 & $1.6649-0.01668i$ & $1.6651- 0.01641i$ & $1.4344-0.004591i$ & $1.4339- 0.004672i$ \\
\hline
Mode 2 & $-0.7200-0.01742i$ & $-0.7201-0.01724i$ & $-0.3580-0.007823i$ & $-0.3585 - 0.007439i$ \\
\hline
\end{tabular}
\label{LeastStableCompCouette}
\end{table}

\begin{figure}[H]
\center
\subfigure[$\alpha=0.5$]{\label{fig:PCFKLSAEvala05GridConv}\includegraphics[trim=10 10 10 10,clip,width=0.45\linewidth]{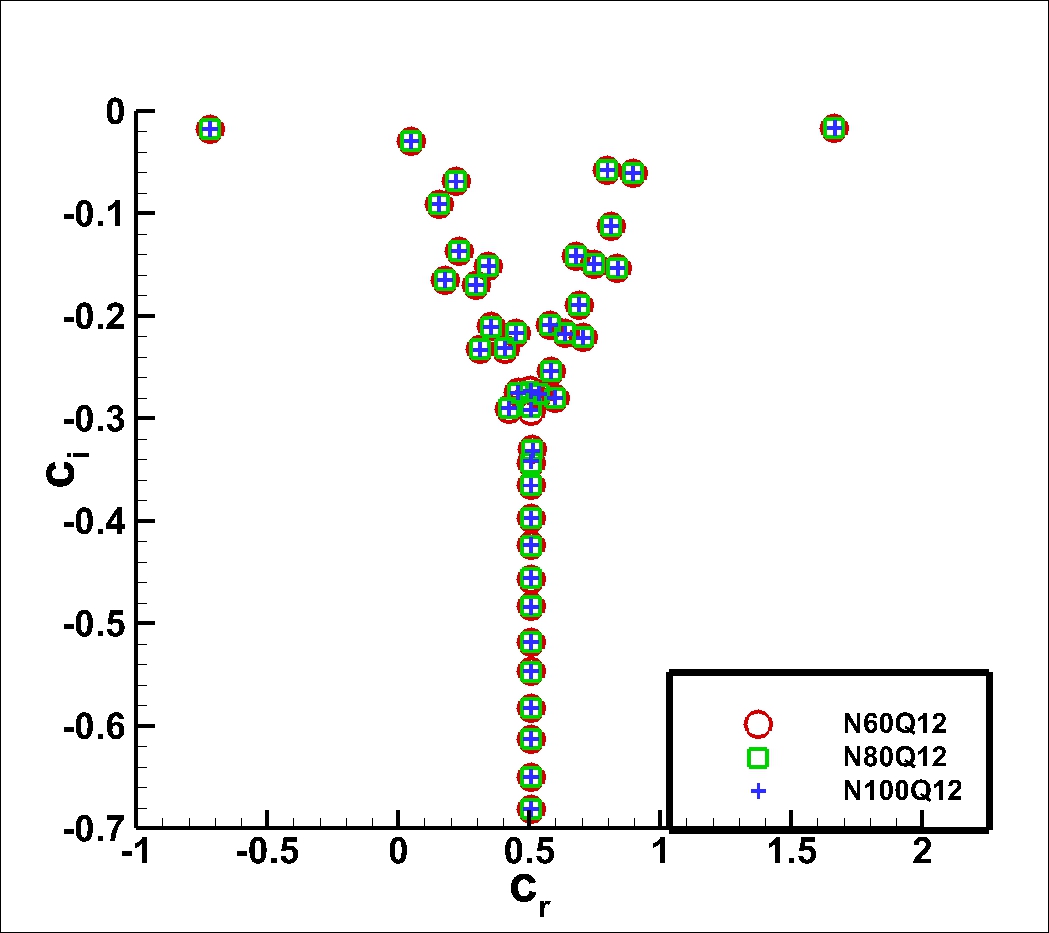}}
\subfigure[$\alpha=6.0$]{\label{fig:PCFKLSAEvala6GridConv}\includegraphics[trim=10 10 10 10,clip,width=0.45\linewidth]{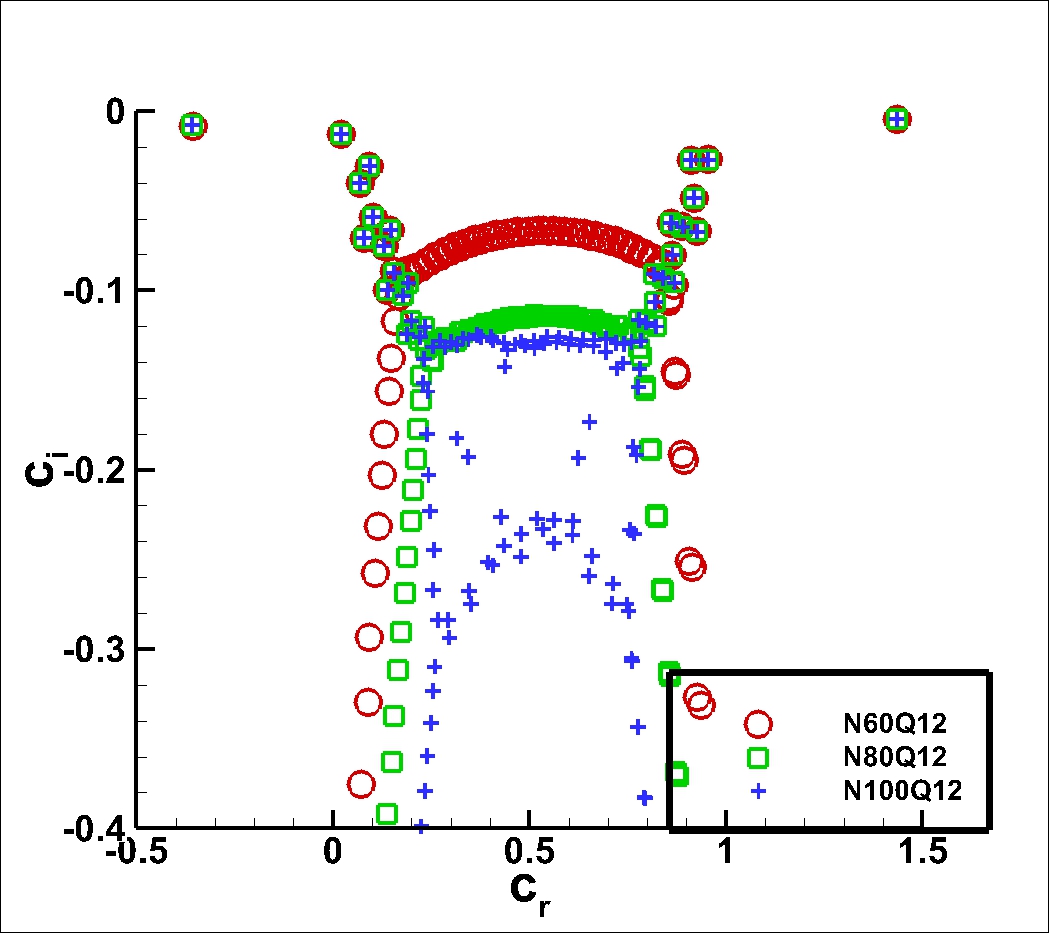}}
\caption{Grid convergence study for M=1, Kn=0.0001 with $\alpha=0.5$ and $\alpha=6.0$.}
\label{fig:MeshConvCouette}
\end{figure}

Lastly, a grid convergence study is needed to fully verify the results obtained by the kLST. Since we are using spatial nodes (N), micro-velocity nodes (QxQ) and Krylov subspaces (k), all three of these variables needs to be checked for convergence and their combinations. Of these, number of Krylov subspaces is the fastest to converge as using $k=100$ to $k=200$ results in exact same eigenvalue solutions, so for the remainder of this work we will use $k=200$. Spatial and microvelocity nodes on the other hand needed more careful consideration and for the case of M=1, Kn=0.0001 with $\alpha=0.5$ and $\alpha=6.0$ convergence is achieved at N=100 and Q=12 as can be seen in Fig. \ref{fig:MeshConvCouette} where the least stable modes are excellent agreement for both wavenumbers. 

\section*{Appendix B: Mesh Convergence Study for Base Flows and kLST for 1D Shocks}\label{sec:MeshConvergence}

The kinetic linear stability framework presented in Sections~\ref{sec:kLSTformulation} and~\ref{sec:NumMethods} rests on BGK base flows and their spatial discretization, so the fidelity of those ingredients must be established before the eigenvalue analysis in Section~\ref{sec:1DShockAnalysis} can be interpreted with confidence.

\begin{figure}[H]
\center
\subfigure[$M=3.0$]{\includegraphics[trim=10 10 10 10,clip,width=0.45\linewidth]{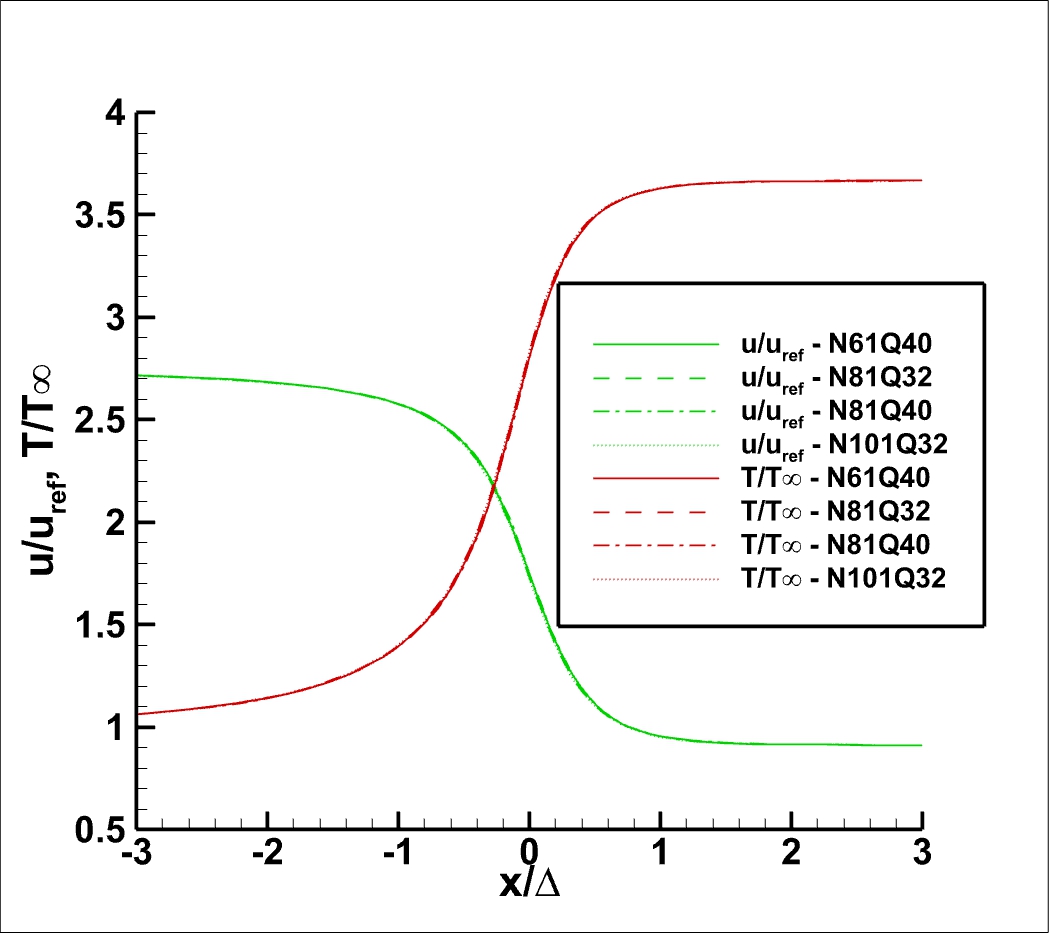}}
\subfigure[$M=4.0$]{\includegraphics[trim=10 10 10 10,clip,width=0.45\linewidth]{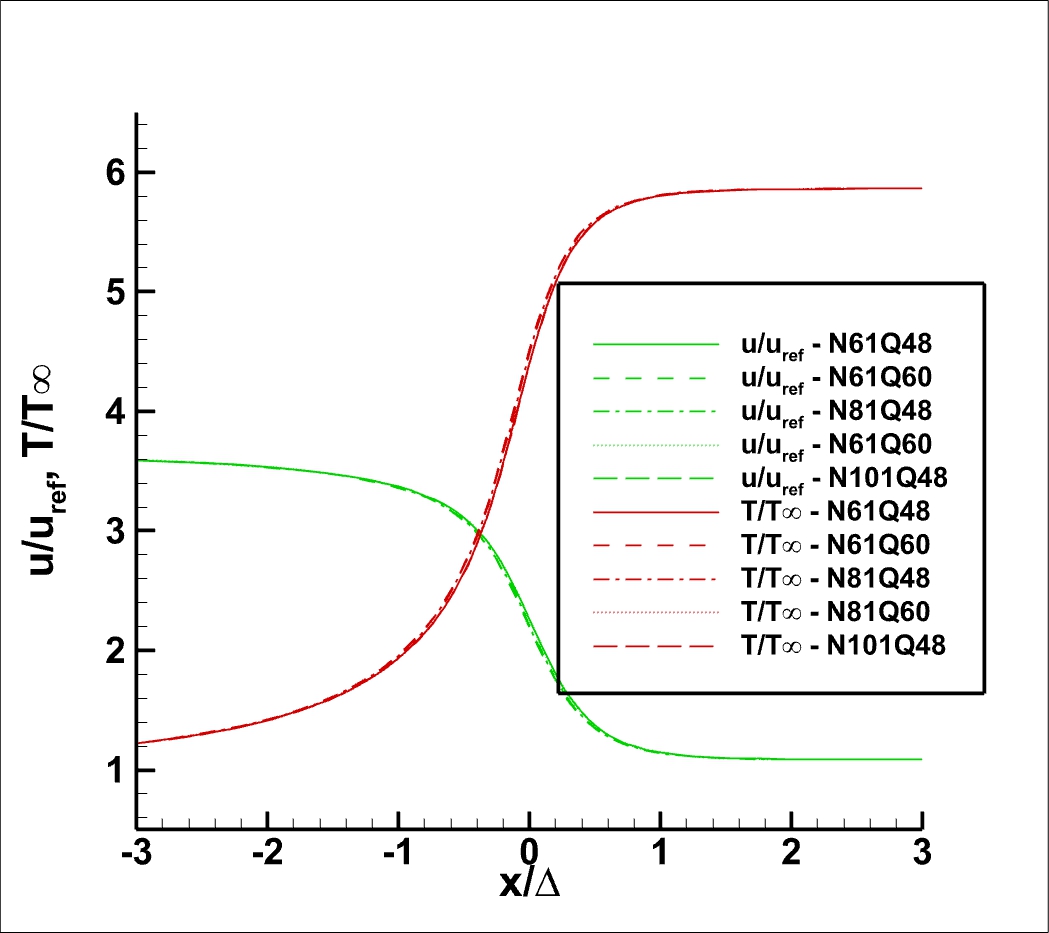}}
\caption{Mesh convergence study for the macroscopic profiles of a 1D normal shock at $M_\infty=3.0$ and $M_\infty=4.0$  utilizing algebraic spatial mapping.}
\label{fig:1DShockProfilesM3MeshConv}
\end{figure}

Achieving accurate base flow predictions at this speed necessitates rigorous spatial grid resolution. Figure~\ref{fig:1DShockProfilesM3MeshConv} demonstrates the spatial and micro-velocity grid convergence, validating the deployment of the algebraic mapping strategy that successfully concentrates collocation points for both $M_\infty=3.0$ and $M_\infty=4.0$ cases. For the $M_\infty=1.2$ case, the base flow was compared against the known G\&P solution in Section~\ref{sec:1DShockAnalysis} (Fig.~\ref{fig:1DShockProfilesM12base}); a separate base-flow grid convergence study was therefore not required.

\begin{figure}[H]
\center
\subfigure[$M_\infty=1.2$]{\label{fig:1DShockProfilesM12Appendix}\includegraphics[trim=10 10 10 10,clip,width=0.45\linewidth]{./FIGURES/M12BaseFlow.jpeg}}
\subfigure[$M_\infty=6.0$]{\label{fig:1DShockProfilesM6Verification}\includegraphics[trim=10 10 10 10,clip,width=0.45\linewidth]{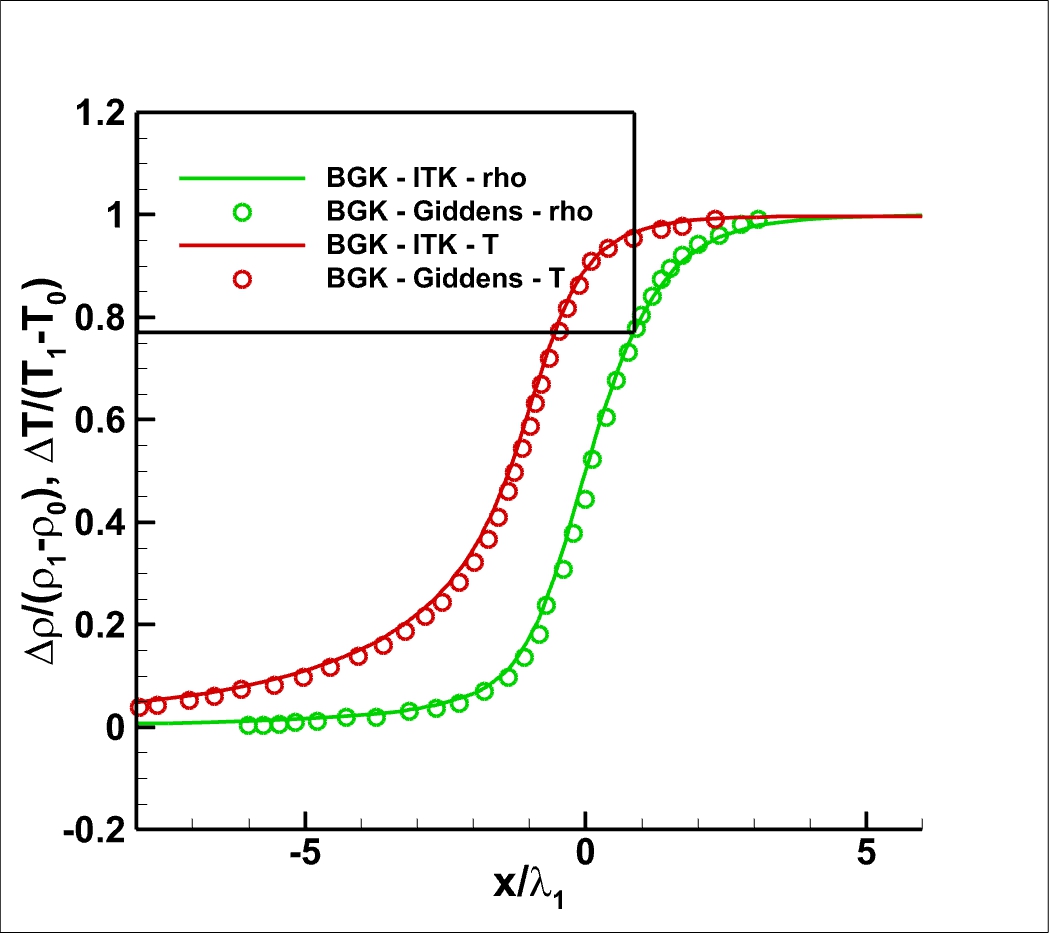}}
\caption{Comparison of the BGK base flow results against known solutions. Note (a) uses the non-dimensional variables defined in Section~\ref{sec:BaseFlowSolvers} whereas (b) uses the non-dimensional parameters from Giddens et al~\cite{Giddens1971}.}
\label{fig:1DShockProfilesM3Verification}
\end{figure}

In addition to the grid convergence study, the base flow BGK solver is also verified against the known G\&P solution at $M_\infty=1.2$, as already shown in Fig.~\ref{fig:1DShockProfilesM12base} and repeated here in Fig.~\ref{fig:1DShockProfilesM3Verification}(a) for convenience. The $M_\infty=6.0$ solution for argon with a viscosity coefficient of 0.71 is compared with the results of Giddens et al.~\cite{Giddens1971} in Fig.~\ref{fig:1DShockProfilesM3Verification}(b). Figure~\ref{fig:1DShockProfilesM3Verification} demonstrates excellent agreement in both cases, verifying the rigor of the BGK base-flow solver. The $M_\infty=6.0$ case was chosen because Giddens et al.~\cite{Giddens1971} did not report results for the exact $M_\infty=3.0$ or $M_\infty=4.0$ cases; demonstrating solver accuracy at a higher Mach number ensures reliability for the lower Mach number regimes considered in this work.

\begin{figure}[H]
\center
\subfigure[$M=1.2, \beta=16$]{\label{fig:M12GridConvBeta16}\includegraphics[trim=10 10 10 10,clip,width=0.45\linewidth]{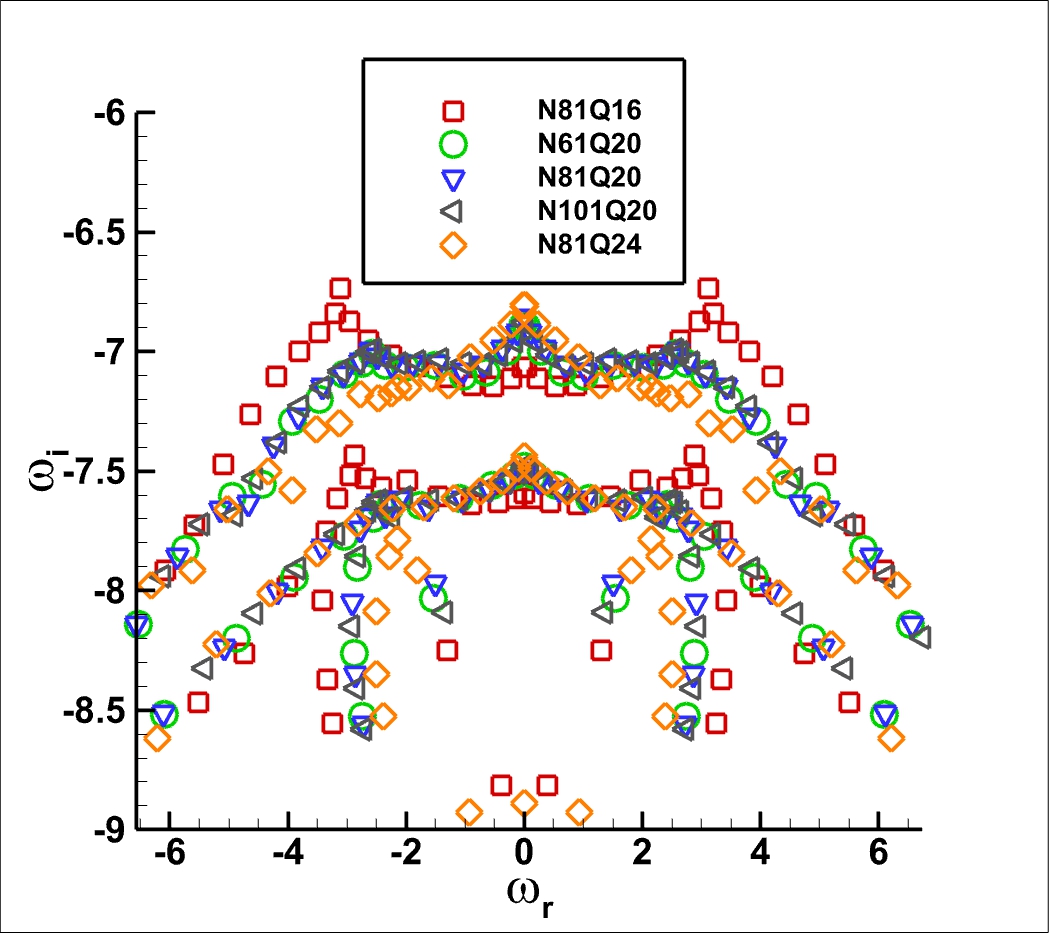}}
\subfigure[$M=3.0, \beta=1$]{\label{fig:M3GridConvBeta1}\includegraphics[trim=10 10 10 10,clip,width=0.32\linewidth]{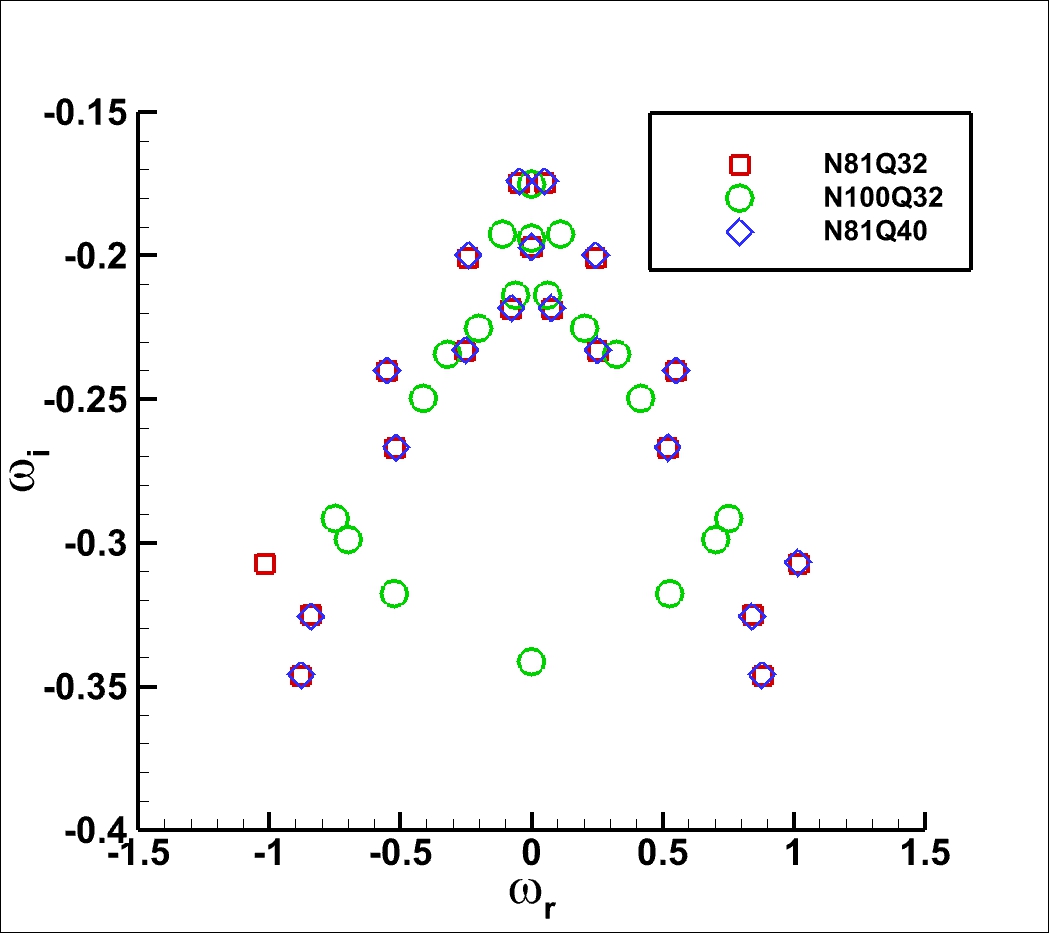}}
\subfigure[$M=3.0, \beta=10$]{\label{fig:M3GridConvBeta10}\includegraphics[trim=10 10 10 10,clip,width=0.32\linewidth]{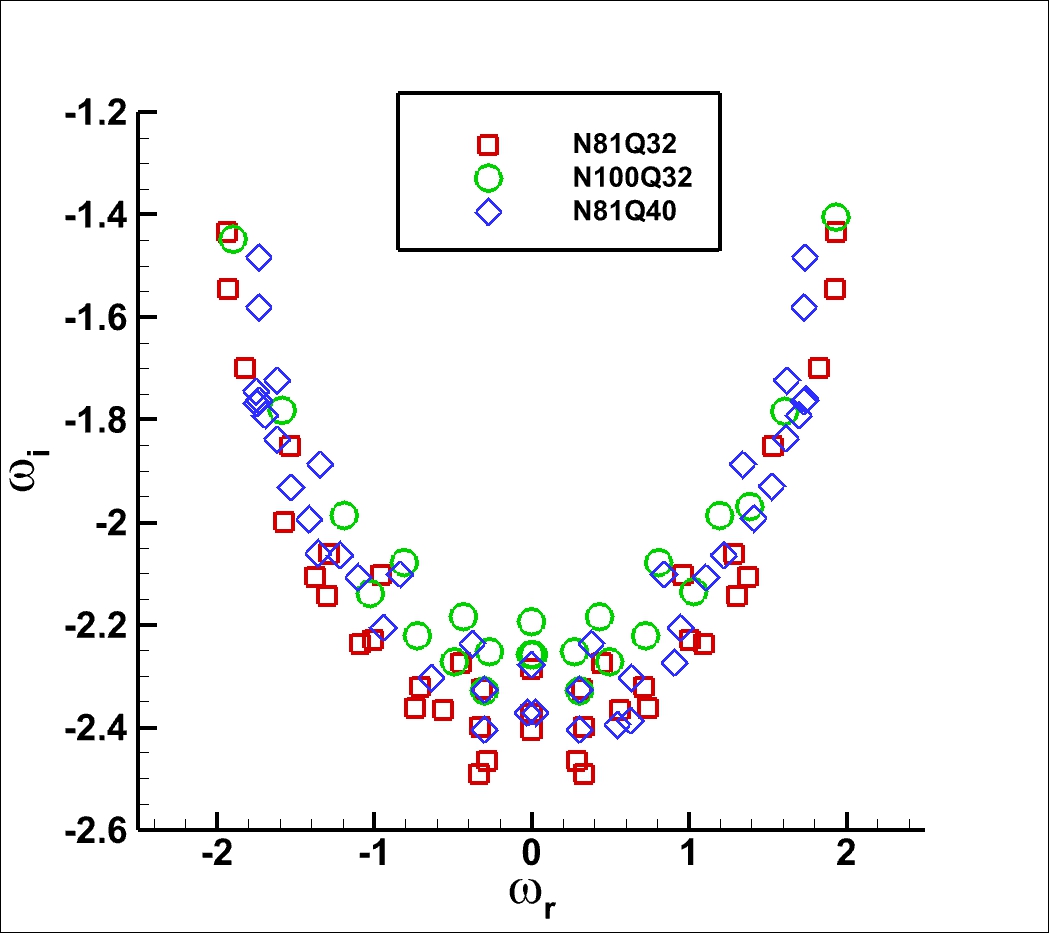}}
\subfigure[$M=3.0, \beta=16$]{\label{fig:M3GridConvBeta16}\includegraphics[trim=10 10 10 10,clip,width=0.32\linewidth]{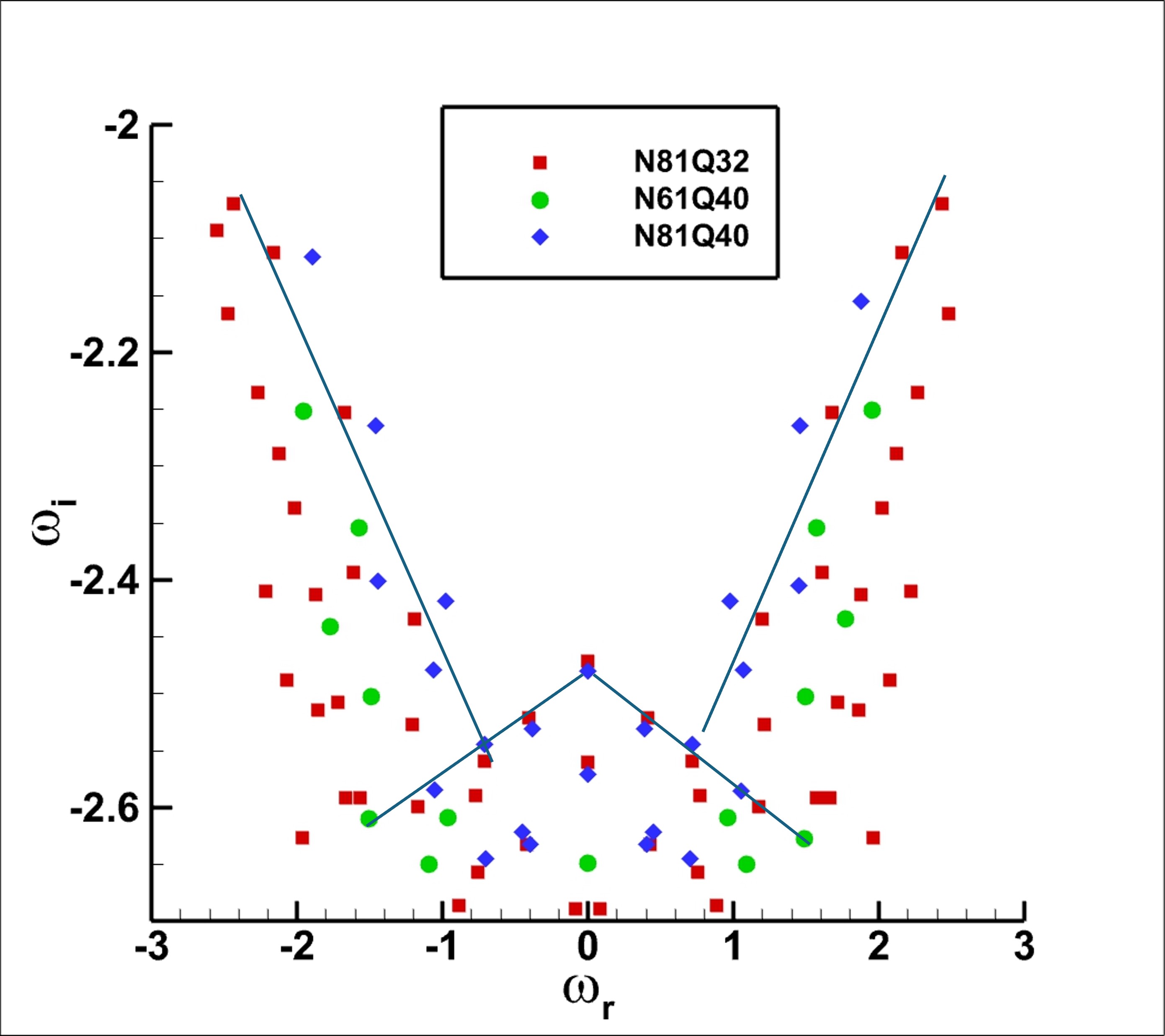}}
\subfigure[$M=4.0, \beta=1$]{\label{fig:M4GridConvBeta1}\includegraphics[trim=10 10 10 10,clip,width=0.32\linewidth]{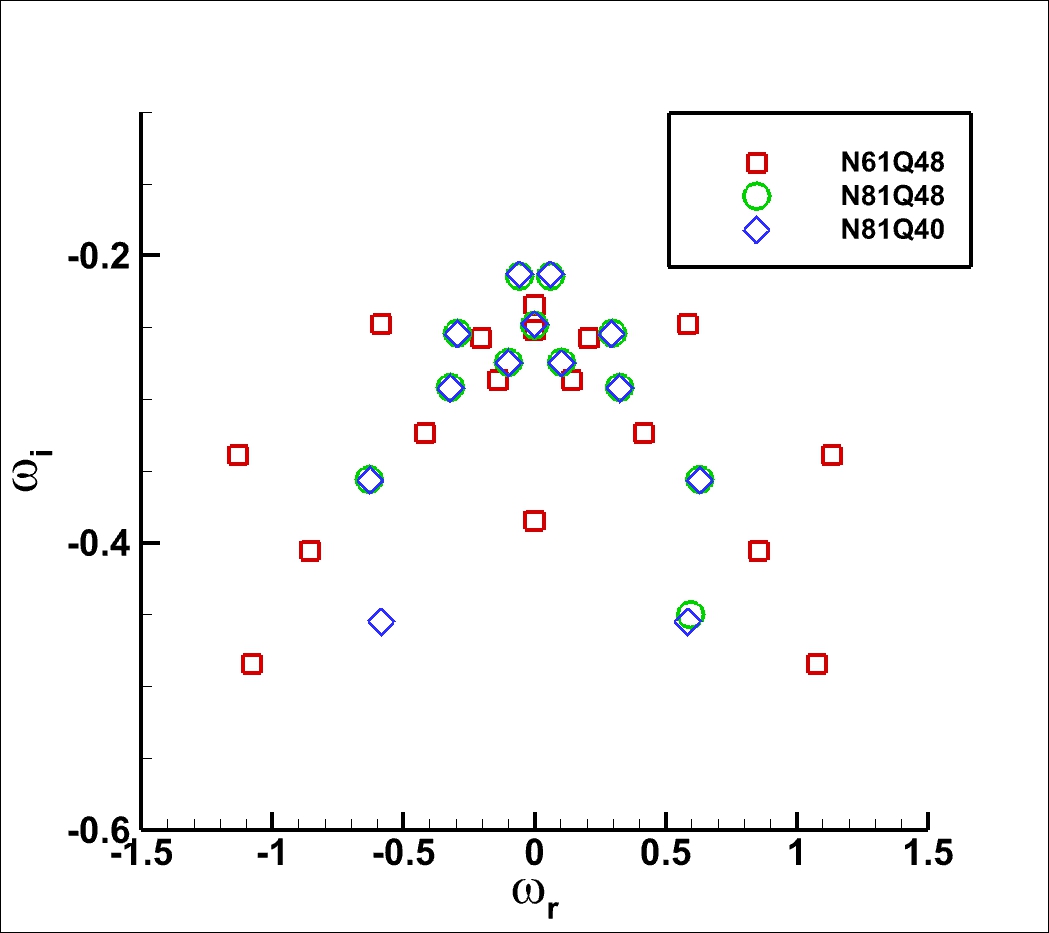}}
\subfigure[$M=4.0, \beta=10$]{\label{fig:M4GridConvBeta10}\includegraphics[trim=10 10 10 10,clip,width=0.32\linewidth]{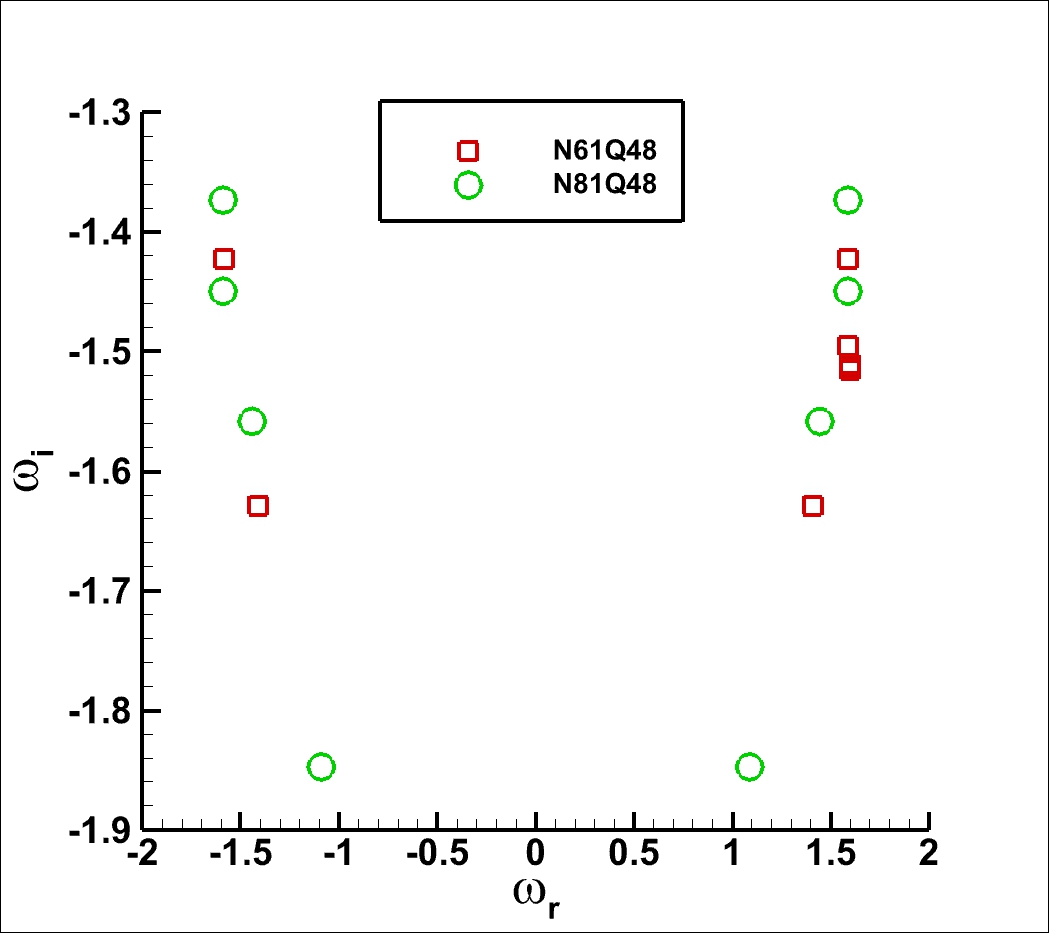}}
\subfigure[$M=4.0, \beta=16$]{\label{fig:M4GridConvBeta16}\includegraphics[trim=10 10 10 10,clip,width=0.32\linewidth]{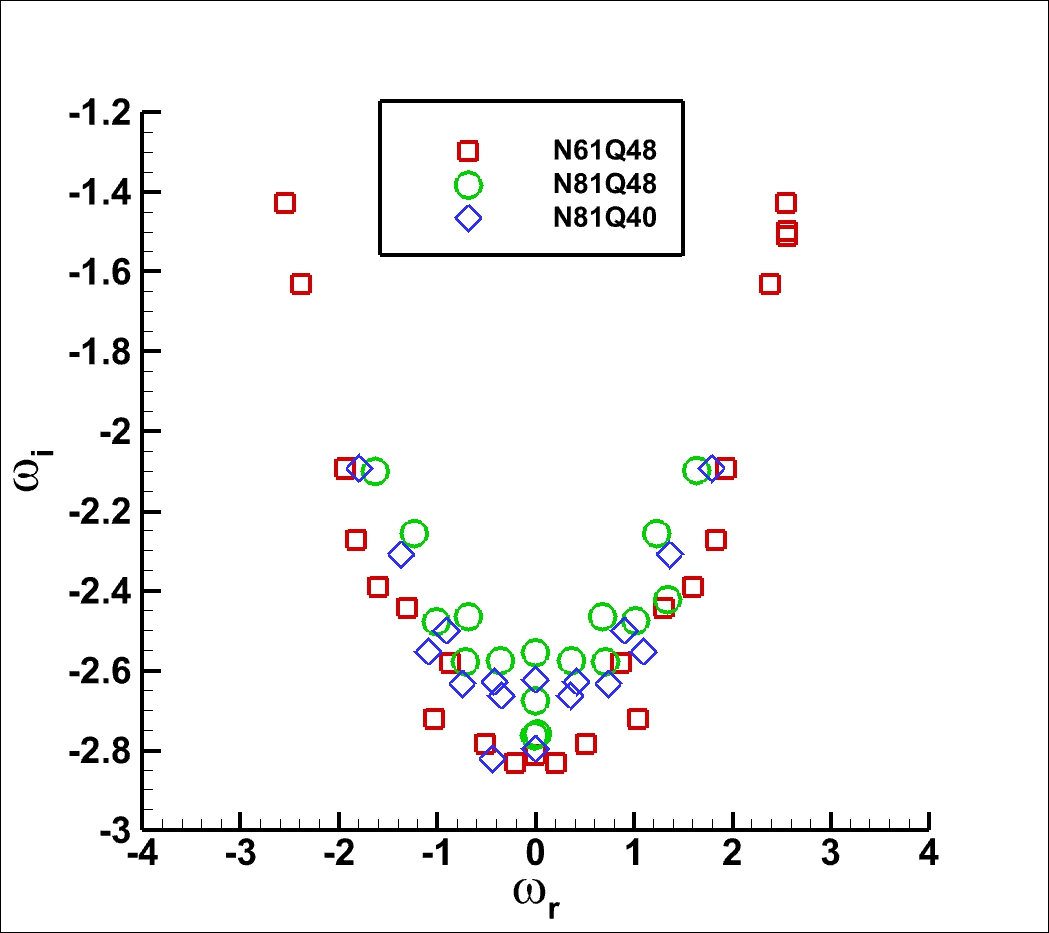}}
\caption{Eigenvalue spectra for a 1D normal shock at $M_\infty=1.2$. $M_\infty=3.0$ and $M_\infty=4.0$ with different spatial (N) and micro-velocity (Q) grid sizes, establishing grid convergence at several $\beta$.}
\label{fig:1DShockEvalGridConvM12}
\end{figure}

To ensure the numerical fidelity of these kLST eigenspectra, Figure~\ref{fig:1DShockEvalGridConvM12} demonstrates the grid independence of the eigenspectra across various spatial ($N$) and micro-velocity ($Q$) resolutions, confirming that the utilized $N=80, Q=20$ grid is sufficient to capture the continuous modal branches accurately for M=1.2 case, $N=80, Q=32$ grid is sufficient to capture the continuous modal branches accurately for M=3.0 case and $N=60, Q=48$ grid is sufficient to capture the continuous modal branches accurately for M=4.0 case. Since the 1D shock linear stability results in continuous eigenmodes, the grid convergence for the linear stability is achived through having modes forming the same continuous branches with varying grid resolution as opposed to having exact same eigenvalues for a discrete eigenspectra. In addition to the representative $\beta=16$ case, the convergence is also shown for $\beta=1$ and $\beta=10$ for the $M_\infty=3.0$ and $M_\infty=4.0$ shocks in Figs.~\ref{fig:M3GridConvBeta1}--\ref{fig:M3GridConvBeta16} and Figs.~\ref{fig:M4GridConvBeta1}--\ref{fig:M4GridConvBeta16}, respectively, demonstrating that the same grid choices preserve the continuous spectral branches across the spanwise wavenumbers used in the main results.

\section*{Appendix C: SLEPc/PETSc Solver Settings for HPC Benchmarks}\label{sec:AppendixSolverSettings}

This appendix lists the complete SLEPc/PETSc option sets used for the single-node and multi-node benchmark studies of Section~\ref{sec:HPCBenchmarks}. All runs request $k=50$ eigenvalues via the command-line option \texttt{-eps\_nev}. Method~1 denotes direct LU factorization with Arnoldi shift-and-invert; Method~2 denotes ILU-preconditioned Jacobi--Davidson. The entries were verified against the \texttt{USER OPTIONS} blocks printed in the corresponding PETSc run logs on Frontera.

\begin{table}[htbp]
\centering
\caption{SLEPc/PETSc solver settings used for the single-node benchmark studies, confirmed from the PETSc run logs. Method~1 denotes direct LU with Arnoldi, while Method~2 denotes ILU-preconditioned Jacobi--Davidson.}
\label{tab:single_node_solver_settings}
\begingroup
\scriptsize
\setlength{\tabcolsep}{2pt}
\begin{tabularx}{\linewidth}{p{4.3cm}p{4.4cm}X}
\toprule
\textbf{Variable name} & \textbf{Definition} & \textbf{Option} \\
\midrule
\multicolumn{3}{l}{\textbf{Common settings}}\\
\texttt{-eps\_nev} & Number of requested eigenvalues & \texttt{50} (command-line option) \\
\texttt{-eps\_ncv} & Dimension of the eigensolver search subspace & \texttt{150} \\
\texttt{-matload\_block\_size} & PETSc binary matrix loading block size & \texttt{1} \\
\texttt{-checkpoint\_every} & Eigenpair checkpoint interval & \texttt{10} eigenvalues \\
\texttt{-bv\_orthog\_type} & SLEPc basis-vector orthogonalization method & \texttt{cgs} \\
\texttt{-eps\_converged\_reason} & Print SLEPc convergence reason & Enabled \\
\midrule
\multicolumn{3}{l}{\textbf{Method~1: direct LU with Arnoldi}}\\
\texttt{-eps\_type} & SLEPc eigensolver type & \texttt{arnoldi} \\
\texttt{-eps\_target} & Target eigenvalue for interior-spectrum extraction & \texttt{0.0+0.0i} \\
\texttt{-st\_type} & SLEPc spectral transformation & \texttt{sinvert} \\
\texttt{-st\_pc\_type} & Preconditioner for the spectral-transformation solve & \texttt{lu} \\
\texttt{-st\_ksp\_type} & KSP solver used inside the spectral transformation & \texttt{preonly} \\
\texttt{-st\_pc\_factor\_mat\_solver\_type} & Direct sparse factorization package & Method~1: \texttt{mumps} \\
\texttt{-st\_mat\_mumps\_icntl\_7} & MUMPS ordering-control parameter & Method~1: \texttt{2} \\
\texttt{-st\_mat\_mumps\_icntl\_22} & MUMPS in-core/out-of-core storage control & Method~1: \texttt{0} (in-core) \\
\texttt{-st\_mat\_mumps\_icntl\_35} & MUMPS BLR-compression control & Method~1: \texttt{0} (disabled) \\
\texttt{-st\_mat\_mumps\_icntl\_14} & MUMPS memory working-space increase & Method~1: \texttt{100}\% \\
\midrule
\multicolumn{3}{l}{\textbf{Method~2: ILU-preconditioned Jacobi--Davidson}}\\
\texttt{-eps\_type} & SLEPc eigensolver type & \texttt{jd} \\
\texttt{-eps\_target} & Target eigenvalue for interior-spectrum extraction & \texttt{1.0e-2+0.0i} \\
\texttt{-eps\_tol} & Eigenpair convergence tolerance & \texttt{1.0e-8} \\
\texttt{-eps\_max\_it} & Maximum outer eigensolver iterations & \texttt{10000} \\
\texttt{-st\_type} & SLEPc spectral transformation & \texttt{precond} \\
\texttt{-st\_shift} & Shift used in the preconditioned JD correction equation & \texttt{1.0e-2} \\
\texttt{-st\_pc\_type} & Preconditioner for the spectral-transformation solve & \texttt{bjacobi} \\
\texttt{-st\_ksp\_type} & KSP solver used inside the spectral transformation & \texttt{gmres} \\
\texttt{-st\_ksp\_rtol}, \texttt{-st\_ksp\_atol} & Relative and absolute tolerances for the inner GMRES solve & \texttt{1.0e-6}, \texttt{1.0e-12} \\
\texttt{-st\_ksp\_max\_it} & Maximum inner GMRES iterations & \texttt{200} \\
\texttt{-st\_sub\_ksp\_type} & Local KSP used inside each block-Jacobi block & Method~2: \texttt{preonly} \\
\texttt{-st\_sub\_pc\_type} & Local preconditioner inside each block-Jacobi block & Method~2: \texttt{ilu} \\
\texttt{-st\_sub\_pc\_factor\_levels} & ILU fill level for each local block & Method~2: \texttt{1} \\
\texttt{-st\_mat\_structure} & PETSc matrix-structure hint for repeated solves & Method~2: \texttt{different} \\
\bottomrule
\end{tabularx}
\endgroup
\end{table}

\begin{table}[htbp]
\centering
\caption{SLEPc/PETSc solver settings used for the multi-node benchmark studies, confirmed from the PETSc run logs. Method~1 denotes direct LU with Arnoldi, while Method~2 denotes ILU-preconditioned Jacobi--Davidson.}
\label{tab:multi_node_solver_settings}
\begingroup
\scriptsize
\setlength{\tabcolsep}{2pt}
\begin{tabularx}{\linewidth}{p{4.3cm}p{4.4cm}X}
\toprule
\textbf{Variable name} & \textbf{Definition} & \textbf{Option} \\
\midrule
\multicolumn{3}{l}{\textbf{Common settings}}\\
\texttt{-eps\_nev} & Number of requested eigenvalues & \texttt{50} (command-line option) \\
\texttt{-eps\_tol} & Eigenpair convergence tolerance & \texttt{1.0e-8} \\
\texttt{-matload\_block\_size} & PETSc binary matrix loading block size & \texttt{1} \\
\texttt{-checkpoint\_every} & Eigenpair checkpoint interval & \texttt{10} eigenvalues \\
\texttt{-bv\_orthog\_type} & SLEPc basis-vector orthogonalization method & \texttt{cgs} \\
\texttt{-eps\_converged\_reason} & Print SLEPc convergence reason & Enabled \\
\midrule
\multicolumn{3}{l}{\textbf{Method~1: direct LU with Arnoldi}}\\
\texttt{-eps\_type} & SLEPc eigensolver type & \texttt{arnoldi} \\
\texttt{-eps\_target} & Target eigenvalue for interior-spectrum extraction & $Q=24$: \texttt{1.0e-2+0.0i}; $Q=32$: \texttt{0.0+0.0i} \\
\texttt{-eps\_ncv} & Dimension of the eigensolver search subspace & \texttt{200} \\
\texttt{-st\_type} & SLEPc spectral transformation & \texttt{sinvert} \\
\texttt{-st\_pc\_type} & Preconditioner for the spectral-transformation solve & \texttt{lu} \\
\texttt{-st\_ksp\_type} & KSP solver used inside the spectral transformation & \texttt{preonly} \\
\texttt{-st\_pc\_factor\_mat\_solver\_type} & Direct sparse factorization package & Method~1: \texttt{mumps} \\
\texttt{-st\_mat\_mumps\_icntl\_7} & MUMPS ordering-control parameter & Method~1: \texttt{5} (METIS) \\
\texttt{-st\_mat\_mumps\_icntl\_22} & MUMPS in-core/out-of-core storage control & Method~1: \texttt{1} (out-of-core enabled) \\
\texttt{-st\_mat\_mumps\_icntl\_35} & MUMPS BLR-compression control & Method~1: \texttt{1} (enabled) \\
\texttt{-st\_mat\_mumps\_cntl\_2} & MUMPS BLR relative tolerance & Method~1: \texttt{1e-4} \\
\texttt{-st\_mat\_mumps\_icntl\_14} & MUMPS memory working-space increase & Method~1 $Q=24$: \texttt{100}\%; Method~1 $Q=32$: \texttt{30}\% \\
\texttt{-st\_mat\_structure} & PETSc matrix-structure hint for repeated solves & \texttt{subset} \\
\midrule
\multicolumn{3}{l}{\textbf{Method~2: ILU-preconditioned Jacobi--Davidson}}\\
\texttt{-eps\_type} & SLEPc eigensolver type & \texttt{jd} \\
\texttt{-eps\_target} & Target eigenvalue for interior-spectrum extraction & \texttt{1.0e-2+0.0i} \\
\texttt{-eps\_max\_it} & Maximum outer eigensolver iterations & \texttt{10000} \\
\texttt{-eps\_ncv} & Dimension of the eigensolver search subspace & \texttt{150} \\
\texttt{-st\_type} & SLEPc spectral transformation & \texttt{precond} \\
\texttt{-st\_shift} & Shift used in the preconditioned JD correction equation & \texttt{1.0e-2} \\
\texttt{-st\_pc\_type} & Preconditioner for the spectral-transformation solve & \texttt{bjacobi} \\
\texttt{-st\_ksp\_type} & KSP solver used inside the spectral transformation & \texttt{gmres} \\
\texttt{-st\_ksp\_rtol}, \texttt{-st\_ksp\_atol} & Relative and absolute tolerances for the inner GMRES solve & \texttt{1.0e-6}, \texttt{1.0e-12} \\
\texttt{-st\_ksp\_max\_it} & Maximum inner GMRES iterations & \texttt{200} \\
\texttt{-st\_sub\_ksp\_type} & Local KSP used inside each block-Jacobi block & Method~2: \texttt{preonly} \\
\texttt{-st\_sub\_pc\_type} & Local preconditioner inside each block-Jacobi block & Method~2: \texttt{ilu} \\
\texttt{-st\_sub\_pc\_factor\_levels} & ILU fill level for each local block & Method~2: \texttt{1} \\
\texttt{-st\_mat\_structure} & PETSc matrix-structure hint for repeated solves & \texttt{different} \\
\bottomrule
\end{tabularx}
\endgroup
\end{table}

\clearpage
\bibliography{MergedRefs}

@BOOK{Bird,
  AUTHOR = {Bird, G. A.},
  ADDRESS = {Oxford, England, U.K.},
  PUBLISHER = {Clarendon},
  TITLE = {Molecular Gas Dynamics and the Direct Simulation of Gas Flows},
  YEAR = {1994},
}

@ARTICLE{sidharth2018onset,
  AUTHOR = {Sidharth, GS and Dwivedi, Anubhav and Candler, Graham V and Nichols, Joseph W},
  PUBLISHER = {APS},
  YEAR = {2018},
  JOURNAL = {Physical Review Fluids},
  NUMBER = {9},
  PAGES = {093901},
  TITLE = {Onset of three-dimensionality in supersonic flow over a slender double wedge},
  VOLUME = {3},
}

@ARTICLE{hao2021occurrence,
  AUTHOR = {Hao, Jiaao and Cao, Shibin and Wen, Chih-Yung and Olivier, Herbert},
  PUBLISHER = {Cambridge University Press},
  YEAR = {2021},
  DOI = {10.1017/jfm.2021.372},
  JOURNAL = {Journal of Fluid Mechanics},
  PAGES = {A4},
  TITLE = {Occurrence of global instability in hypersonic compression corner flow},
  VOLUME = {919},
}

@ARTICLE{sawant_etal_2022,
  AUTHOR = {Sawant, Saurabh S. and Theofilis, V. and Levin, D.A.},
  PUBLISHER = {Cambridge University Press},
  YEAR = {2022},
  DOI = {10.1017/jfm.2022.276},
  JOURNAL = {Journal of Fluid Mechanics},
  PAGES = {A7},
  TITLE = {On the synchronisation of three-dimensional shock layer and laminar separation bubble instabilities in hypersonic flow over a double wedge},
  VOLUME = {941},
}

@TECHREPORT{Edney,
  AUTHOR = {Edney, Barry},
  INSTITUTION = {The Aeronautical Research Institute of Sweden},
  NUMBER = {115},
  URL = {https://www.osti.gov/biblio/4480948},
  YEAR = {1968},
  DOI = {10.2172/4480948},
  TITLE = {Anomalous heat transfer and pressure distributions on blunt bodies at hypersonic speeds in the presence of an impinging shock.},
}

@ARTICLE{sawant_POF,
  AUTHOR = {Sawant, Saurabh S. and Levin, Deborah A. and Theofilis, Vassilios},
  URL = {https://doi.org/10.1063/5.0065971},
  YEAR = {2021-10},
  DOI = {10.1063/5.0065971},
  EPRINT = {https://pubs.aip.org/aip/pof/article-pdf/doi/10.1063/5.0065971/16113325/104106\_1\_online.pdf},
  ISSN = {1070-6631},
  JOURNAL = {Physics of Fluids},
  NUMBER = {10},
  PAGES = {104106},
  TITLE = {A kinetic approach to studying low-frequency molecular fluctuations in a one-dimensional shock},
  VOLUME = {33},
}

@ARTICLE{StabilityBGK,
  AUTHOR = {Zou, Sen and Zhong, Chengwen and Bi, Lin and Yuan, Xianxu and Tang, Zhigong},
  URL = {https://doi.org/10.1063/5.0131135},
  YEAR = {2022-12},
  DOI = {10.1063/5.0131135},
  EPRINT = {https://pubs.aip.org/aip/pof/article-pdf/doi/10.1063/5.0131135/16607850/124114\_1\_online.pdf},
  ISSN = {1070-6631},
  JOURNAL = {Physics of Fluids},
  NUMBER = {12},
  PAGES = {124114},
  TITLE = {A new linear stability analysis approach for microchannel flow based on the Boltzmann Bhatnagar–Gross–Krook equation},
  VOLUME = {34},
}

@ARTICLE{zou_bi_zhong_yuan_tang_2023,
  AUTHOR = {Zou, Sen and Bi, Lin and Zhong, Chengwen and Yuan, Xianxu and Tang, Zhigong},
  PUBLISHER = {Cambridge University Press},
  YEAR = {2023},
  DOI = {10.1017/jfm.2023.230},
  JOURNAL = {Journal of Fluid Mechanics},
  PAGES = {A33},
  TITLE = {A novel linear stability analysis method for plane Couette flow considering rarefaction effects},
  VOLUME = {963},
}

@ARTICLE{HirokiBGK,
  AUTHOR = {Yoshida, Hiroaki and Aoki, Kazuo},
  PUBLISHER = {American Physical Society},
  URL = {https://link.aps.org/doi/10.1103/PhysRevE.73.021201},
  YEAR = {2006-02},
  DOI = {10.1103/PhysRevE.73.021201},
  ISSUE = {2},
  JOURNAL = {Phys. Rev. E},
  PAGES = {021201},
  TITLE = {Linear stability of the cylindrical Couette flow of a rarefied gas},
  VOLUME = {73},
}

@INPROCEEDINGS{chigullapalli2009modeling,
  AUTHOR = {Chigullapalli, Sruti and Venkattraman, Ayyaswamy and Alexeenko, Alina},
  BOOKTITLE = {47th AIAA Aerospace Sciences Meeting including The New Horizons Forum and Aerospace Exposition},
  YEAR = {2009},
  PAGES = {1317},
  TITLE = {Modeling of viscous shock tube using ES-BGK model kinetic equations},
}

@ARTICLE{MaBELSE,
  AUTHOR = {Ma, Qiang and Lv, Jianxin and Bi, Lin},
  URL = {https://www.sciencedirect.com/science/article/pii/S1007570423005580},
  YEAR = {2024},
  DOI = {https://doi.org/10.1016/j.cnsns.2023.107637},
  ISSN = {1007-5704},
  JOURNAL = {Communications in Nonlinear Science and Numerical Simulation},
  PAGES = {107637},
  TITLE = {The equivalence between BE-LSE and NS-LSEs under continuum assumption},
  VOLUME = {128},
}

@INPROCEEDINGS{DuckBalakumarNormalShock,
  AUTHOR = {Duck, P. W. and Balakumar, P.},
  EDITOR = {Hussaini, M. Y. and Kumar, A. and Streett, C. L.},
  ADDRESS = {New York, NY},
  PUBLISHER = {Springer New York},
  BOOKTITLE = {Instability, Transition, and Turbulence},
  YEAR = {1992},
  ISBN = {978-1-4612-2956-8},
  PAGES = {253--265},
  TITLE = {On the Stability of Normal Shock Waves},
}

@ARTICLE{NMKuznetsov_1989,
  AUTHOR = {Kuznetsov, N M},
  URL = {https://dx.doi.org/10.1070/PU1989v032n11ABEH002777},
  YEAR = {1989-11},
  DOI = {10.1070/PU1989v032n11ABEH002777},
  JOURNAL = {Soviet Physics Uspekhi},
  NUMBER = {11},
  PAGES = {993},
  TITLE = {Stability of shock waves},
  VOLUME = {32},
}

@ARTICLE{SwanFowlesShock,
  AUTHOR = {Swan, G. W. and Fowles, G. R.},
  URL = {https://doi.org/10.1063/1.860989},
  YEAR = {1975-01},
  DOI = {10.1063/1.860989},
  EPRINT = {https://pubs.aip.org/aip/pfl/article-pdf/18/1/28/12705833/28\_1\_online.pdf},
  ISSN = {0031-9171},
  JOURNAL = {The Physics of Fluids},
  NUMBER = {1},
  PAGES = {28--35},
  TITLE = {Shock wave stability},
  VOLUME = {18},
}

@ARTICLE{SwanFowlesShock2,
  AUTHOR = {Fowles, G. R. and Swan, G. W.},
  PUBLISHER = {American Physical Society},
  URL = {https://link.aps.org/doi/10.1103/PhysRevLett.30.1023},
  YEAR = {1973-05},
  DOI = {10.1103/PhysRevLett.30.1023},
  ISSUE = {21},
  JOURNAL = {Phys. Rev. Lett.},
  PAGES = {1023--1025},
  TITLE = {Stability of Plane Shock Waves},
  VOLUME = {30},
}

@ARTICLE{Erpenbeck1,
  AUTHOR = {Erpenbeck, Jerome J.},
  URL = {https://doi.org/10.1063/1.1706503},
  YEAR = {1962-10},
  DOI = {10.1063/1.1706503},
  EPRINT = {https://pubs.aip.org/aip/pfl/article-pdf/5/10/1181/12448903/1181\_1\_online.pdf},
  ISSN = {0031-9171},
  JOURNAL = {The Physics of Fluids},
  NUMBER = {10},
  PAGES = {1181--1187},
  TITLE = {Stability of Step Shocks},
  VOLUME = {5},
}

@ARTICLE{Gardner,
  AUTHOR = {Gardner, C. S.},
  URL = {https://doi.org/10.1063/1.1706917},
  YEAR = {1963-09},
  DOI = {10.1063/1.1706917},
  EPRINT = {https://pubs.aip.org/aip/pfl/article-pdf/6/9/1366/12284619/1366\_1\_online.pdf},
  ISSN = {0031-9171},
  JOURNAL = {The Physics of Fluids},
  NUMBER = {9},
  PAGES = {1366--1367},
  TITLE = {Comment on ``Stability of Step Shocks''},
  VOLUME = {6},
}

@ARTICLE{Erpenbeck2,
  AUTHOR = {Erpenbeck, Jerome J.},
  URL = {https://doi.org/10.1063/1.1706918},
  YEAR = {1963-09},
  DOI = {10.1063/1.1706918},
  EPRINT = {https://pubs.aip.org/aip/pfl/article-pdf/6/9/1368/12284617/1368\_1\_online.pdf},
  ISSN = {0031-9171},
  JOURNAL = {The Physics of Fluids},
  NUMBER = {9},
  PAGES = {1368--1368},
  TITLE = {Reply to Comments by Gardner},
  VOLUME = {6},
}

@PHDTHESIS{sawantPhDThesis,
  AUTHOR = {Sawant, Saurabh Satish},
  SCHOOL = {University of Illinois, Urbana Champaign},
  YEAR = {2022},
  TITLE = {The development of kinetic models and simulation methods to study molecular fluctuations, modal response, shock-laminar separation bubble instabilities},
}

@ARTICLE{BGK,
  AUTHOR = {Bhatnagar, P. L. and Gross, E. P. and Krook, M.},
  PUBLISHER = {American Physical Society},
  URL = {https://link.aps.org/doi/10.1103/PhysRev.94.511},
  YEAR = {1954-05},
  DOI = {10.1103/PhysRev.94.511},
  ISSUE = {3},
  JOURNAL = {Phys. Rev.},
  PAGES = {511--525},
  TITLE = {A Model for Collision Processes in Gases. I. Small Amplitude Processes in Charged and Neutral One-Component Systems},
  VOLUME = {94},
}

@ARTICLE{QZ,
  AUTHOR = {Moler, C. B. and Stewart, G. W.},
  PUBLISHER = {Society for Industrial and Applied Mathematics},
  URL = {http://www.jstor.org/stable/2156353},
  YEAR = {1973},
  ISSN = {00361429},
  JOURNAL = {SIAM Journal on Numerical Analysis},
  NUMBER = {2},
  PAGES = {241--256},
  TITLE = {An Algorithm for Generalized Matrix Eigenvalue Problems},
  URLYEAR = {2024-03-23},
  VOLUME = {10},
}

@BOOK{gottlieb1977numerical,
  AUTHOR = {Gottlieb, David and Orszag, Steven A},
  PUBLISHER = {SIAM},
  YEAR = {1977},
  TITLE = {Numerical analysis of spectral methods: theory and applications},
}

@BOOK{canutospectral,
  AUTHOR = {Canuto, Claudio and Hussaini, M Yousuff and Quarteroni, Alfio and Zang, Thomas A and others},
  PUBLISHER = {Berlin, Heidelberg: Springer Berlin Heidelberg,},
  YEAR = {1988},
  TITLE = {Spectral Methods in Fluid Dynamics [electronic resource]},
}

@BOOK{cheb1853,
  AUTHOR = {Chebyshev, P.L.},
  PUBLISHER = {Imprimerie de l'Académie impériale des sciences},
  URL = {https://books.google.com.tr/books?id=o7AwkgAACAAJ},
  YEAR = {1853},
  NUMBER = {1. c.},
  TITLE = {Théorie des mécanismes connus sous le nom de parallélogrammes},
}

@ARTICLE{TheofilisLecture2014,
  AUTHOR = {Juniper, Matthew P. and Hanifi, Ardeshir and Theofilis, Vassilios},
  URL = {https://doi.org/10.1115/1.4026604},
  YEAR = {2014-03},
  DOI = {10.1115/1.4026604},
  EPRINT = {https://asmedigitalcollection.asme.org/appliedmechanicsreviews/article-pdf/66/2/024804/6073990/amr\_066\_02\_024804.pdf},
  ISSN = {0003-6900},
  JOURNAL = {Applied Mechanics Reviews},
  NUMBER = {2},
  PAGES = {024804},
  TITLE = {Modal Stability Theory: Lecture notes from the FLOW-NORDITA Summer School on Advanced Instability Methods for Complex Flows, Stockholm, Sweden, 2013},
  VOLUME = {66},
}

@ARTICLE{Giddens1971,
  AUTHOR = {Giddens, D. P. and Huang, A. B. and Young, V. Y. C.},
  URL = {https://doi.org/10.1063/1.1693387},
  YEAR = {1971-12},
  DOI = {10.1063/1.1693387},
  EPRINT = {https://pubs.aip.org/aip/pfl/article-pdf/14/12/2645/12700805/2645\_1\_online.pdf},
  ISSN = {0031-9171},
  JOURNAL = {The Physics of Fluids},
  NUMBER = {12},
  PAGES = {2645--2651},
  TITLE = {Evaluation of Two Statistical Models Using the Shock Structure Problem},
  VOLUME = {14},
}

@BOOK{MatlabSpectral,
  AUTHOR = {Trefethen, Lloyd N.},
  PUBLISHER = {Society for Industrial and Applied Mathematics},
  URL = {https://epubs.siam.org/doi/abs/10.1137/1.9780898719598},
  YEAR = {2000},
  DOI = {10.1137/1.9780898719598},
  EPRINT = {https://epubs.siam.org/doi/pdf/10.1137/1.9780898719598},
  TITLE = {Spectral Methods in MATLAB},
}

@ARTICLE{ZouetalCompCouetteBGK2024,
  AUTHOR = {Zou, Sen and Bi, Lin and Zhong, Chengwen and Yuan, Xianxu and Tang, Zhigong},
  URL = {https://doi.org/10.1063/5.0187318},
  YEAR = {2024-01},
  DOI = {10.1063/5.0187318},
  EPRINT = {https://pubs.aip.org/aip/pof/article-pdf/doi/10.1063/5.0187318/18703596/014121\_1\_5.0187318.pdf},
  ISSN = {1070-6631},
  JOURNAL = {Physics of Fluids},
  NUMBER = {1},
  PAGES = {014121},
  TITLE = {A normal-mode approach for high-speed rarefied plane Couette flow},
  VOLUME = {36},
}

@ARTICLE{Arnoldi1951,
  AUTHOR = {Arnoldi, W. E.},
  PUBLISHER = {Brown University},
  URL = {http://www.jstor.org/stable/43633863},
  YEAR = {1951},
  ISSN = {0033569X, 15524485},
  JOURNAL = {Quarterly of Applied Mathematics},
  NUMBER = {1},
  PAGES = {17--29},
  TITLE = {The principle of minimized iterations in the solution of the matrix eigenvalue problem},
  URLYEAR = {2024-12-03},
  VOLUME = {9},
}

@ARTICLE{Lanczos1952,
  AUTHOR = {Lanczos, Cornelius},
  URL = {https://api.semanticscholar.org/CorpusID:7484650},
  YEAR = {1952},
  JOURNAL = {Journal of research of the National Bureau of Standards},
  PAGES = {33},
  TITLE = {Solution of Systems of Linear Equations by Minimized Iterations1},
  VOLUME = {49},
}

@INCOLLECTION{Sorensen1997,
  AUTHOR = {Sorensen, Danny C.},
  EDITOR = {Keyes, David E. and Sameh, Ahmed and Venkatakrishnan, V.},
  ADDRESS = {Dordrecht},
  PUBLISHER = {Springer Netherlands},
  URL = {https://doi.org/10.1007/978-94-011-5412-3_5},
  BOOKTITLE = {Parallel Numerical Algorithms},
  YEAR = {1997},
  DOI = {10.1007/978-94-011-5412-3_5},
  ISBN = {978-94-011-5412-3},
  PAGES = {119--165},
  TITLE = {Implicitly Restarted Arnoldi/Lanczos Methods for Large Scale Eigenvalue Calculations},
}

@PHDTHESIS{RadkePhD1996,
  AUTHOR = {Radke, R.J.},
  SCHOOL = {Rice University},
  YEAR = {1996},
  TITLE = {A Matlab implementation of the Implicitly Restarted Arnoldi Method for solving large-scale eigenvalue problems},
}

@ARTICLE{GILBARGPAOLUCCI1953,
  AUTHOR = {Gilbarg, D. and Paolucci, D.},
  PUBLISHER = {Indiana University Mathematics Department},
  URL = {http://www.jstor.org/stable/24900350},
  YEAR = {1953},
  ISSN = {19435282, 19435290},
  JOURNAL = {Journal of Rational Mechanics and Analysis},
  PAGES = {617--642},
  TITLE = {The Structure of Shock Waves in the Continuum Theory of Fluids},
  URLYEAR = {2025-04-14},
  VOLUME = {2},
}

@INPROCEEDINGS{KarpuzcukLSTScitech2025,
  AUTHOR = {Karpuzcu, Irmak Taylan and Levin, Deborah},
  BOOKTITLE = {AIAA SCITECH 2025 Forum},
  PUBLISHER = {American Institute of Aeronautics and Astronautics},
  YEAR = {2025},
  DOI = {10.2514/6.2025-1313},
  PAGES = {AIAA Paper 2025--1313},
  TITLE = {Linear Stability Analysis of Canonical Flows Using Boltzmann BGK Equation},
}

@ARTICLE{karpuzcu2024linear,
  AUTHOR = {Karpuzcu, Irmak T. and Theofilis, Vassilis and Levin, Deborah A.},
  YEAR = {2024},
  DOI = {10.48550/arXiv.2405.06775},
  EPRINT = {2405.06775},
  EPRINTTYPE = {arxiv},
  JOURNAL = {arXiv preprint},
  TITLE = {On linear stability of supersonic flow over a short compression corner at large ramp angles},
}

@BOOK{boyd2013chebyshev,
  AUTHOR = {Boyd, J.P.},
  PUBLISHER = {Dover Publications},
  URL = {https://books.google.com/books?id=b4TCAgAAQBAJ},
  YEAR = {2013},
  ISBN = {9780486141923},
  SERIES = {Dover Books on Mathematics},
  TITLE = {Chebyshev and Fourier Spectral Methods: Second Revised Edition},
}

@ARTICLE{slepc,
  AUTHOR = {Hernandez, V. and Roman, J. E. and Vidal, V.},
  YEAR = {2005},
  JOURNAL = {ACM Transactions on Mathematical Software},
  NUMBER = {3},
  PAGES = {351--362},
  TITLE = {{SLEPc}: A Scalable and Flexible Toolkit for the Solution of Eigenvalue Problems},
  VOLUME = {31},
}

@TECHREPORT{petsc,
  AUTHOR = {Balay, Satish and Abhyankar, Shrirang and Adams, Mark F. and Benson, Steven and Brown, Jed and Brune, Peter and Buschelman, Kris and Constantinescu, Emil M. and Dalcin, Lisandro and Dener, Alp and Eijkhout, Victor and Faibussowitsch, Jacob and Gropp, William D. and Hapla, Václav and Isaac, Tobin and Jolivet, Pierre and Karpeev, Dmitry and Kaushik, Dinesh and Knepley, Matthew G. and Kong, Fande and Kruger, Scott and Arabinda, Arabinda and Zhang, Junchao and Zhang, Hong and Zhang, Hong},
  INSTITUTION = {Argonne National Laboratory},
  YEAR = {2023},
  NUMBER = {ANL-21/39 - Revision 3.20},
  TITLE = {{PETSc/TAO} Users Manual},
  TYPE = {techreport},
}

@ARTICLE{mumps,
  AUTHOR = {Amestoy, P. R. and Duff, I. S. and Koster, J. and L'Excellent, J.-Y.},
  YEAR = {2001},
  JOURNAL = {SIAM Journal on Matrix Analysis and Applications},
  NUMBER = {1},
  PAGES = {15--41},
  TITLE = {A Fully Asynchronous Multifrontal Solver Using Distributed Dynamic Scheduling},
  VOLUME = {23},
}

@ARTICLE{metis,
  AUTHOR = {Karypis, George and Kumar, Vipin},
  YEAR = {1998},
  JOURNAL = {SIAM Journal on Scientific Computing},
  NUMBER = {1},
  PAGES = {359--392},
  TITLE = {A Fast and High Quality Multilevel Scheme for Partitioning Irregular Graphs},
  VOLUME = {20},
}

@MISC{Frontera,
  HOWPUBLISHED = {\url{https://www.tacc.utexas.edu/systems/frontera}},
  NOTE = {Accessed: 2022-08-26},
  TITLE = {Frontera Supercomputer},
  YEAR = {2022},
}

@article{Alsmeyer1976,
  author  = {Alsmeyer, H.},
  title   = {Density profiles in argon and nitrogen shock waves measured by the absorption of an electron beam},
  journal = {Journal of Fluid Mechanics},
  year    = {1976},
  volume  = {74},
  number  = {3},
  pages   = {497--513},
  doi     = {10.1017/S0022112076001912}
}

@article{DiPernaLions1989,
  author  = {DiPerna, R. J. and Lions, P. L.},
  title   = {On the {C}auchy Problem for {B}oltzmann Equations: Global Existence and Weak Stability},
  journal = {Annals of Mathematics},
  year    = {1989},
  volume  = {130},
  number  = {2},
  pages   = {321--366},
  doi     = {10.2307/1971423}
}

@article{Guo2003VMB,
  author  = {Guo, Yan},
  title   = {The {V}lasov--{M}axwell--{B}oltzmann System Near {M}axwellians},
  journal = {Inventiones Mathematicae},
  year    = {2003},
  volume  = {153},
  number  = {3},
  pages   = {593--630},
  doi     = {10.1007/s00222-003-0301-z}
}

@article{SaadSchultz1986,
  author    = {Youcef Saad and Martin H. Schultz},
  title     = {{GMRES}: A Generalized Minimal Residual Algorithm for Solving Nonsymmetric Linear Systems},
  journal   = {SIAM Journal on Scientific and Statistical Computing},
  volume    = {7},
  number    = {3},
  pages     = {856--869},
  year      = {1986},
  doi       = {10.1137/0907058}
}

@article{Schmidt1969,
  author  = {Schmidt, B.},
  title   = {Electron beam density measurements in shock waves in argon},
  journal = {Journal of Fluid Mechanics},
  year    = {1969},
  volume  = {39},
  number  = {2},
  pages   = {361--373},
  doi     = {10.1017/S0022112069002229}
}

@article{SleijpenVanDerVorst1996,
  author    = {Gerard L. G. Sleijpen and Henk A. Van der Vorst},
  title     = {A {Jacobi--Davidson} Iteration Method for Linear Eigenvalue Problems},
  journal   = {SIAM Journal on Matrix Analysis and Applications},
  volume    = {17},
  number    = {2},
  pages     = {401--425},
  year      = {1996},
  doi       = {10.1137/S0895479894270427}
}

@article{Ukai1974,
  author  = {Ukai, Seiji},
  title   = {On the Existence of Global Solutions of Mixed Problem for Non-linear {B}oltzmann Equation},
  journal = {Proceedings of the Japan Academy},
  year    = {1974},
  volume  = {50},
  number  = {3},
  pages   = {179--184},
  doi     = {10.3792/pja/1195519027}
}

@article{vanLeer1979MUSCL,
  author    = {Bram van Leer},
  title     = {Towards the Ultimate Conservative Difference Scheme. {V}. {A} Second-Order Sequel to {G}odunov's Method},
  journal   = {Journal of Computational Physics},
  volume    = {32},
  number    = {1},
  pages     = {101--136},
  year      = {1979},
  doi       = {10.1016/0021-9991(79)90145-1}
}

@article{padovan2025resolvent4py,
  author   = {Padovan, Alberto and Anantharaman, Vishal and Rowley, Clarence W. and Vollmer, Blaine and Colonius, Tim and Bodony, Daniel J.},
  title    = {Resolvent4py: A parallel Python package for analysis, model reduction and control of large-scale linear systems},
  journal  = {SoftwareX},
  volume   = {31},
  pages    = {102286},
  year     = {2025},
  doi      = {10.1016/j.softx.2025.102286}
}

@article{campos2021nep,
  author   = {Campos, Carmen and Roman, Jose E.},
  title    = {NEP: A Module for the Parallel Solution of Nonlinear Eigenvalue Problems in SLEPC},
  journal  = {ACM Transactions on Mathematical Software (TOMS)},
  volume   = {47},
  number   = {3},
  articleno= {23},
  pages    = {1--29},
  year     = {2021},
  doi      = {10.1145/3447544}
}

@article{campos2016parallel,
  author   = {Campos, Carmen and Roman, Jose E.},
  title    = {Parallel Krylov Solvers for the Polynomial Eigenvalue Problem in SLEPC},
  journal  = {SIAM Journal on Scientific Computing},
  volume   = {38},
  number   = {5},
  pages    = {S385--S411},
  year     = {2016},
  doi      = {10.1137/15M1022458}
}

@article{romero2014parallel,
  author   = {Romero, Eloy and Roman, Jose E.},
  title    = {A Parallel Implementation of Davidson Methods for Large-Scale Eigenvalue Problems in SLEPC},
  journal  = {ACM Transactions on Mathematical Software (TOMS)},
  volume   = {40},
  number   = {2},
  articleno= {13},
  pages    = {1--29},
  year     = {2014},
  doi      = {10.1145/2543696}
}

@techreport{hernandez2009survey,
  author      = {Hern{\'a}ndez, Vicente and Rom{\'a}n, Jose E. and Tom{\'a}s, Andr{\'e}s and Vidal, Vicente},
  title       = {A Survey of Software for Sparse Eigenvalue Problems},
  institution = {Universidad Polit{\'e}cnica de Valencia},
  type        = {SLEPc Technical Report},
  number      = {STR-6},
  year        = {2009},
  url         = {http://www.grycap.upv.es/slepc}
}

@article{tzounas2020comparison,
  author   = {Tzounas, Georgios and Dassios, Ioannis and Liu, Muyang and Milano, Federico},
  title    = {Comparison of Numerical Methods and Open-Source Libraries for Eigenvalue Analysis of Large-Scale Power Systems},
  journal  = {Applied Sciences},
  volume   = {10},
  number   = {21},
  pages    = {7592},
  year     = {2020},
  doi      = {10.3390/app10217592}
}

@inproceedings{ramalingam2012improving,
  author    = {Ramalingam, Shreyas and Hall, Mary and Chen, Chun},
  title     = {Improving High-Performance Sparse Libraries Using Compiler-Assisted Specialization: A PETSc Case Study},
  booktitle = {2012 IEEE 26th International Parallel and Distributed Processing Symposium Workshops \& PhD Forum},
  pages     = {487--496},
  year      = {2012},
  doi       = {10.1109/IPDPSW.2012.63}
}

@article{jezequel2012solving,
  author   = {Jezequel, Fabienne and Couturier, Rapha{\"e}l and Denis, Christophe},
  title    = {Solving large sparse linear systems in a grid environment: the GREMLINS code versus the PETSc library},
  journal  = {The Journal of Supercomputing},
  volume   = {59},
  number   = {3},
  pages    = {1517--1532},
  year     = {2012},
  doi      = {10.1007/s11227-011-0563-y}
}

@article{shivaraj2025stability,
  author   = {Shivaraj, D. L.},
  title    = {Stability of Natural Convection in a Vertical Porous Layer of Viscoelastic Navier-Stokes-Voigt Fluid},
  journal  = {ASME Journal of Heat and Mass Transfer},
  volume   = {147},
  number   = {10},
  pages    = {102601},
  year     = {2025},
  doi      = {10.1115/1.4069013}
}

@article{ThiranietalJTHT2025,
author = {Thirani, Shubham and Karpuzcu, Irmak T. and Levin, Deborah A.},
title = {Modeling of Nitric Oxide Vibrational Level Populations for High Mach Number Flows},
journal = {Journal of Thermophysics and Heat Transfer},
volume = {39},
number = {4},
pages = {835-849},
year = {2025},
doi = {10.2514/1.T7111}
}

\end{document}